\documentclass[journal,onecolumn]{IEEEtran}
\usepackage{epsfig,amssymb,psfrag,latexsym}
\usepackage{amsmath}
\usepackage{booktabs}
\usepackage{longtable}

\usepackage{multirow}
\usepackage{lineno}
\usepackage{color}
\usepackage{nomencl}
\usepackage{nomentbl}
\usepackage[normalem]{ulem}
\usepackage{booktabs}
\usepackage{rotating}
\usepackage{ulem}
\usepackage{enumerate}
\usepackage{cite}

\usepackage{algorithm}
\usepackage{algpseudocode}
\usepackage{graphicx}
\usepackage{subfigure}

\usepackage{epsfig}
\usepackage{array}

\usepackage{bbding}

\newcommand{\be}{\begin{equation}}
\newcommand{\ee}{\end{equation}}
\newcommand{\bea}{\begin{eqnarray}}
\newcommand{\eea}{\end{eqnarray}}

\def\QEDclosed{\mbox{\rule[0pt]{1.3ex}{1.3ex}}}
\def\QED{\QEDclosed}
\def\endproof{\hspace*{\fill}~\QED\par\endtrivlist\unskip}

\newtheorem{theorem}{Theorem}
\newtheorem{remark}{Remark}

\ifCLASSINFOpdf
\else
\fi
\begin{document}
%
\title{Secure Control of Networked Inverted Pendulum Visual Servo System with Adverse Effects of Image Computation (Extended Version)}
%
%
%

\author{Dajun Du,
        Changda Zhang,
        Qianjiang Lu,
        Minrui Fei,
        and Huiyu Zhou
\thanks{The work of Dajun Du, Changda Zhang, Qianjiang Lu and Minrui Fei was supported in part by the National Science Foundation of China under Grant 92067106, Grant 61633016, Grant 61773253, Grant 61803252, and Grant 61833011; in part by the 111 Project under Grant D18003; and in part by the Project of Science and Technology Commission of Shanghai Municipality under Grant 20JC1414000, Grant 19500712300, Grant 19510750300, and Grant 21190780300.
}
\thanks{Dajun Du, Changda Zhang, Qianjiang Lu and Minrui Fei are with Shanghai Key Laboratory of Power Station Automation Technology,
School of Mechatronic Engineering and Automation, Shanghai University, Shanghai 200444, China (e-mail: ddj@i.shu.edu.cn; lqj18490@shu.edu.cn; changdazhang@shu.edu.cn; mrfei@staff.shu.edu.cn).}
\thanks{Huiyu Zhou is with School of Computing and Mathematical Sciences, University of Leicester, Leicester LE1 7RH, U.K (e-mail: hz143@leicester.ac.uk).}
}

%
%

\markboth{}
{Shell \MakeLowercase{\textit{et al.}}: Bare Demo of IEEEtran.cls for IEEE Journals}
%



\maketitle

\begin{abstract}
When visual image information is transmitted via communication networks, it easily suffers from image attacks, leading to system performance degradation or even crash. This paper investigates secure control of networked inverted pendulum visual servo system (NIPVSS) with adverse effects of image computation. Firstly, the image security limitation of the traditional NIPVSS is revealed, where its stability will be destroyed by eavesdropping-based image attacks. Then, a new NIPVSS with the fast scaled-selective image encryption (F2SIE) algorithm is proposed, which not only meets the real-time requirement by reducing the computational complexity, but also improve the security by reducing the probability of valuable information being compromised by eavesdropping-based image attacks. Secondly, adverse effects of the F2SIE algorithm and image attacks are analysed, which will produce extra computational delay and errors. Then, a closed-loop uncertain time-delay model of the new NIPVSS is established, and a robust controller is designed to guarantee system asymptotic stability. Finally, experimental results of the new NIPVSS demonstrate the feasibility and effectiveness of the proposed method.
\end{abstract}

\IEEEpeerreviewmaketitle

\section{Introduction}
Networked visual servo control \cite{NVSS1} has been a class of new control system technology along with rapid development of visual sensors (e.g., camera \cite{CAMERA} and radar \cite{RADAR}) and communication network. It has been gradually employed in some innovative scenarios such as autonomous vehicles, mobile robots, unmanned helicopter and intelligent manufacturing. However, it also brings some new problems (e.g., long image processing time, non-ignorable image computational error and potential image attacks), which will lead to system performance degradation or even crash. To solve these new problems, some new control methods and technologies of networked visual servo control needs to be developed, which also needs further be validated on proper experimental platforms. The traditional inverted pendulum system \cite{1,3} is a typical experimental platform, but it cannot directly work for validation. It must be reformed, which is reconstructed as networked inverted pendulum visual servo system (NIPVSS) \cite{7,8}.

Communication networks are introduced into NIPVSS, where the system becomes open to the public. It will inevitably suffer from cyber attacks \cite{9} such as false data injection attacks and denial-of-service attacks, which causes the system to be deteriorate or crash. The current research mainly focuses on non-visual information under cyber attacks. However, when visual image information is transmitted via communication networks, it suffers from eavesdropping-based image attacks such as salt and pepper attack, shearing attack, Gaussian attack. These images attacks tamper with the information after eavesdropping the images, which will further make the receiver be unable to obtain the correct and integral image information, leading to system performance degradation and even instability. Thus, how to guarantee the image security during network transmission is a critical issue. To achieve the image security, the most popular method is chaotic image encryption. For example, an image encryption method is proposed by combining dynamic DNA sequence encryption and hyper-chaotic maps \cite{12}; A colour image encryption algorithm with higher security based on the chaotic system is proposed to ensure safe transmission of image information\cite{13}. These chaotic image encryption methods are mainly employed to encrypt non-real-time images, which do not consider real-time requirements in industrial control applications.

Image attacks will bring extra computational errors, while the image encryption methods will produce extra computational delay, further declining system performance. However, the most existing NIPVSS do not consider the image attacks and not consider computational delay or errors either \cite{15}. But there exists several studies that begin to consider the impact of errors on the inverted pendulum system (IPS). For example, the influence of pendulum angle errors on the IPS is considered \cite{16} and the errors are taken as an energy finite disturbance and ${H_\infty }$ controller is designed to suppress this disturbance \cite{8}. Besides the error problems, the delay related to the networked control system has been reported. For instance, a new controller based on sliding mode estimation is designed to handle the time delay system with different input matrices \cite{20}; Based on a new event triggering mechanism, T-S fuzzy event triggering control is employed to support the distributed delay system \cite{21}; Stabilization of time-delay systems under delay-dependent impulse control is studied \cite{22}.

\begin{figure}[!t]
     \centering
       \includegraphics[width=0.485\textwidth]{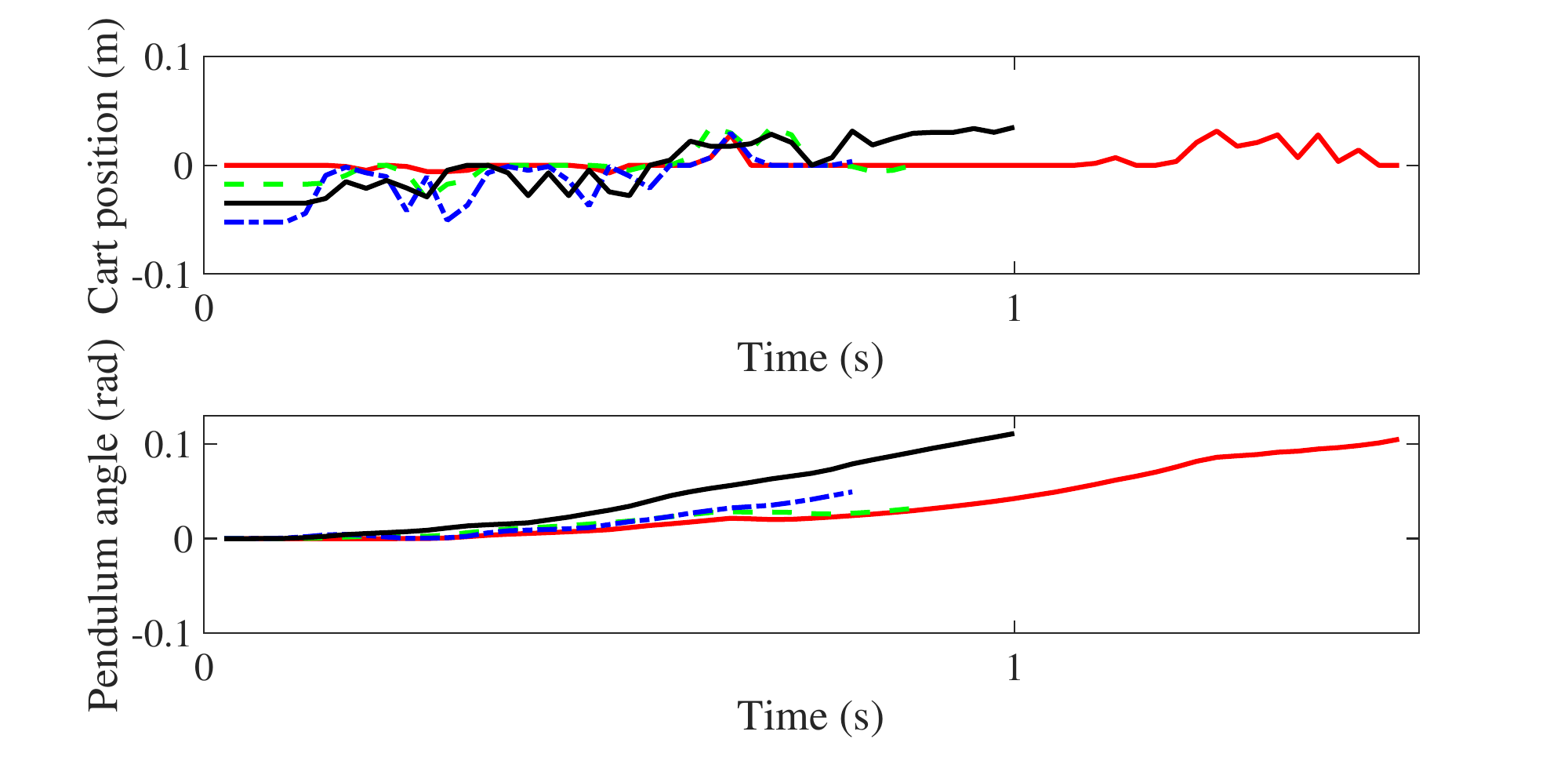}
     \caption{Cart position and pendulum angle of traditional NIPVSS under shearing attacks with different shearing rates: \textcolor[rgb]{1,0,0}{---}, 1$\%$ shearing rate; \textcolor[rgb]{0,1,0}{- -}, 2$\%$ shearing rate; \textcolor[rgb]{0,0,1}{$\cdot  \cdot  \cdot  \cdot  \cdot $}, 4$\%$ shearing rate;  \textcolor[rgb]{0,0,0}{---},  5$\%$ shearing rate.}
     \label{figA1}
\end{figure}

Furthermore, the image attacks (e.g., shearing attacks) are operated for the most existing NIPVSS, experimental results of Fig.~\ref{figA1} shows that traditional NIPVSS runs for a short time and then collapses. This is because some areas of images are cut off intentionally so that accurate system state cannot be obtained. Therefore, based on these observations, we have analysed two challenge problems that need to be addressed:
\begin{enumerate}
\item
What is the image security limitation of the traditional NIPVSS? How to establish a new NIPVSS with image encryption to not only meet the real-time requirement but also guarantee image security?
\item
What are adverse effects of image information security? How to build a closed-loop model under the adverse effects and design a robust controller with strong system stability to tolerate computational errors and delays?
\end{enumerate}

To deal with these challenges, this paper investigates secure control of NIPVSS with adverse effects of image computation. Comparative analysis between contributions of this paper and those of existing image encryption and control methods is shown in Tab.~\ref{tabcom}. Some existing advanced image encryption algorithms mainly solve security problems of still images but usually have high computational complexity, and some other existing references on NIPVSS mainly provide controller design methods and stability criteria under network and image processing computation constraints but cannot consider image attacks. In this paper, we have revealed image security limitation of traditional NIPVSS, proposed a fast image encryption algorithm to meet real-time requirement, established a new model including extra computational delay and errors produced by image encryption and image attacks, and designed a new robust controller to guarantee system stability. The main contributions of this paper include:
\begin{enumerate}
\item[1)]
The image security limitation of traditional NIPVSS is revealed, where its stability will be destroyed by eavesdropping-based image attacks. To overcome the limitation, a new NIPVSS with a fast scale-selective image encryption (F2SIE) algorithm is proposed, which not only meets real-time requirement by reducing computational complexity, but also improves security of new NIPVSS by reducing probability of valuable information being compromised by eavesdropping-based image attacks.
\item[2)]
Adverse effects of the F2SIE algorithm and image attacks are analysed, which will produce extra computational delay and extra computational errors. A closed-loop uncertain time-delay model of the new NIPVSS is then established, and a robust controller is designed to guarantee system asymptotic stability.
\end{enumerate}
\begin{table}[!t]
\centering
\caption{Comparative Analysis between This Paper and Existing References}
\label{tabcom}
\begin{tabular}{lccccccc}
\toprule
References & IAE$^1$ & RR$^2$ & CD$^3$ & CE$^4$ & ND$^5$ & ECD$^6$ & ECE$^7$\\
\midrule
\cite{12,13} & \CheckmarkBold  & \XSolidBrush & \XSolidBrush& \XSolidBrush &\XSolidBrush & \CheckmarkBold&\XSolidBrush\\
\cite{15} & \XSolidBrush & \XSolidBrush & \XSolidBrush & \XSolidBrush & \XSolidBrush & \XSolidBrush & \XSolidBrush\\
\cite{16} & \XSolidBrush & \XSolidBrush & \XSolidBrush &\CheckmarkBold & \XSolidBrush & \XSolidBrush & \XSolidBrush\\
\cite{8}  & \XSolidBrush & \XSolidBrush & \CheckmarkBold & \CheckmarkBold & \CheckmarkBold & \XSolidBrush & \XSolidBrush\\
\cite{20,21,22}& \XSolidBrush & \XSolidBrush & \XSolidBrush & \XSolidBrush & \CheckmarkBold & \XSolidBrush & \XSolidBrush\\
This paper & \CheckmarkBold & \CheckmarkBold & \CheckmarkBold & \CheckmarkBold & \CheckmarkBold & \CheckmarkBold & \CheckmarkBold\\
\bottomrule
\multicolumn{8}{l}{$^1$Image attack and image encryption. $^2$Real-time requirement for image}\\
\multicolumn{8}{l}{encryption.$^3$Computational delay from image processing.}\\
\multicolumn{8}{l}{$^4$Computational error from image processing. $^5$Network-induced delay}\\
\multicolumn{8}{l}{from network transmission. $^6$Extra computational delay from image}\\
\multicolumn{8}{l}{encryption. $^7$Extra computational error from image attack.}
\end{tabular}
\end{table}

The remainder of this paper is organized as follows. In Section \uppercase\expandafter{\romannumeral2}, the image security limitation of traditional NIPVSS is firstly analysed, and new NIPVSS with an F2SIE algorithm is then proposed. In Section \uppercase\expandafter{\romannumeral3}, adverse effects caused by F2SIE algorithm and image attacks are analysed and thus a new closed-loop NIPVSS model is established. Furthermore, the control design of new NIPVSS with F2SIE algorithm is presented . Section \uppercase\expandafter{\romannumeral4} presents experimental results. Conclusions and future work are given in Section \uppercase\expandafter{\romannumeral5}.

\section{The Traditional and New NIPVSS}
\subsection{Image Security Limitation of the Traditional NIPVSS}
\begin{figure}[!t]
  \centering
     \includegraphics[width=0.485\textwidth]{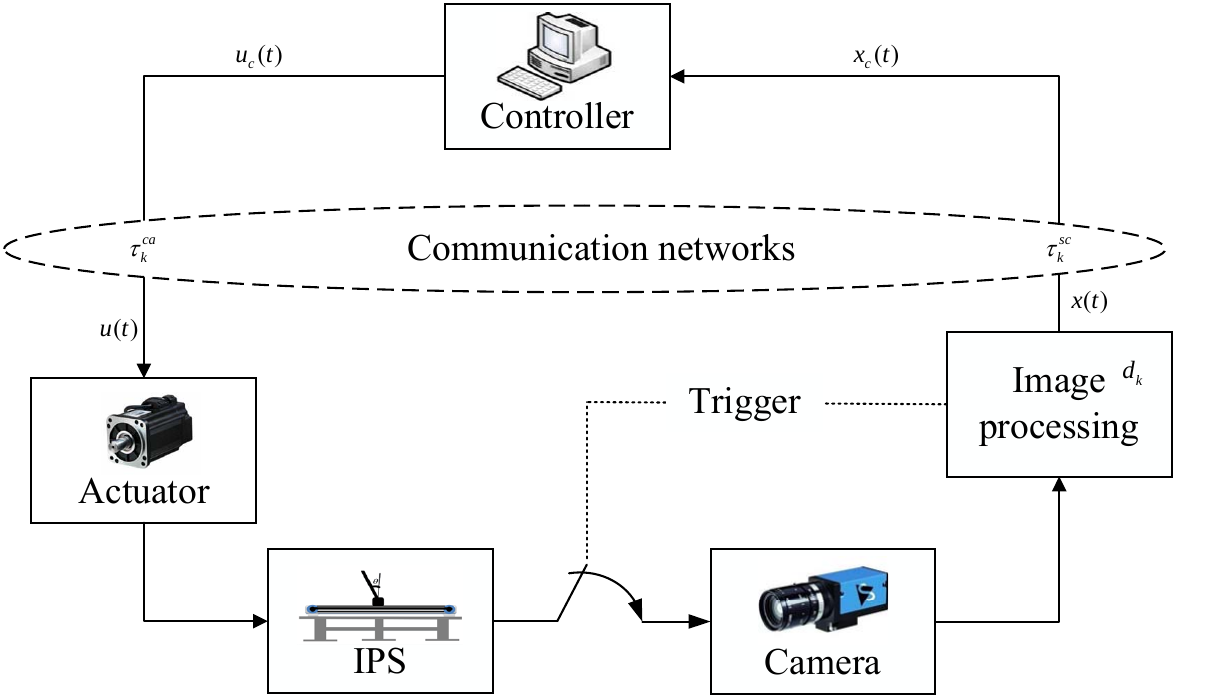}
     \caption{The structure of traditional NIPVSS \cite{8}.}
  \label{fig1}
\end{figure}

The structure of traditional NIPVSS \cite{8} is shown in Fig.~{\ref{fig1}}, where real-time moving images of IPS captured by industrial cameras based on an event-triggered sampling strategy are sent to image processing unit to extract system state, and the acquired system state is sent to remote controller for calculating control input. Finally, the corresponding control input is sent to actuator for achieving stable control of IPS.

\begin{figure*}[!t]
     \centering
       \includegraphics[width=0.7\textwidth]{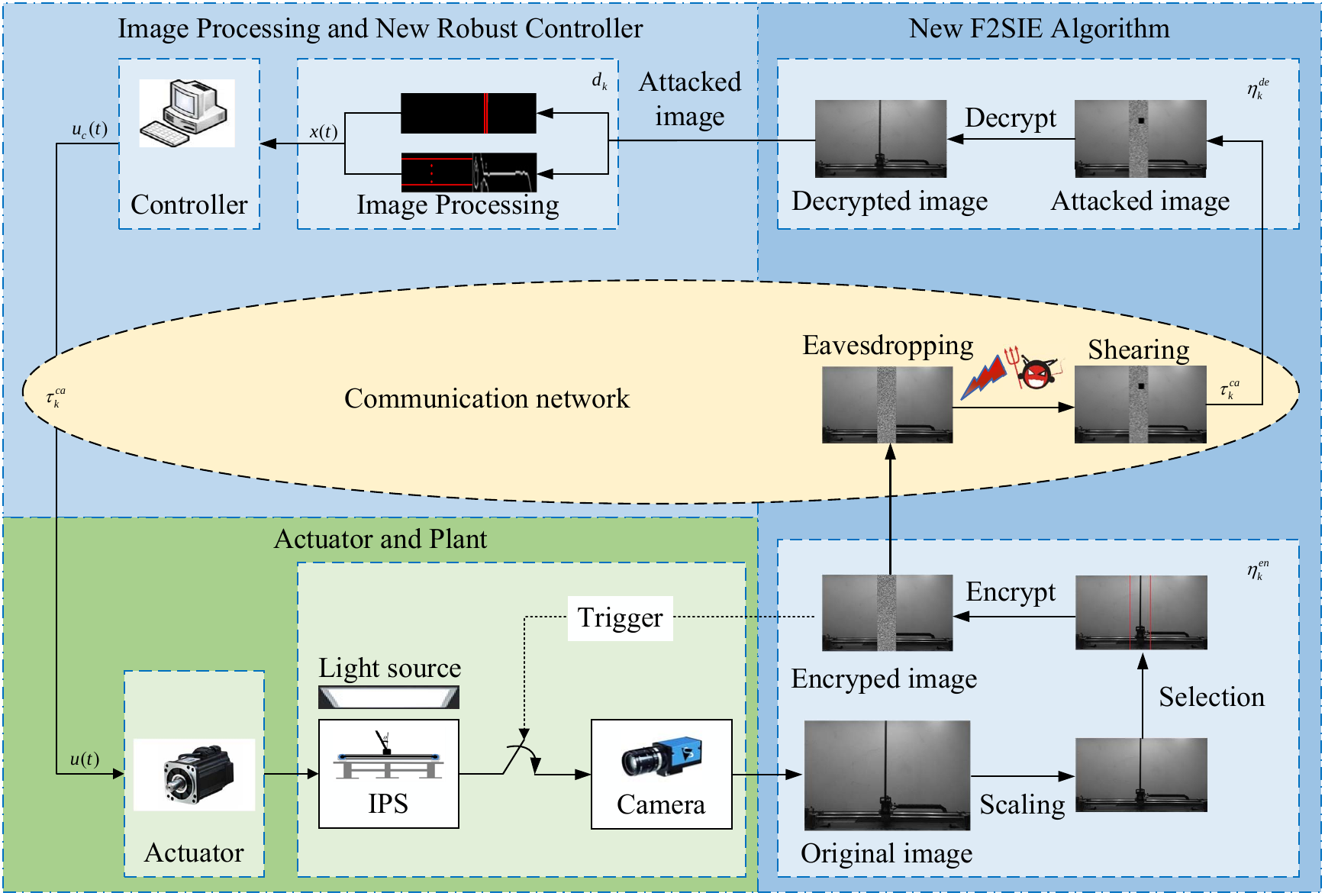}
     \caption{The structure of the new NIPVSS combined with the proposed F2SIE algorithm and robust controller. An example of the new NIPVSS under a slight eavesdropping-based shearing attack (e.g., with shearing rate 1\%) is shown. In this case, the image decrypted by the F2SIE algorithm will be with small errors. This indicates that the F2SIE algorithm can improve security of the new NIPVSS against eavesdropping-based image attacks.}
     \label{fig3}
\end{figure*}

In view of both communication and computational constraints in Fig.~{\ref{fig1}}, an $H_\infty$ controller $u_c(t)=Kx(t)$ is designed to achieve system stability in \cite{8} where $K$ is feedback gain. The closed-loop model of traditional NIPVSS is
\be
\left\{ \begin{gathered}
  \dot x\left( t \right) = Ax\left( t \right) + BKx\left( {t - d\left( t \right) - \tau \left( t \right)} \right) + {B_\omega }\omega (t), \hfill \\
  t \in [{t_k} + {d_k} + \tau _k^{sc} + \tau _k^{ca},{t_{k + 1}} + {d_{k + 1}} + \tau _{k + 1}^{sc} + \tau _{k + 1}^{ca}), \hfill \\
\end{gathered}  \right.
\label{eq1}
\ee
where $x(t) = [\alpha (t),\theta (t), \dot \alpha (t),\dot \theta (t)]$ is system state, $\alpha (t)$, $\theta (t)$, $\dot \alpha (t)$ and $\dot \theta (t)$ are cart position, pendulum angle, cart and angular velocity respectively; $d(t)$ is image-induced delay; $\tau(t)$ is network-induced delay; $t_k$ is image sampling instant; $d_k$ is image processing time; $\tau _k^{sc}$ is network transmission time from sensor to controller; $\tau _k^{ca}$ is network transmission time from controller to actuator; $B_\omega \omega (t)$ is computational errors; $A$ and $B$ are constant matrices.

In the above traditional NIPVSS, images may be attacked in an open network, which will lead to system performance degradation or even crash (see an example of Fig.~\ref{figA1}). 

\subsection{The New NIPVSS with an F2SIE algorithm}
To cope with the above problem of image attacks, a fast scaled-selective image encryption (F2SIE) algorithm will be proposed, which can ensure image security while meet real-time requirement. The structure of new NIPVSS with F2SIE algorithm is shown in Fig.~\ref{fig3}. The difference between Figs.~\ref{fig1} and \ref{fig3} lies on that the image is encrypted at the local end and the corresponding image via network transmission is decrypted at the remote end so as to guarantee image security. After decryption, the system state are obtained and then sent to controller. Furthermore, the control input is calculated in terms of system state, which is finally transmitted to actuator via network and to make IPS run stably.

The idea of F2SIE algorithm is that image is scaled and the scaled image is then encrypted by replacement and diffusion. The encryption steps are as follows:
\begin{enumerate}
\item
\textbf{Image scaling and selection:} To reduce image transmission and encryption time, a bilinear difference method is employed to scale original image of IPS (e.g., 40\%), maintaining continuity of the generated pixel value to ensure image smooth. To further reduce amount of image data, Hough transform is used to determine position of pendulum and the width is set as 100 pixels to cover cart and pendulum. The final selected area is shown in Fig.~\ref{figA2}.
\begin{figure}[!t]
\centering
\includegraphics[width=0.3\textwidth]{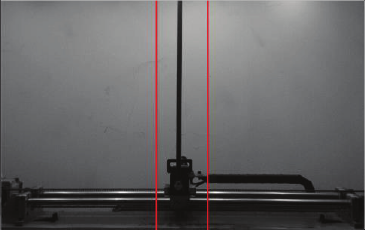}
\caption{The selected area of the inverted pendulum image.}
\label{figA2}
\end{figure}
\item
\textbf{Generation of initial values:} After image scaling and selection, next task is to encrypt key areas, B{\"u}lban mapping is used to generate random numbers as follows
\be
{y_{n + 1}} = {y_n} \times \sqrt {\frac{a}{{{y_n} - b}}},
\label{eqA1}
\ee
where $a$ and $b$ are artificially selected parameters, respectively. The chaotic system \eqref{eqA1} is a one-dimensional chaotic system, which can effectively reduce generation time in comparison with complex chaotic systems. Fig.~\ref{figA3} shows Bifurcation and Lyapunov indexes analysis of B{\"u}lban mapping when $a = 0.5$ and $b = 2$, where it can be seen that system has a large chaotic range with few or no non-chaotic window. The size of the encrypted area is $P$, where ${P_i}$($i \in \{1,2, \cdots ,M\}$) and ${P_j}$($j \in \{1,2, \cdots ,N\}$) represent rows and columns of the encrypted area respectively. B{\"u}lban mapping operates 500 times with initial value ${y_0}$ to ensure its chaos.
\begin{figure}[!t]
\centering
\subfigure[Bifurcation index.]{\includegraphics[width=3.8cm]{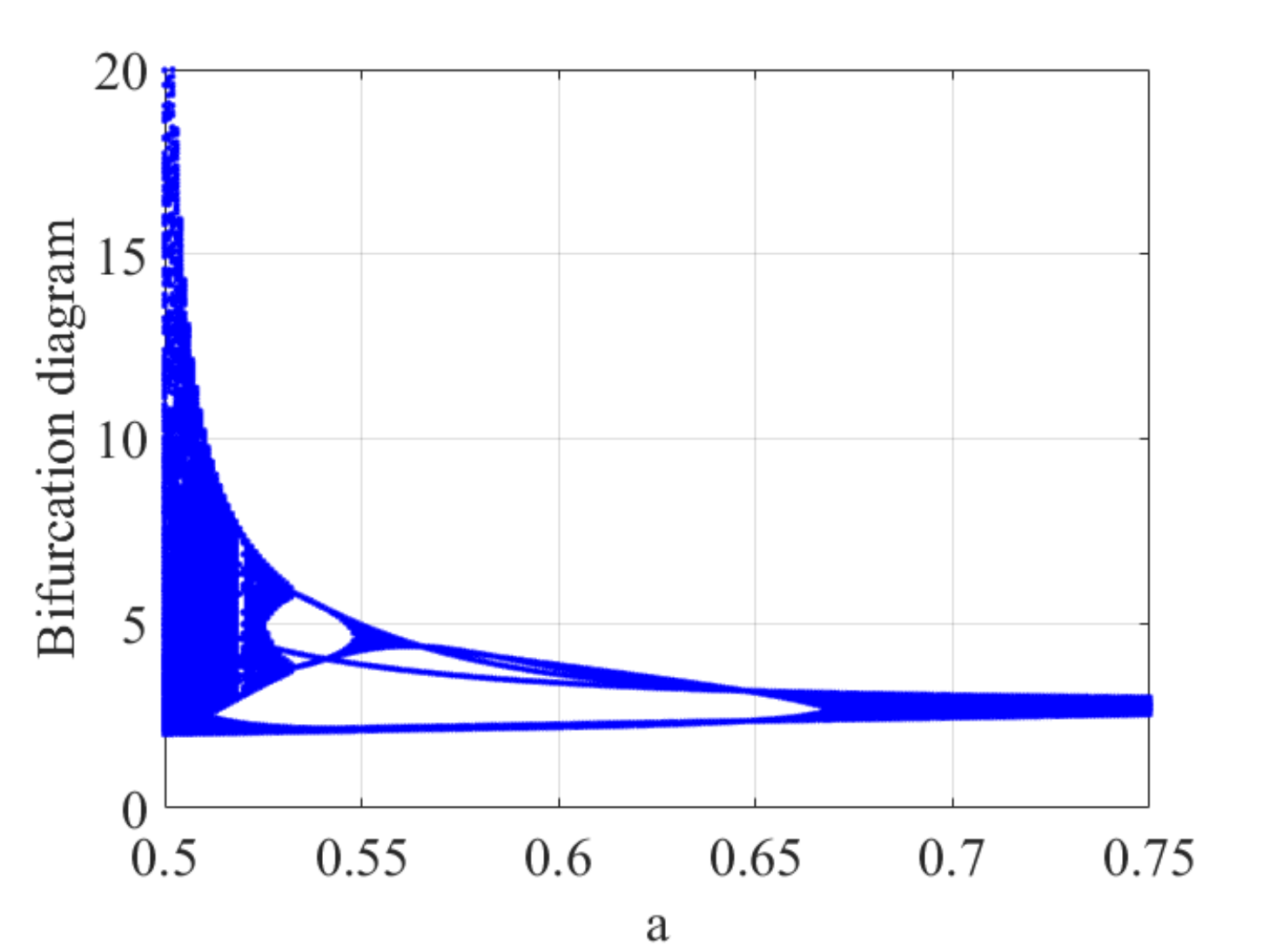}}\quad
\subfigure[Lyapunov index.]{\includegraphics[width=3.8cm]{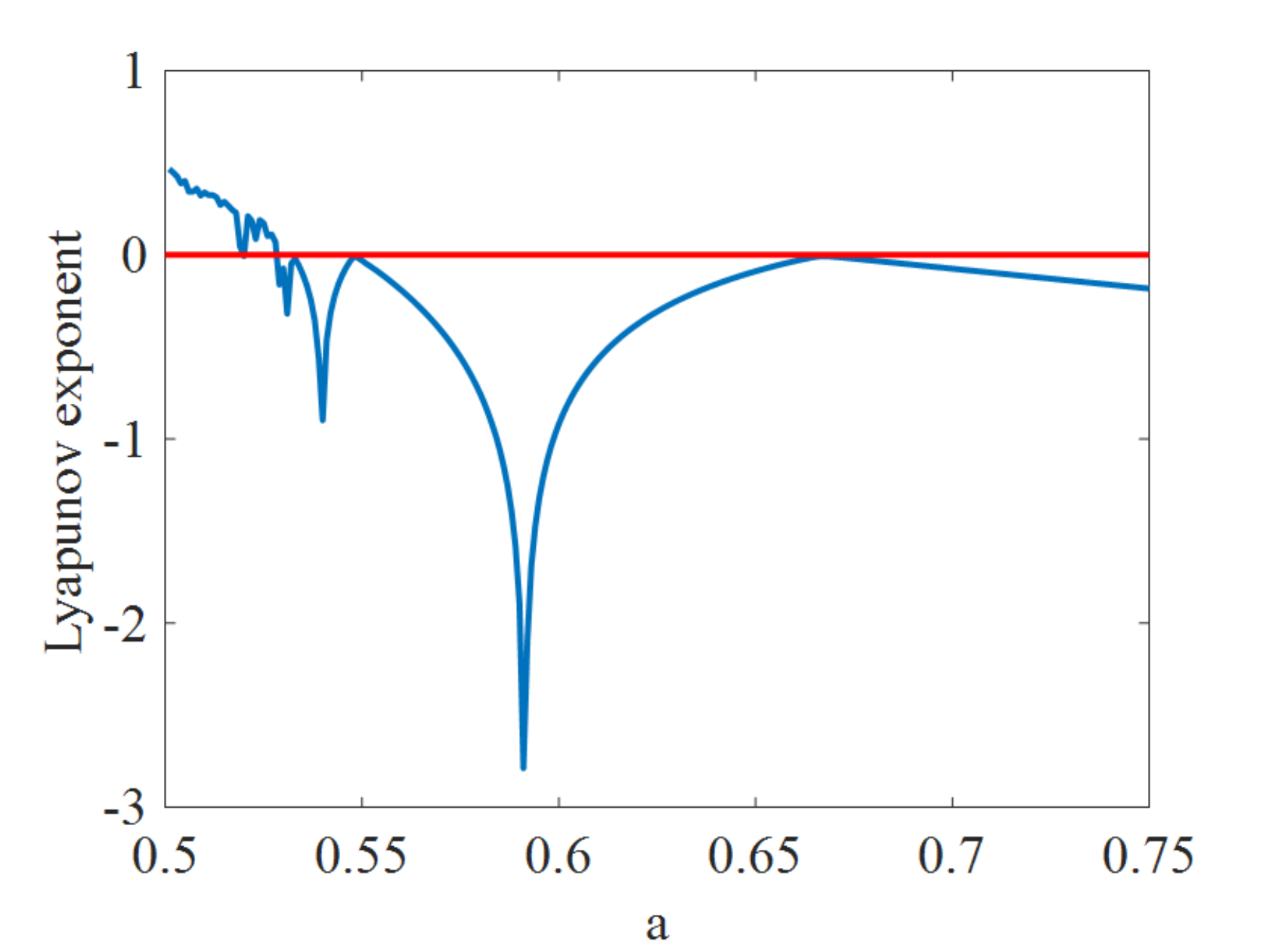}}
\caption{Bifurcation and Lyapunov indexes of B{\"u}lban map ($b = 2$).}
\label{figA3}
\end{figure}
\item
\textbf{Replacement algorithm:} After chaotic system parameters is set, they can be used to generate random numbers for replacement and diffusion. B{\"u}lban mapping is used to generate two real number sequences $PR=\{PR_i\}$ and $PC=\{PC_j\}$ where $i \in \{1,2, \cdots ,M\}$, $j \in \{1,2, \cdots ,N\}$. Then, real number sequences $PR$ and $PC$ are converted into unsigned integer number sequences
\begin{align}
PR^{un} &= \{ P{R_i^{un}}|P{R_i^{un}} = P{R_i} \times {10^5}\bmod M\}\label{eqA2} \hfill \\
PC^{un} &= \{ P{C_j^{un}}|P{C_j^{un}} = P{C_j} \times {10^5}\bmod N\}.\label{eqA3} \hfill
\end{align}
$PR^{un}$ corresponds to each row in $P$, and the pixel values of each row in $P$ are cyclically shifted according to $PR^{un}$ so that $P$ becomes $P^{sr}$, i.e.,
\be
P_{ij}^{sr} = \left\{ \begin{gathered}
{P_{i,j - PR_j^{un}}},if\ j > PR_j^{un} \hfill \\
{P_{i,j + N - PR_j^{un}\bmod N}},if\ j \leqslant PR_j^{un} \hfill \\
\end{gathered}  \right.
\label{eqA4}
\ee
Similarly, $PC^{un}$ corresponds to each column in $P^{sr}$, and the pixel values of each row in $P^{sr}$ are cyclically shifted based on $PC^{un}$ so that $P^{sr}$ becomes $P^{sc}$, i.e.,
\be
P_{ij}^{sc} = \left\{ \begin{gathered}
P_{i - PC_i^{un}, j}^{sr},if\ i > PC_i^{un} \hfill \\
P_{i + M - PC_i^{un}\bmod M, j}^{sr},if\ i \leqslant PC_i^{un} \hfill \\
\end{gathered}  \right.
\label{eqA5}
\ee
An example is shown in Fig.~\ref{figA4}.
\begin{figure}[!t]
\centering
\includegraphics[width=0.46\textwidth]{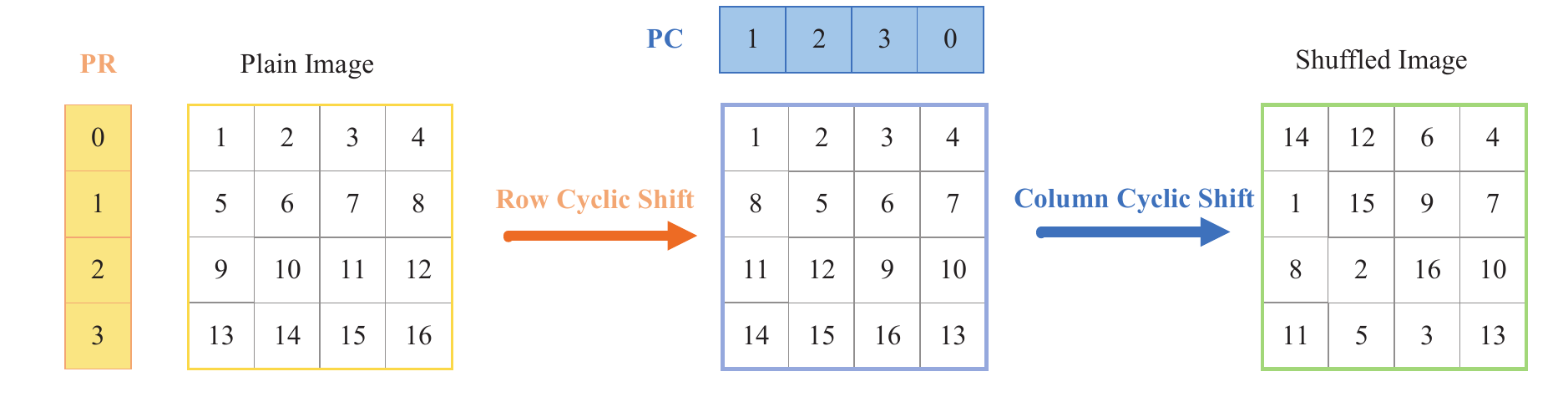}
\caption{Image replacement example.}
\label{figA4}
\end{figure}
\item
\textbf{Diffusion algorithm:} After replacement, it is necessary to change pixel value to improve encryption. B{\"u}lban mapping is used to generate real sequence $\mathcal{K} = \left\{ {{\mathcal{K}_{ij}}} \right\}$, where $i \in \left\{ {1,2, \cdots ,M} \right\}$ and $j \in \left\{ {1,2, \cdots ,N} \right\}$. Then, $\mathcal{K}$ is converted into unsigned integer number sequence
\be
\mathcal{K}^{un} = \left\{ {\mathcal{K}_{ij}^{un}|\mathcal{K}_{ij}^{un} = \mathcal{K}_{ij}\times {{10}^5}\bmod 256} \right\}.
\label{eqA6}
\ee
To improve security of encryption algorithm, the ciphertext feedback mechanism is integrated into diffusion stage, which is shown in Fig.~\ref{figA5}, while the bidirectional ciphertext feedback mode is employed to avoid feedback effect of the first or the last pixel from one-way feedback mechanism. $P^{sc}$ with positive feedback is denoted as $P^{po}=\{P_{ij}^{po}\}$, and $P^{po}$ with negative feedback is denoted as $P^{ne}=\{P_{ij}^{ne}\}$. $P^{po}$ and $P^{ne}$ can be described by
\begin{equation}\label{eqA7}
\begin{gathered}
{P_{ij}^{po}} =  \hfill \\
\left\{ \begin{gathered}
({P_{00}^{sc}} \otimes {\mathcal{K}_{00}^{un}} + {\mathcal{K}_{00}^{un}})\bmod 256, if{\text{ }}i = 0,j = 0; \hfill \\
({P_{ij}^{sc}} \otimes {\mathcal{K}_{ij}^{un}} + {\mathcal{K}_{i - 1,N}^{un}})\bmod 256 \otimes {P^{sc}_{i - 1,N}}, \hfill \\
if\ i \ne 0,j = 0; \hfill \\
({P_{ij}^{sc}} \otimes {\mathcal{K}_{ij}^{un}} + {\mathcal{K}_{i,j - 1}^{un}})\bmod 256 \otimes {P^{sc}_{i,j - 1}}, \hfill \\
if\ i = 0,j \ne 0. \hfill \\
\end{gathered}  \right. \hfill \\
\end{gathered}
\end{equation}
\begin{equation}\label{eqA8}
\begin{gathered}
{P_{ij}^{ne}} =  \hfill \\
\left\{ \begin{gathered}
({P_{MN}^{po}} \otimes {\mathcal{K}_{MN}^{un}} + {\mathcal{K}_{MM}^{un}})\bmod 256, \hfill \\
if\ i = M,j = N; \hfill \\
({P_{ij}^{po}} \otimes {\mathcal{K}_{ij}^{un}} + {\mathcal{K}_{i + 1,0}^{un}})\bmod 256 \otimes {P_{i + 1,0}^{po}}, \hfill \\
if\ i \ne M,j = N; \hfill \\
({P_{ij}^{po}} \otimes {\mathcal{K}_{ij}^{un}} + {\mathcal{K}_{i,j + 1}^{un}})\bmod 256 \otimes {P_{i,j + 1}^{po}}, \hfill \\
if\ i = M,j \ne N. \hfill \\
\end{gathered}  \right. \hfill \\
\end{gathered}
\end{equation}
\begin{figure}[!t]
\centering
\includegraphics[width=0.38\textwidth]{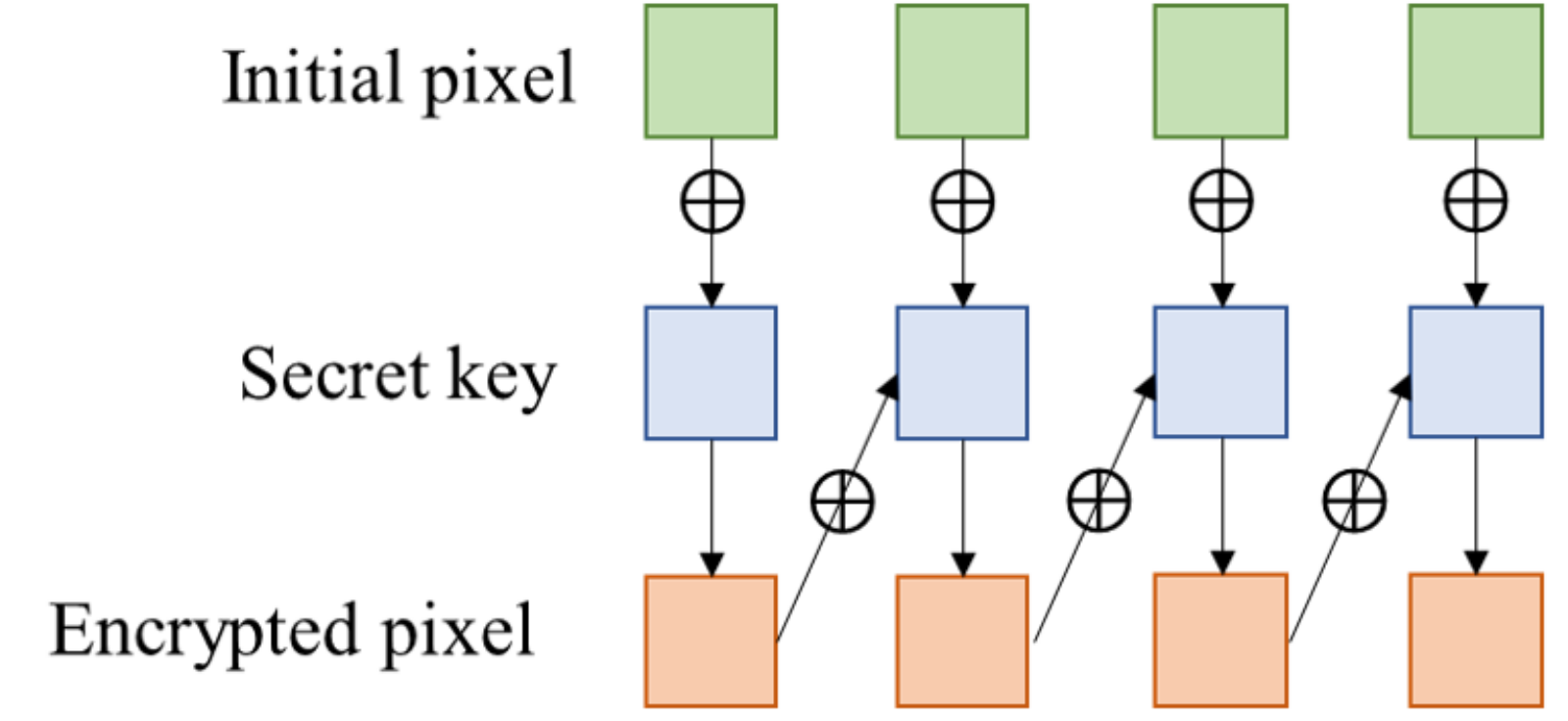}
\caption{Image diffusion example.}
\label{figA5}
\end{figure}
\end{enumerate}

The whole image encryption process is summarized in Algorithm~\ref{alg:Framwork}, while the decryption process is a reverse process of the encryption so it is omitted.
\begin{algorithm}[!t]
\centering
	\caption{F2SIE Algorithm for NIPVSS.}
	\label{alg:Framwork}
	\begin{algorithmic}[1]
		\Require
		Image $P$, cycle period $T$ and initial value $y_0$;
		\Ensure
		The encrypted image $C$;
		\State Scale image and select the encryption area $C_1$ as $M \times N$;
		\State Set cycle count $r \leftarrow 1$;
		\While {$r<T$}
		\State Generate two real chaotic sequences $PR^{un}$ and $PC^{un}$ using (3) and (4);
		\State Generate real chaotic sequences $\mathcal{K}^{un}$ using (7);
		\For {$i=1$ to $M$}
		\State Cyclic shift these pixels in row $i$ of $C_1$ to right with step size $PR_{ij}^{sr}$ using (5);
		\EndFor
		\State Denote the row shift results as $C_2$;
		\For {$i=1$ to $N$}
		\State Cyclic shift these pixels in column $i$ of $C_2$ to down with step size $PC_i^{sc}$ using (6);
		\EndFor
		\State Denote the column shift results as $C_3$;
        \State Positive feedback diffusion using (8);
        \State Reverse feedback diffusion using (9);
        \State Denote the diffuse results as $C$;
        \State Update the cycle count $r \leftarrow  r+1$;
		\EndWhile
	\end{algorithmic}
\end{algorithm}

The new NIPVSS with F2SIE algorithm has been proposed, which not only improves security of new NIPVSS by reducing probability of valuable information being compromised by eavesdropping-based image attacks, but also meets real-time requirement by reducing computational complexity. Therefore, the challenge 1 is solved.

\begin{figure}[!t]
\centering
\subfigure[Original image histogram.]{\includegraphics[width=3.8cm]{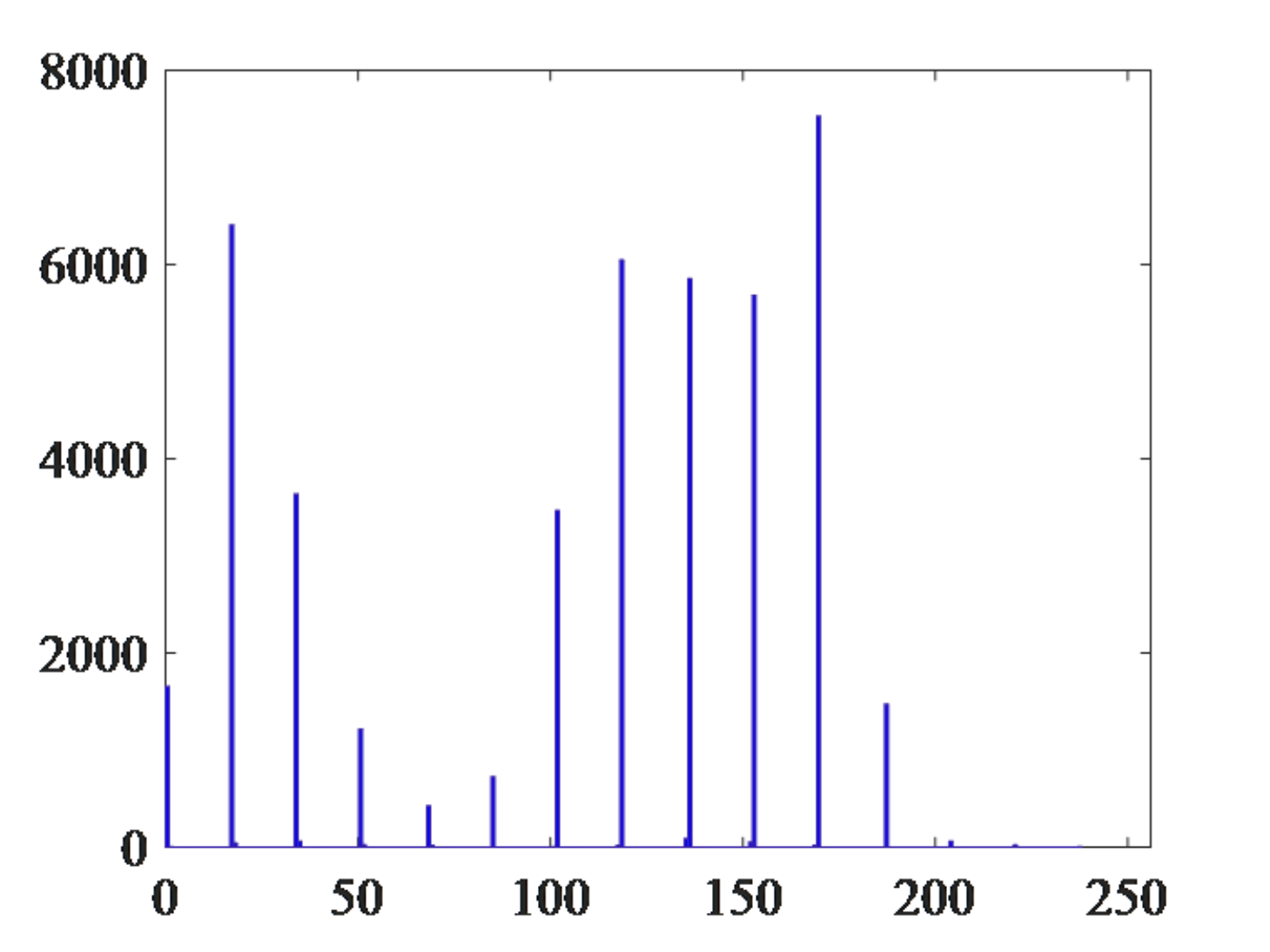}\label{figA10a}}\quad
\subfigure[Encrypted image histogram.]{\includegraphics[width=3.8cm]{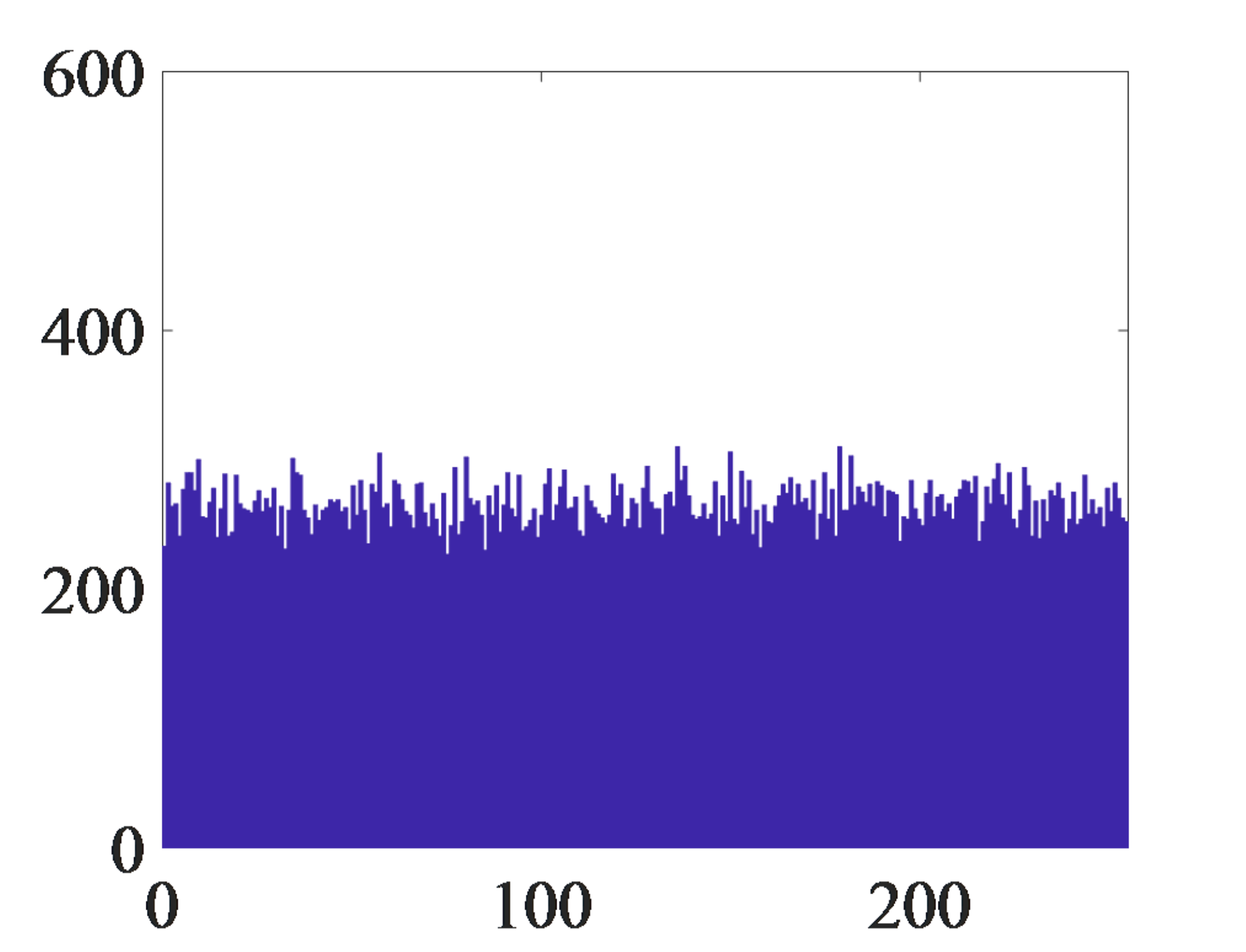}\label{figA10b}}
\caption{Histogram of Original and encrypted image}
\label{figA10}
\end{figure}

\begin{remark}
The F2SIE algorithm is used to improve security of new NIPVSS against eavesdropping-based image attacks. Take shearing attack as an example. When F2SIE algorithm is not used, shearing attacks can shear valuable information (e.g., a part of the pendulum) from the eavesdropped unencrypted image with probability 1, so that pendulum angle cannot be obtained and NIPVSS will lose its stability. When there exists F2SIE algorithm, valuable information is uniformly distributed in image \cite{UDC} (see an example in Fig.~\ref{figA10}). For simplicity, consider a case where image is with $\mathcal{N}_i \in \mathbb{Z}_{+}$ pixels and the sheared area is with $\mathcal{N}_s \in \mathbb{Z}_{+}$ pixels. In this case, the probability that the sheared area of the eavesdropped encrypted image contains all valuable information (e.g., a part of pendulum) is $1/ \mathcal{C}_{\mathcal{N}_i}^{\mathcal{N}_s}$ due to uniformly distributed characteristics of the encrypted image, where $\mathcal{C}$ is combination number. Even if $\mathcal{N}_i$ is small (e.g., $\mathcal{N}_i=100$), when shearing rate is small (e.g., 4\% that is $\mathcal{N}_s=0.04\times \mathcal{N}_i=4$), the probability $1/ \mathcal{C}_{\mathcal{N}_i}^{\mathcal{N}_s}$ is very small ($1/ \mathcal{C}_{\mathcal{N}_i}^{\mathcal{N}_s}=2.6034\times 10^{-7}$). Therefore, the decrypted image under shearing attacks with small shearing rates will have small error with the unencrypted image. It has been verified in real-world experiments of Section IV that cart position and pendulum angle can be obtained with small errors under eavesdropping-based image attacks and the proposed controller hereinafter can be used to stabilize new NIPVSS.
\end{remark}

\begin{remark}
Compared with existing advanced image encryption methods (e.g., \cite{IEM1,IEM2}) with high computational complexity, F2SIE algorithm is with low computational complexity. Specifically, for encryption of an $M\times N$ ($M\in \mathbb{Z}_{+}$, $N \in \mathbb{Z}_{+}$) image, F2SIE algorithm only needs $O(MN+M+N)$ iterations of floating point numbers multiplication, while \cite{IEM1} and \cite{IEM2} need $O(3MN)$ and $O(2MN)$ iterations. Moreover, in real-world experiments, it is verified that for encryption and decryption of an $100\times 480 $ image, F2SIE algorithm consumes $0.014 s$, while \cite{IEM1} and \cite{IEM2} consume $0.021s$ and $0.019s$. Therefore, existing advanced image encryption methods in \cite{IEM1,IEM2} cannot meet real-time requirement of new NIPVSS (i.e., they cannot work in new NIPVSS), but F2SIE algorithm can meet high real-time requirement of new NIPVSS.
\end{remark}

\begin{remark}
In real-world industrial applications, the (private or public) key used for encryption and decryption can be fixed \cite{KM2} or random \cite{KM4}, however random key is more difficult to be decrypted and thus provides better security than fixed key. Furthermore, the usage of random private key in F2SIE algorithm requires that encryption and decryption device keep synchronization, which can be implemented by key synchronization algorithm \cite{KM5}. The detailed process of key synchronization algorithm is as follows: The key used in F2SIE algorithm is firstly from chaotic sequences. Then, when initial value and parameters in \eqref{eqA1} are the same, encryption and decryption device can generate two identical chaotic sequences. Next, according to the agreed choosing order (e.g., choosing from the 100-th number in chaotic sequence), encryption and decryption device can keep same key from the same chaotic sequences. Therefore, the synchronization of random ley can be guaranteed.
\end{remark}

\begin{remark}
When images are severely compromised by image attacks (e.g., shearing attacks with big shearing rates) in new NIPVSS, some new techniques such as authentication \cite{DIW} and signature \cite{DIS} need to be adopted to prevent from information leakage. In the worst case after the above techniques are invalid, attack detection and data compensation \cite{IADAC} can be used to improve system security.
\end{remark}

\section{Adverse Effects Analysis and Robust Controller Design}
\subsection{Adverse Effects of F2SIE Algorithm and Image Attacks}
\begin{figure}[!t]
\centering
\includegraphics[width=0.46\textwidth]{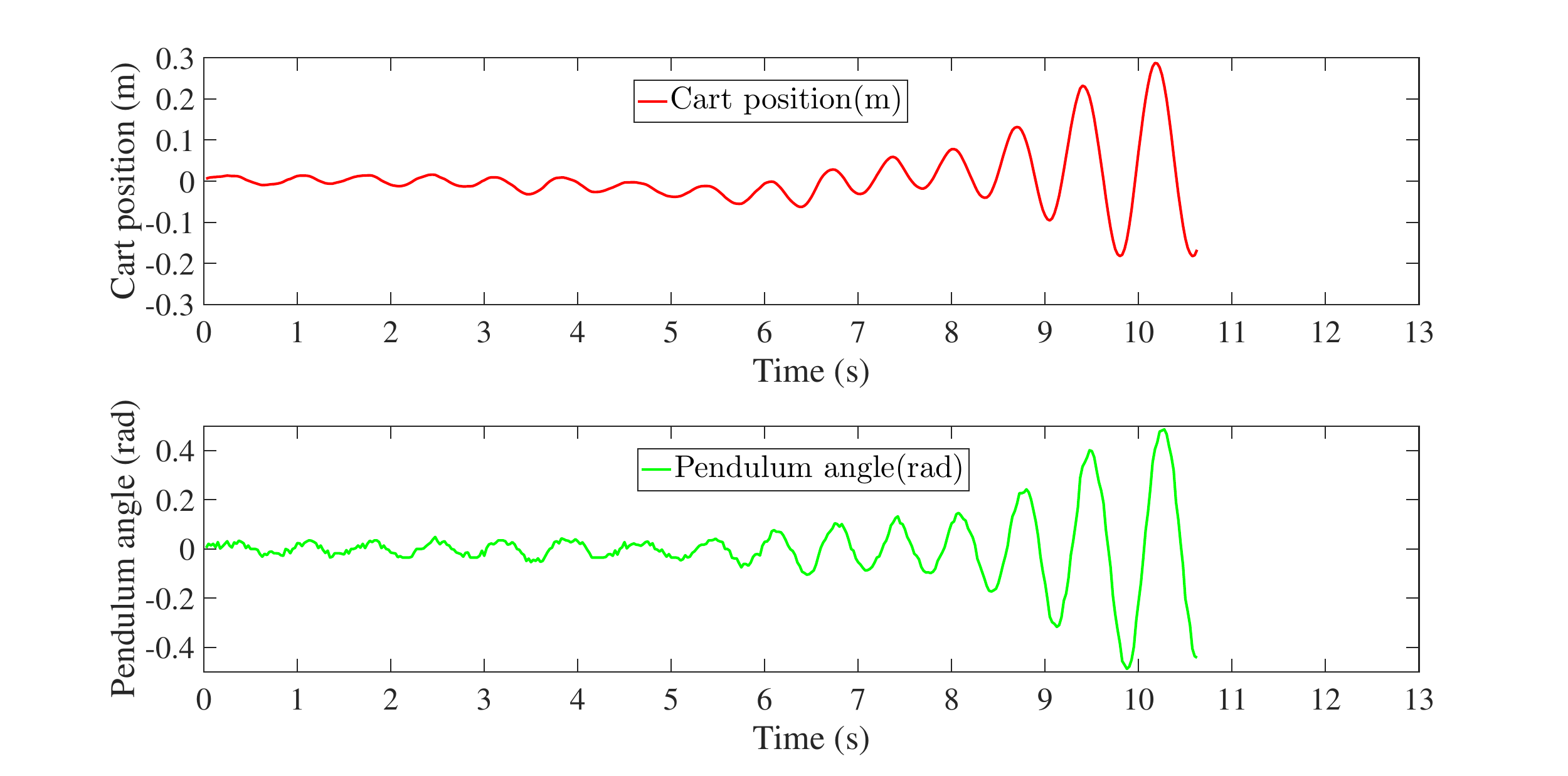}
\caption{Cart position and pendulum angle of the new NIPVSS with the F2SIE algorithm under the $H_\infty$ controller in [7].}
\label{figA6}
\end{figure}

The above has designed a new NIPVSS with F2SIE algorithm to guarantee image security. However, this will bring some adverse effects, i.e., extra computational times from F2SI2 algorithm and extra computational errors from image attacks. It will degrade system performance or even drive system collapse. E.g., Fig.~\ref{figA6} shows experimental results of new NIPVSS with F2SIE algorithm under the $H_{\infty}$ controller in \cite{8}, where new NIPVSS is unstable after the images have been encrypted and decrypted. Hence, these adverse effects must be analysed.

\subsubsection{Extra Computational Times  from F2SIE Algorithm}
\begin{figure}[!t]
\centering
\label{figA7a}\subfigure[Image encryption time.]{\includegraphics[width=3.8cm]{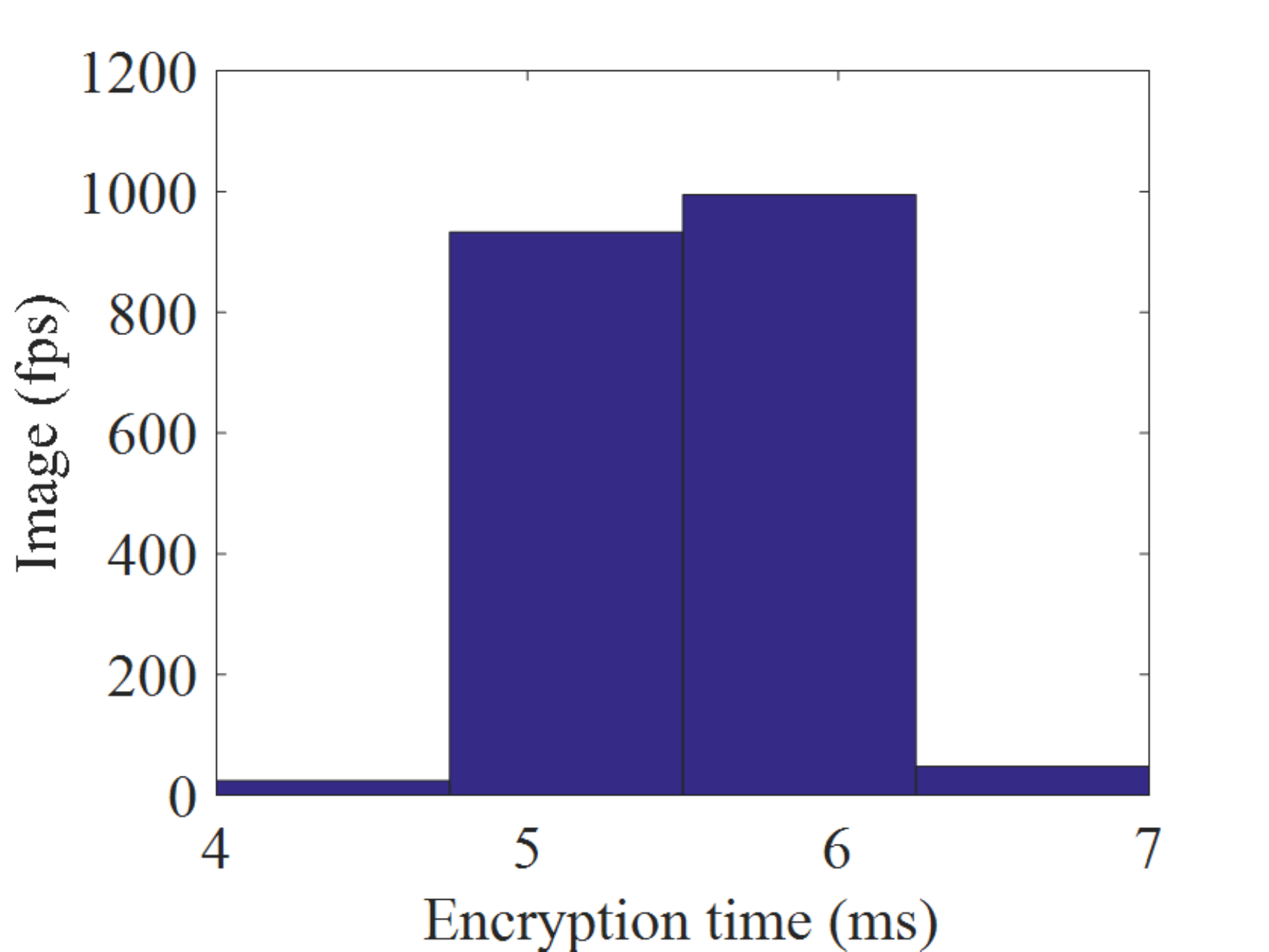}}\quad
\label{figA7b}\subfigure[Image decryption time.]{\includegraphics[width=3.8cm]{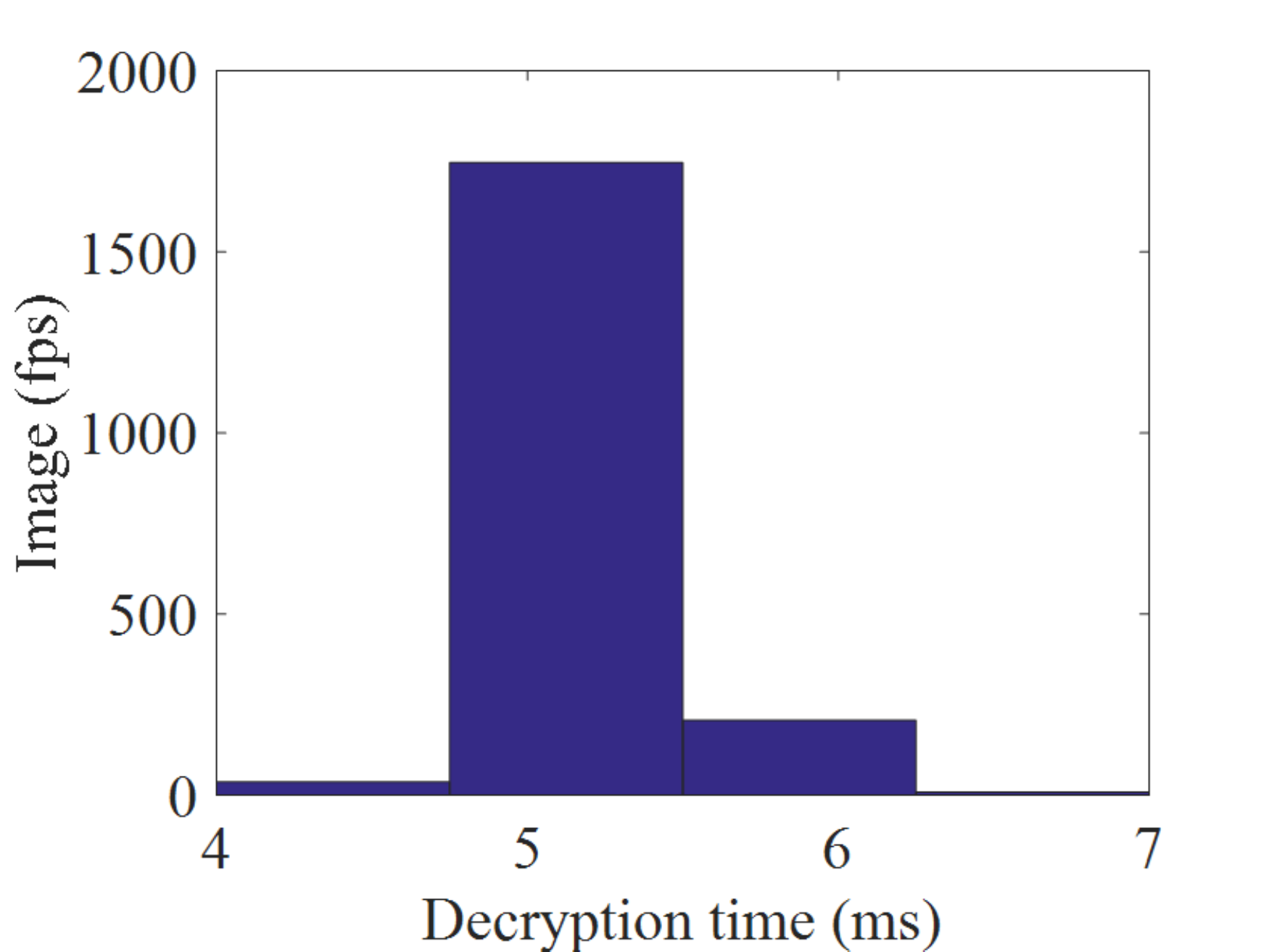}}
\caption{2000 set of image encryption and decryption times.}
\label{figA7}
\end{figure}

There already exist ${\tau _k^{sc}}\in [\underline \tau ^{sc}, \bar \tau ^{sc}]=[0,0.005s]$, ${\tau _k^{ca}}\in [\underline \tau ^{ca}, \bar \tau ^{ca}]=[0,0.005s]$ and $d_k \in [\underline d, \bar d]=[0.009s,0.019s]$ in traditional NIPVSS \cite{8}. After F2SIE algorithm is introduced, its consumed time cannot be ignored, which cause system unstable as shown in Fig.~\ref{figA6}. Therefore, two extra computational times (i.e., image encryption time $\eta_k^{en}$ and image decryption time $\eta_k^{de}$) are to be analysed as follows. Due to introduction of encryption and decryption, NIPVSS will inevitably consume computational time. To determine encryption and decryption time, 2000 images are encrypted and decrypted respectively. According to Fig.~\ref{figA7}, the upper and lower bounds of image encryption and decryption are
\begin{align}
\eta_k^{en} &\in [\underline \eta^{en},\bar\eta^{en}],\underline \eta^{en}=0.004s,\bar\eta^{en}=0.007s \label{eqA8} \hfill \\
\eta_k^{de} &\in [\underline \eta^{de},\bar\eta^{de}],\underline \eta^{de}=0.004s,\bar\eta^{de}=0.007s\label{eqA9} \hfill
\end{align}
From the above experimental statistical results, their upper and lower bounds are ${\eta^{en}_k} \in [ {\underline \eta^{en},\bar \eta^{en}}] = [ {0.004,0.007} ]s$ and ${\eta^{de}_k} \in [ {\underline \eta^{de},\bar \eta^{de}}] = [ {0.004,0.007} ]s$.

\begin{remark}
To meet high real-time requirement of new NIPVSS, image scaling and selection are adopted in F2SIE algorithm to reduce delay. Therefore, original image processing time $d_k \in [0.009,0.019]s$ in \cite{8} is reduced to scaled-selective image processing time $d_k \in [0.007,0.009]s$ in this paper.
\end{remark}

\begin{remark}
When image attacks begin to enter new NIPVSS, they require a certain injection time $\Delta \eta_{k}$ to modify the encrypted pixel. The values of $\Delta \eta_{k}$ of different image attacks are shown in Tab.~\ref{tabAI}, where it can be seen that: $\Delta \eta_{k}$ of slight shearing attack and salt and pepper attack is almost zero, and only 0.001s with increase of attack intensity; $\Delta \eta_{k}$ of Gaussian attack is generally 0.001s. Therefore, $\Delta \eta_{k}$ is treated as a constant, i.e., $\Delta \eta_{k}=\Delta \eta=0.001s$. Although injection time of these attacks is short, it must be considered when designing a robust controller.
\end{remark}
\begin{table}[!t]
\caption{The values of $\Delta \eta_k$ of Different Image Attacks.}
\label{tabAI}
\centering
\begin{tabular}{cccccc}
  \toprule
  Shearing rate & 1\%  & 2\% & 4\% & 6\% & 8\%\\
  $\Delta \eta_k$ (s) & - & - & - & - & 0.001\\
  \midrule
  Intensity & 1\%  & 2\% & 4\% & 6\% & 8\%  \\
  $\Delta \eta_k$ (s) & - & - & - & 0.001 & 0.001 \\
  \midrule
  $(\mu ,\sigma )$ & (0,1) & (0,5) & (2,2) & (2,5) & (5,5) \\
  $\Delta \eta_k$ (s) & 0.001  & 0.001 & 0.001 & 0.001 & 0.001 \\
  \bottomrule
  \multicolumn{6}{l}{Shearing rate represents shearing rate of shear attacks}\\
  \multicolumn{6}{l}{Intensity represents intensity of salt and pepper attack}\\
  \multicolumn{6}{l}{$\mu$ and $\sigma$ represent mean and variance of Gaussian attack}\\
  \multicolumn{6}{l}{- represents injection time is almost 0s, which can be ignored}\\
\end{tabular}
\end{table}

\subsubsection{Extra Computational Errors from Image Attacks}
The F2SIE algorithm brings extra computational delay, but does not cause extra computational errors. This is because encryption process is completely opposite to decryption process so that image can be decrypted with zero errors. However, it has been analysed in \cite{8} that environmental noise will cause computational errors. Moreover, attacks on the encrypted images will produce extra computational errors. These computational errors will destroy system stability, which is well illustrated in Fig.~\ref{figA1}.

\begin{figure}[!t]
\centering
\subfigure[]{\includegraphics[width=3.8cm]{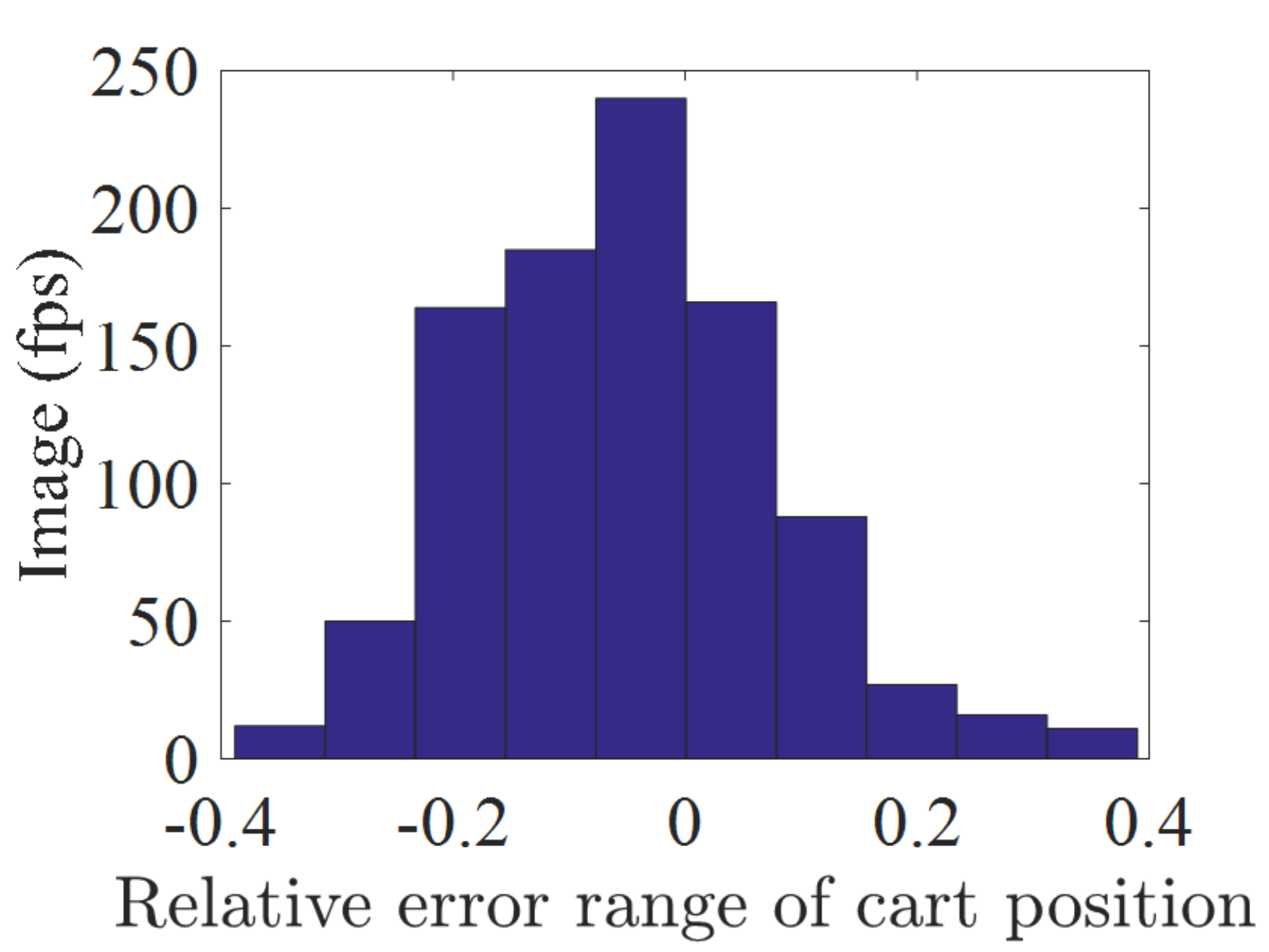}\label{figA8a}}\quad
\subfigure[]{\includegraphics[width=3.8cm]{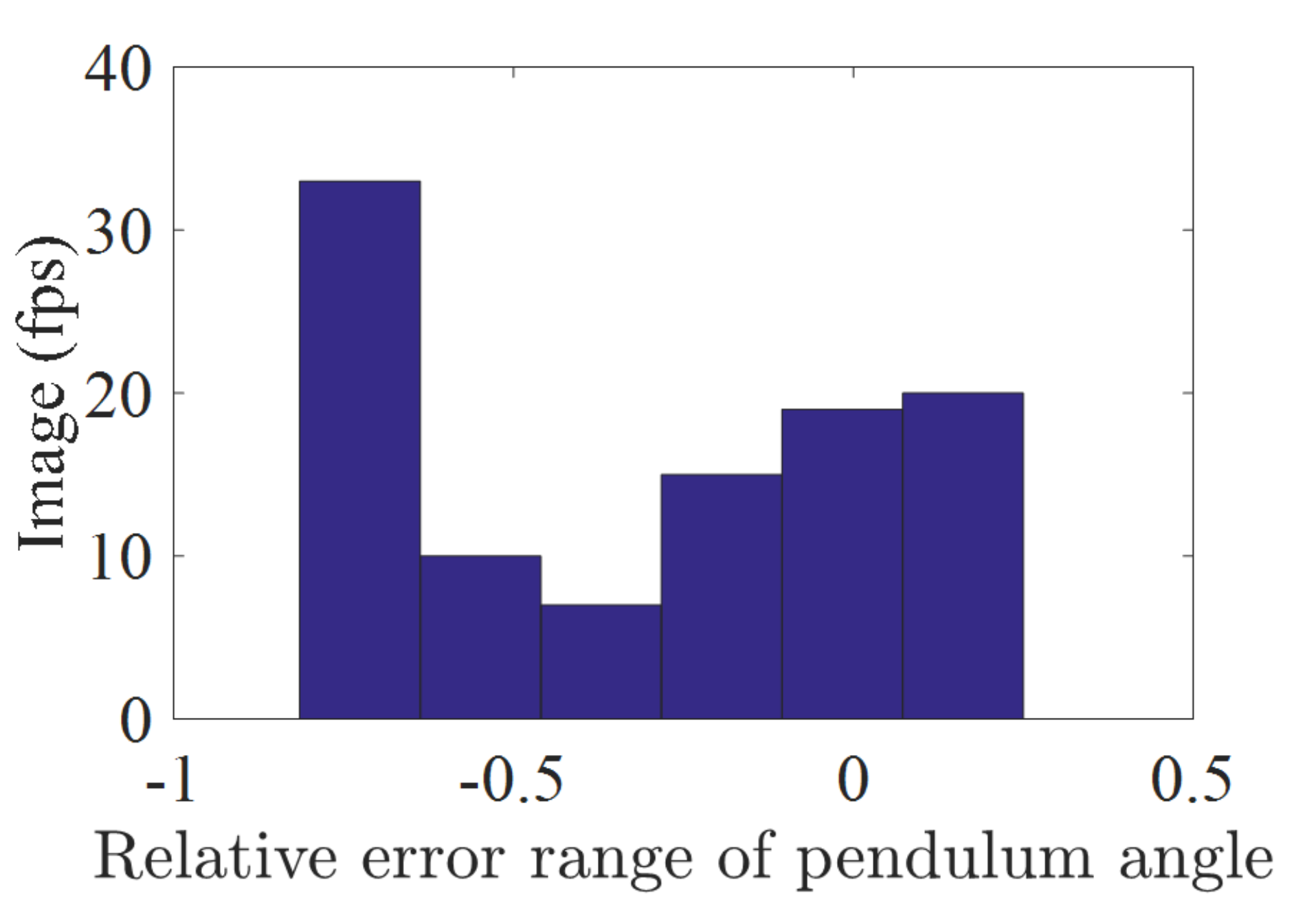}\label{figA8b}}
\caption{Relative error range of each state information.}
\label{figA8}
\end{figure}

The computational errors from environmental noise and image attacks need be converted into form of parameter uncertainty, where conversion process is as follows. 
The computational errors are treated as disturbance [7], ${H_\infty }$ disturbance attenuation level $\gamma $ is employed to evaluate computational errors. However, to analyze how much computational errors system can tolerate, it is modeled as model uncertainty
\be
diag\left\{{\Delta _1},{\Delta _2},{0},{0}\right\}x(t) = {B_\omega }\omega (t)
\label{eqA11}
\ee
where ${\Delta _1}$, ${\Delta _2}$ are parameter uncertainty of cart position and pendulum angle caused by environmental \& background noise; ${B_w}\omega (t) = [ {err_1};{err_2};err_3;{err_4}]$, $err_i(i = 1,2,3,4)$ are relative error of states. Since cart fluctuates at equilibrium point, cart velocity and angular velocity change rapidly. So, uncertainty of cart and angular velocity are not considered. According to \eqref{eqA11}, it follows that
\be
\begin{gathered}
{\Delta _1}\alpha  = er{r_1},{\Delta _2}\theta  = er{r_2}.
\end{gathered}
\label{eqA12}
\ee
Next, relative error of each state is shown in Fig.~\ref{figA8}, and range of ${\Delta _1}$ and ${\Delta _2}$ are
${\Delta _1} \in \left[ { - 0.4,0.4} \right]$ and ${\Delta _2} \in \left[ { - 0.82,0.82} \right]$. The above analysis does not take into account uncertainties caused by cyber attacks, they can be regarded as extra uncertainties ${\Delta _{1,a}}$ and ${\Delta _{2,a}}$, and added into ${\Delta _1}$ and ${\Delta _2}$ respectively. The final conversion result is
\be
\Delta A = DF(t)E,
\label{eq2}
\ee
where $\Delta A = diag\left\{ {\Delta _1+\Delta_{1,a},\Delta _2+\Delta_{2,a},0,0} \right\}$ is the whole parameter uncertainty matrix, ${\Delta _1} \in [ { - 0.4,0.4} ]$ and ${\Delta _2} \in [ { - 0.82,0.82} ]$ are parameter uncertainty of cart position and pendulum angle caused by environmental noise respectively, ${\Delta _{1,a}} \in [ {\underline \Delta  }_{1,a},{{\overline \Delta  }_{1,a}} ]$ and ${\Delta _{2,a}} \in [ {{{\underline \Delta  }_{2,a}},{{\overline \Delta  }_{2\_a}}} ]$ are parameter uncertainty of cart position and pendulum angle caused by image attacks respectively; $D=I$; $F(t) = diag\left\{ {{r_1}(t),{r_2}(t),0,0} \right\}$, ${r_1}(t) \in [ - 1,1]$, ${r_2}(t) \in [ - 1,1]$;  $E = diag\left\{ {0.4 + \Delta^M_{1,a},0.82 + \Delta^M _{2,a},0,0} \right\}$, $\Delta^M_{1,a}=max\left\{ |{\underline \Delta  }_{1,a} |, | {\overline \Delta  }_{1,a} |\right\}$ and $\Delta^M_{2,a}=max\left\{ |{\underline \Delta  }_{2,a} |, | {\overline \Delta  }_{2,a} |\right\}$.

\begin{remark}
When no attack, both ${\Delta _{1,a}}$ and ${\Delta _{2,a}}$ are zero. When the encrypted images are attacked, $\Delta_{1,a}$ and $\Delta_{2,a}$ from image attacks are added into $\Delta_{1}$ and $\Delta_{2}$ from environmental noise, leading to system performance decrease or even crash.
\end{remark}

\subsection{Closed-Loop Uncertain Model of New NIPVSS}
The adverse effects (i.e., extra computational times and extra computational errors) from F2SIE algorithm and image attacks in new NIPVSS have been analysed. Next, to design a new robust controller $u_c(t)=Kx(t)$, the closed-loop model of new NIPVSS with F2SIE algorithm under image attacks need to be established. The detailed modelling process is as follows.

The whole signal transmission process is shown in Fig.~\ref{figA9}, which is divided into the following 9 steps:
\begin{figure}[!t]
     \centering
       \includegraphics[width=0.48\textwidth]{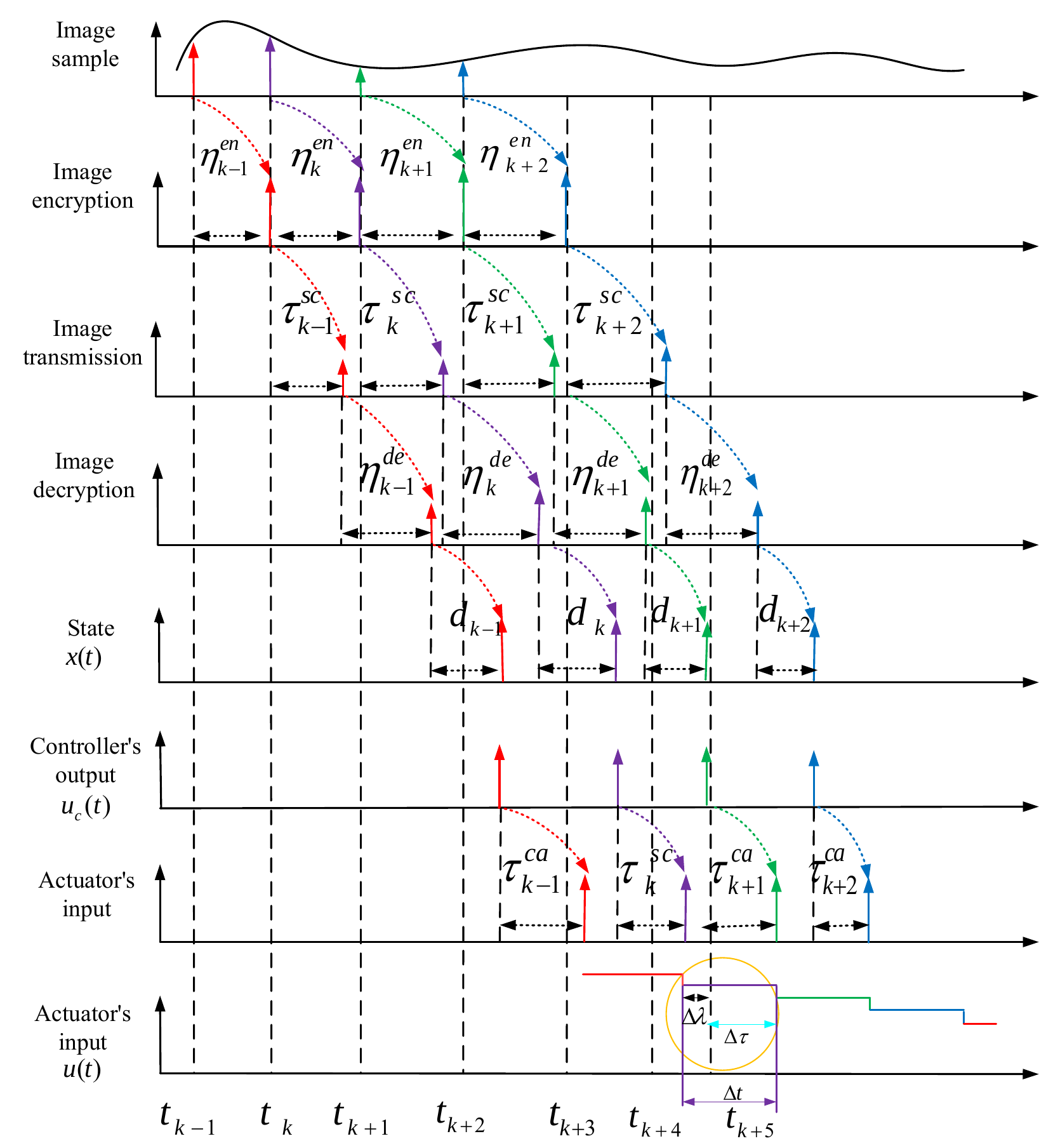}
     \caption{Signal's timing diagram of the NIPVSS with image encryption.}
     \label{figA9}
\end{figure}

\begin{enumerate}
\item[S1:]
The real-time images of IPS are captured at instants $\left\{ {{t_{k - 1}},{t_k},{t_{k + 1}}, \cdots } \right\}$ based on event-triggered mechanism.
\item[S2:]
For each captured image (e.g., at sampling instant $t_k$), it will takes encryption time before network transmission. Hence, the captured image is encrypted at instant ${t_k} + \eta _k^{en}$.
\item[S3:]
The encrypted image is sent to remote image processing unit, which will consume time $\tau _k^{sc}$. Hence, the encrypted image is received at instant ${t_k} + \eta _k^{en} + \tau _k^{sc}$.
\item[S4:]
It takes time $\eta _k^{de}$ to decrypt after the encrypted image is received in remote image processing unit, so image decryption is finished at instant ${t_k} + \eta _k^{en} + \tau _k^{sc} + \eta _k^{de}$.
\item[S5:]
After image is decrypted, it takes time ${d_k}$ to calculate state $x(t)$, so
\[\begin{gathered}
  x(t) = x({t_k}), \hfill \\
  t \in \left\{ {{t_k} + \eta _k^{en} + \tau _k^{sc} + \eta _k^{de} + {d_k},k = 0,1,2, \cdots } \right\} \hfill \\
\end{gathered}\]
\item[S6:]
The event-triggered mechanism is employed in controller, it will receive $x(t)$ at the same instant, i.e.,
\[\begin{gathered}
  \tilde x(t) = x({t_k}) \hfill \\
  t \in \left\{ {{t_k} + \eta _k^{en} + \tau _k^{sc} + \eta _k^{de} + {d_k},k = 0,1,2, \cdots } \right\} \hfill \\
\end{gathered}\]
\item[S7:]
The output signal of controller is expressed as
\[\begin{gathered}
  u(t) = Kx({t_k}) \hfill \\
  t \in \left\{ {{t_k} + \eta _k^{en} + \tau _k^{sc} + \eta _k^{de} + {d_k},k = 0,1,2, \cdots } \right\} \hfill \\
\end{gathered}\]
\item[S8:]
The control signal arrive at actuator by $\tau _k^{ca}$, i.e.,
\[\begin{gathered}
  \tilde u(t) = Kx({t_k}) \hfill \\
  t \in \left\{ {{t_k} + \eta _k^{en} + \tau _k^{sc} + \eta _k^{de} + {d_k} + \tau _k^{ca},k = 0,1,2, \cdots } \right\} \hfill \\
\end{gathered} \]
\item[S9:]
The actuator receives control signal and saves it in zero order holder and acts on plant till next control signal arrives at instant ${t_{k + 1}} + \eta _{k + 1}^{en} + \tau _{k + 1}^{sc} + \eta _{k + 1}^{de} + {d_{k + 1}} + \tau _{k + 1}^{ca}$. It means that effective time of control signal is
\[\begin{gathered}
  \left[ {{t_k} + \eta _k^{en} + \tau _k^{sc} + \eta _k^{de} + {d_k} + \tau _k^{ca},} \right. \hfill \\
  \left. {{t_{k + 1}} + \eta _{k + 1}^{en} + \tau _{k + 1}^{sc} + \eta _{k + 1}^{de} + {d_{k + 1}} + \tau _{k + 1}^{ca}} \right) \hfill \\
\end{gathered} \]
\end{enumerate}

According to the above analysis and considering image attacks which consume extra time $\Delta {\eta _k}$,  the control signal is
\begin{align}
  &\label{eqA13} u_c(t) = Kx(t) \hfill \\
  &t \in \left[ {{t_k} + \eta _k^{en} +\Delta {\eta _k}+ \tau _k^{sc} + \eta _k^{de} + {d_k} + \tau _k^{ca},} \right. \nonumber \hfill \\
  &\left. {{t_{k + 1}} + \eta _{k + 1}^{en}+\Delta {\eta _{k+1}} + \tau _{k + 1}^{sc} + \eta _{k + 1}^{de} + {d_{k + 1}} + \tau _{k + 1}^{ca}} \right) \nonumber \hfill
\end{align}
The relationship between sampling time $t_k$ and delay $\lambda(t)$ and $\tau(t)$ can be re-written as
\begin{align}
  &\label{eqA14}{t_k} = t - (t - {t_k}) = t - \lambda \left( t \right) - \tau \left( t \right), \hfill \\
  &t \in \left[ {{t_k} + \eta _k^{en} +\Delta {\eta _k}+ \tau _k^{sc} + \eta _k^{de} + {d_k} + \tau _k^{ca},} \right. \nonumber \hfill \\
  &\left. {{t_{k + 1}} + \eta _{k + 1}^{en}+\Delta {\eta _{k+1}} + \tau _{k + 1}^{sc} + \eta _{k + 1}^{de} + {d_{k + 1}} + \tau _{k + 1}^{ca}} \right) \nonumber \hfill
\end{align}
where ${\underline \lambda}  \leqslant \lambda (t) < {\overline \lambda}$ and $0 \leqslant \tau (t) < {\overline \tau}$ are both time-varying delays with upper and lower bounds, and
\[\begin{gathered}
  {\underline \lambda} = \mathop {\min }\limits_{k \in N} \left( {\eta _k^{en} +\Delta {\eta _k}+ \eta _k^{de} + \tau _k^{sc} + {d_k}} \right) \hfill \\
  {\overline \lambda}= \hfill \\
  \mathop {\max }\limits_{k \in N} \left( {\eta _k^{en}+\Delta {\eta _k} + \eta _{k + 1}^{en}+\Delta {\eta _{k+1}} + \tau _{k + 1}^{sc} + \eta _{k + 1}^{de} + {d_{k + 1}}} \right) \hfill \\
  {\overline \tau}  = \mathop {\max }\limits_{k \in N} \left( {\tau _{k}^{ca}} \right) \hfill \\
\end{gathered} \]
Considering the above \eqref{eqA14} and parameter uncertainty in \eqref{eq2}, finally a new closed-loop NIPVSS model with parameter uncertainty and multiple time-varying delays can be established as
\be
\left\{ \begin{gathered}
  \dot x\left( t \right) = (A + DF(t)E)x\left( t \right) + BKx\left( {t - \lambda \left( t \right) - \tau \left( t \right)} \right), \hfill \\
  t \in \left[ {{t_k} + \eta _k^{en}+\Delta {\eta _k} + \tau _k^{sc} + \eta _k^{de} + {d_k} + \tau _k^{ca},} \right. \hfill \\
  \left. {{t_{k + 1}} + \eta _{k + 1}^{en} +\Delta {\eta _{k+1}}+ \tau _{k + 1}^{sc} + \eta _{k + 1}^{de} + {d_{k + 1}} + \tau _{k + 1}^{ca}} \right), \hfill \\
\end{gathered}  \right.
\label{eq3}
\ee
where $\tau(t) \in [0,\bar \tau]$ is network-induced delay and $\bar \tau = {{\bar \tau }^{ca}}$; $\lambda(t) \in [\underline \lambda ,\bar \lambda]$ is the new image-induced delay, ${\underline \lambda} = {\underline \eta^{en}}  + {\Delta \eta} + {\underline \tau }^{sc} + {\underline \eta^{de}} + {\underline d}$ and ${\bar\lambda} = 2{{\bar \eta }^{en}} + 2{\Delta \eta}  + {{\bar \tau }^{sc}} + {\bar \eta^{de}} + \bar d$.

\begin{remark}
The problem of network communication and image processing has been well handled in \cite{8}, but problem of image attacks is not involved. The image attacks can degrade system performance or even cause system crash.  To cope with image attacks, new F2SIE algorithm has been designed in the above. However, F2SIE algorithm brings some new problems that the controller in \cite{8} with F2SIE algorithm cannot keep system stable as shown in experimental results of Fig.~\ref{figA6}. To solve these problems, a new closed-loop uncertain time-delay model \eqref{eq3} has been established, including that 1)  extra computational delays in $\lambda (t)$ caused by F2SIE algorithm running time $\eta_k^{en}$, $\eta_k^{de}$ and image attacks injection time $\Delta \eta _k$ are considered in \eqref{eq3}, 2) extra computational errors caused by image attacks are considered, and the whole computational errors caused by environment  noise and image attacks are treated as parameter uncertainty $DF(t)E$ in \eqref{eq3}. In comparison with time-delay model \eqref{eq1} in \cite{8}, the controller in \cite{8} is invalid and a new robust controller must be designed for new model \eqref{eq3}.
\end{remark}

\subsection{Robust Controller Design}
As a typical networked visual servo control application scenario, new NIPVSS with F2SIE algorithm under image attacks has been established as the above closed-loop model \eqref{eq3}, which is regarded as an uncertain time-delay system. For this system, it can be handled based on the idea of robust networked control \cite{8,HINF,GCC}, and a new robust controller is designed in the following Theorem 1.

\begin{theorem}
For given constants $0 < {\overline \tau } $, $0 < {\underline \lambda} < {\overline \lambda  }$, $0 < {\varepsilon _1}$, $0 < {\varepsilon _2}$ and ${\theta _j}(j = 1,2,3,4)$, if there exist $0 < \epsilon $ and real symmetric matrices $X$, ${\tilde Q_i}$ $(i = 1,2, \cdots ,7)$, ${\tilde Z_i}$ $(i = 1,2,3,4)$ with appropriate dimensions, such that
\be
\left[ {\begin{array}{*{20}{c}}
  {{\Phi _{11}}}&{{\Phi _{12}}}&{{\Phi _{13}}} \\
   * &{{\Phi _{22}}}&0 \\
   * & * &{ - \epsilon I}
\end{array}} \right] < 0,
\label{eq4}
\ee
holds, where ${\Phi _{11}} = \left[ {\begin{array}{*{20}{c}}
  {\Phi _{11}^{11}}&{\Phi _{11}^{12}} \\
   * &{{\Theta _{8,8}}}
\end{array}} \right]$ and\\
$\Phi _{11}^{11} = \left[ {\begin{array}{*{20}{c}}
  {{\Theta _{1,1}}}&{{{\tilde Z}_1}}&0&0&{{{\tilde Z}_3}}&0&{BY} \\
   * &{{\Theta _{2,2}}}&{{{\tilde Z}_2}}&0&0&0&{{{\tilde Z}_4}} \\
   * & * &{{\Theta _{3,3}}}&{{{\tilde Z}_2}}&0&0&0 \\
   * & * & * &{{\Theta _{4,4}}}&0&0&0 \\
   * & * & * & * &{{\Theta _{5,5}}}&{{{\tilde Z}_3}}&0 \\
   * & * & * & * & * &{{\Theta _{6,6}}}&0 \\
   * & * & * & * & * & * &{{\Theta _{7,7}}}
\end{array}} \right],$\\
$\Phi _{11}^{12} = {[0,0,0,0,0,0,{\tilde Z}_4]},$\\
$\begin{gathered}
{\Phi _{12}} = \left[ {\begin{array}{*{20}{c}}
  {{\Theta _{1,9}}}&{{\Theta _{1,10}}}&{{\Theta _{1,11}}}&{{\Theta _{1,12}}} \\
  0&0&0&0 \\
  0&0&0&0 \\
  0&0&0&0 \\
  0&0&0&0 \\
  0&0&0&0 \\
  {{\underline \lambda}{Y^T}{B^T}}&{{\lambda_2}{Y^T}{B^T}}&{{\overline \tau} {Y^T}{B^T}}&{{\lambda _1}{Y^T}{B^T}} \\
  0&0&0&0
\end{array}} \right]
\end{gathered},$\\
${\Phi _{13}} = [E,0,0,0,0,0,0,0]^T,$\\
$\begin{gathered}
{\Phi _{2,2}} = \left[ {\begin{array}{*{20}{c}}
  {{\Theta _{9,9}}}&{{\Theta _{9,10}}}&{{\Theta _{9,11}}}&{{\Theta _{9,12}}} \\
  *&{{\Theta _{10,10}}}&{{\Theta _{10,11}}}&{{\Theta _{10,12}}} \\
  *&*&{{\Theta _{11,11}}}&{{\Theta _{11,12}}} \\
  *&*&*&{{\Theta _{12,12}}}
\end{array}} \right]
\end{gathered},$\\
${\lambda _1} = \bar \lambda   + \bar \tau  ,{\lambda _2} = \bar \lambda   - \underline \lambda  ,{\lambda _3} = {\lambda _1} - \underline \lambda,$\\
${\Theta _{1,1}} = AX + X{A^T} + \sum\limits_i^7 {{{\tilde Q}_i}}  - {\tilde Z_1} - {\tilde Z_3} + \epsilon D{D^T},$\\
${\Theta _{2,2}} =  - {\tilde Q_1} - {\tilde Z_1} - {\tilde Z_2} - {\tilde Z_4},{\Theta _{3,3}} =  - \left( {1 - {\varepsilon _1}} \right){\tilde Q_2} - 2{\tilde Z_2},$\\
${\Theta _{4,4}} =  - {\tilde Q_3} - {\tilde Z_2},{\Theta _{5,5}} =  - \left( {1 - {\varepsilon _2}} \right){\tilde Q_4} - 2{\tilde Z_3},$\\
${\Theta _{6,6}} =  - {\tilde Q_5} - {\tilde Z_3},{\Theta _{7,7}} =  - 2{\tilde Z_4},{\Theta _{8,8}} =  - {\tilde Q_7} - {\tilde Z_4},$\\
${\Theta _{1,9}} = {\underline \lambda}X{A^T} + \epsilon {\underline \lambda}D{D^T},{\Theta _{1,10}} = {\lambda _2}X{A^T} + \epsilon {\lambda _2}D{D^T},$\\
${\Theta _{1,11}} = {\overline \tau} X{A^T} + \epsilon {\overline \tau} D{D^T},{\Theta _{1,12}} = {\lambda _3}X{A^T} + \epsilon {\lambda _3}D{D^T},$\\
${\Theta _{9,9}} =  - 2{\theta _1}X + \theta _1^2{\tilde Z_1} + \epsilon {\lambda_1^2}D{D^T},{\Theta _{9,10}} = \epsilon {\underline \lambda}{\lambda_{2}}D{D^T}$\\
${\Theta _{9,11}} = \epsilon {\underline \lambda}{\overline \tau} D{D^T},{\Theta _{9,12}} = \epsilon {\underline \lambda}{\lambda _3}D{D^T},$\\
${\Theta _{10,10}} =  - 2{\theta _2}X + \theta _2^2{\tilde Z_2} + \epsilon {\lambda _2}^2D{D^T},{\Theta _{10,11}} = \epsilon {\lambda _2}{\overline \tau }D{D^T},$\\
${\Theta _{10,12}} = \epsilon {\lambda _2}{\lambda _3}D{D^T},{\Theta _{11,11}} =  - 2{\theta _3}X + \theta _3^2{\tilde Z_3} + \epsilon {{\overline \tau } ^2}D{D^T},$\\
${\Theta _{11,12}} = \epsilon {\overline \tau }{\lambda _3}D{D^T},{\Theta _{12,12}} =  - 2{\theta _4}X + \theta _4^2{\tilde Z_4} + \epsilon {\lambda _3}^2D{D^T},$\\
then the closed-loop system \eqref{eq3} with gain $K = Y{X^{ - 1}}$ is asymptotically stable for parametric uncertainty satisfying $\left\| {F(t)} \right\| \leqslant 1$ and time delays in \eqref{eq3}.
\end{theorem}

\emph{Proof:} First, we present the following lemmas, which forms foundation for deriving major Theorem.

\emph{Lemma 1 \cite{S1}:} For any constant matrix $R \in {\mathbb{R} ^{n \times n}}, R = {R^T} > 0$, scalar $\gamma  > 0$ , and vector function $\dot x:\left[ { - \gamma ,0} \right] \to {\mathbb{R}^n}$, such that the following integration is well defined, then
\be
\begin{gathered}
   - \gamma \int_{t - \gamma }^t {{{\dot x}^T}\left( \omega  \right)} R\dot x\left( \omega  \right)d\omega  \hfill \\
   \leqslant {\left( {\begin{array}{*{20}{c}}
  {x\left( t \right)} \\
  {x\left( {t - \gamma } \right)}
\end{array}} \right)^T}\left( {\begin{array}{*{20}{c}}
  { - R}&R \\
  *&{ - R}
\end{array}} \right)\left( {\begin{array}{*{20}{c}}
  {x\left( t \right)} \\
  {x\left( {t - \gamma } \right)}
\end{array}} \right) \hfill \\
\end{gathered}
\label{eqA15}
\ee
\emph{Lemma 2 \cite{S2}:} For any constant matrix $R \in {\mathbb{R} ^{n \times n}}, R = {R^T} > 0$, scalar $0 < {d_1} \leqslant d\left( t \right) \leqslant {d_2}$, and vector function $\dot x:\left[ { - {d_2}, - {d_1}} \right] \to {\mathbb{R}^n}$, such that the following integration is well defined, then
\be
- \left( {{d_2} - {d_1}} \right)\int_{t - {d_2}}^{t - {d_1}} {{{\dot x}^T}\left( \omega  \right)R\dot x\left( \omega  \right)d\omega }  \leqslant {H^T}\left( t \right)\Omega H\left( t \right)
\label{eqA16}
\ee
where
\[H\left( t \right) = \left[ \begin{gathered}
  x\left( {t - {d_1}} \right) \hfill \\
  x\left( {t - d\left( t \right)} \right) \hfill \\
  x\left( {t - {d_2}} \right) \hfill \\
\end{gathered}  \right],{\text{ }}\Omega  = \left[ {\begin{array}{*{20}{c}}
  { - R}&R&0 \\
  *&{ - 2R}&R \\
  *&*&{ - R}
\end{array}} \right]\]

Next, we will prove Theorem 1. An Lyapunov-Krasovskii functional is defined as
\be
V\left( {x\left( t \right)} \right) = {V_1} + {V_2} + {V_3} + {V_4} + {V_5} + {V_6} + {V_7}
\label{eqA17}
\ee
where\\
${V_1} = {x^T}\left( t \right)Px\left( t \right)$\\
$\begin{gathered}
  {V_2} = \int_{t - {\underline \lambda}}^t {{x^T}} \left( s \right){Q_1}x\left( s \right)ds \hfill \\
   + \int_{t - \lambda \left( t \right)}^t {{x^T}} \left( s \right){Q_2}x\left( s \right)ds + \int_{t - {\overline \lambda}}^t {{x^T}} \left( s \right){Q_3}x\left( s \right)ds \hfill \\
\end{gathered} $\\
$\begin{gathered}
  {V_3} = \int_{ - {\underline \lambda}}^0 {\int_{t + \theta }^t {{\underline \lambda}{{\dot x}^T}\left( s \right){Z_1}\dot x\left( s \right)dsd\theta } }  \hfill \\
   + \int_{ - {\overline \lambda}}^{ - {\underline \lambda}} {\int_{t + \theta }^t {{\lambda _{1}}} } {{\dot x}^T}\left( s \right){Z_2}\dot x\left( s \right)dsd\theta  \hfill \\
\end{gathered} $\\
${V_4} = \int_{t - \tau \left( t \right)}^t {{x^T}} \left( s \right){Q_4}x\left( s \right)ds + \int_{t - {\overline \tau } }^t {{x^T}} \left( s \right){Q_5}x\left( s \right)ds$\\
${V_5} = \int_{ - {\overline \tau }  }^0 {\int_{t + \theta }^t {\overline \tau }   } {\dot x^T}\left( s \right){Z_3}\dot x\left( s \right)dsd\theta $\\
${V_6} = \int_{t - \lambda \left( t \right) - \tau \left( t \right)}^t {{x^T}\left( s \right){Q_6}x\left( s \right)} ds + \int_{t - {\lambda _2}}^t {{x^T}\left( s \right){Q_7}x\left( s \right)} ds$\\
${V_7} = \int_{ - {\lambda _2}}^{ - {\underline \lambda}} {\int_{t + \theta }^t {{\lambda _{3}}} } {\dot x^T}\left( s \right){Z_4}\dot x\left( s \right)dsd\theta $\\
${\lambda _1} = \overline \lambda   + \overline \tau  ,{\lambda _2} = \overline \lambda   - \underline \lambda  ,{\lambda _3} = {\lambda _1} - \underline \lambda$

Taking derivative of $V\left( {x\left( t \right)} \right)$ with respect to $t$ and using Lemmas 1 and 2 lead to
\be
\dot V\left( {x\left( t \right)} \right) \leqslant {\xi ^T}\left( t \right)\Lambda \xi \left( t \right)
\label{eqA18}
\ee
where\\
$\begin{gathered}
  {\xi ^T}\left( t \right) = \left[ {\begin{array}{*{20}{c}}
  {{x^T}\left( t \right)}&{{x^T}\left( {t - {\underline \lambda}} \right)}&{{x^T}\left( {t - \lambda \left( t \right)} \right)}
\end{array}} \right. \hfill \\
  \begin{array}{*{20}{c}}
  {{x^T}\left( {t - {\overline \lambda}} \right)}&{{x^T}\left( {t - \tau \left( t \right)} \right)}&{{x^T}\left( {t - {\overline \tau} } \right)}
\end{array} \hfill \\
  \left. {\begin{array}{*{20}{c}}
  {{x^T}\left( {t - \lambda \left( t \right) - \tau \left( t \right)} \right)}&{{x^T}\left( {t - {\underline \lambda}} \right)}
\end{array}} \right] \hfill \\
\end{gathered}$\\
$\Lambda  = \left[ {\begin{array}{*{20}{c}}
  {{\Lambda _{1,1}}}&{{\Lambda _{1,2}}} \\
   * &{{\Lambda _{2,2}}}
\end{array}} \right]$\\
${\Lambda _{1,1}} = \left[ {\begin{array}{*{20}{c}}
  {{\Psi _{1,1}}}&{{Z_1}}&0&0&{{Z_3}}&0&{{\Psi _{1,7}}} \\
   * &{{\Psi _{2,2}}}&{{Z_2}}&0&0&0&{{Z_4}} \\
   * & * &{{\Psi _{3,3}}}&{{Z_2}}&0&0&0 \\
   * & * & * &{{\Psi _{4,4}}}&0&0&0 \\
   * & * & * & * &{{\Psi _{5,5}}}&{{Z_3}}&0 \\
   * & * & * & * & * &{{\Psi _{6,6}}}&0 \\
   * & * & * & * & * & * &{{\Psi _{7,7}}}
\end{array}} \right]$\\
${\Lambda _{12}} = {\left[ {\begin{array}{*{20}{c}}
  0&0&0&0&0&0&{{Z_4}}
\end{array}} \right]^T}$\\
${\Lambda _{22}} = {\Psi _{8,8}} =  - {Q_7} - {Z_4}$\\
${\Psi _{1,1}} = P\bar A + {\bar A^T}P + \sum\limits_1^7 {{Q_i}}  - {Z_1} - {Z_3} + {\bar A^T}\bar Z{\text{ }}\bar A$\\
${\Psi _{2,2}} =  - {Q_1} - {Z_1} - {Z_2} - {Z_4}$\\
${\Psi _{3,3}} =  - \left( {1 - {\varepsilon_1}} \right){Q_2} - 2{Z_2}$\\
${\Psi _{4,4}} =  - {Q_3} - {Z_2}$\\
${\Psi _{5,5}} =  - \left( {1 - {\varepsilon _2}} \right){Q_4} - 2{Z_3}$\\
${\Psi _{6,6}} =  - {Q_5} - {Z_3}$\\
${\Psi _{7,7}} =  - 2{Z_4} + {K^T}{B^T}\bar ZBK$\\
$\bar A = DF(t)E$\\
$\bar Z = {\underline \lambda}^2{Z_1} + \lambda _{2}^2{Z_2} + {{\overline \tau} ^2}{Z_3} + \lambda _{3}^2{Z_4}$\\
If $\Lambda  < 0$, closed-loop system (11) is asymptotically stable. Furthermore, we can get
\be
\overset{\lower0.5em\hbox{$\smash{\scriptscriptstyle\frown}$}}{\Lambda }  + {\Delta ^T}\overset{\lower0.5em\hbox{$\smash{\scriptscriptstyle\frown}$}}{Z} \Delta  < 0
\label{eqA19}
\ee
where ${\Psi _{1,1}}$ is replaced by ${\overset{\lower0.5em\hbox{$\smash{\scriptscriptstyle\frown}$}}{\Psi } _{1,1}} = P\bar A + {\bar A^T}P + \sum\limits_1^7 {{Q_i}}  - {Z_1} - {Z_3}$,
${\Psi _{1,7}}$ is replaced by ${\overset{\lower0.5em\hbox{$\smash{\scriptscriptstyle\frown}$}}{\Psi } _{1,1}} = PBK$ while elements in other positions remain unchanged in matrix $\Lambda $, and\\
$\Delta  = \left[ {\begin{array}{*{20}{c}}
  {{\Delta _1}}&{{{\mathbf{0}}_{4 \times 5}}}&{{\Delta _3}}&{{{\mathbf{0}}_{4 \times 1}}}
\end{array}} \right]$\\
$\Delta _1^T = \left[ {{\underline \lambda}{{\bar A}^T}{Z_1}{\text{ }}{\lambda _{2}}{{\bar A}^T}{Z_2}{\text{ }}{\overline \tau} {{\bar A}^T}{Z_3}{\text{ }}{\lambda _{3}}{{\bar A}^T}{Z_4}} \right]$\\
$\Delta _3^T = \left[ {{\underline \lambda}{K^T}{B^T}{Z_1}{\text{ }}{\lambda _{2}}{K^T}{B^T}{Z_2}{\text{ }}{\overline \tau} {K^T}{B^T}{Z_3}{\text{ }}{\lambda _{3}}{K^T}{B^T}{Z_4}} \right]$\\
$\overset{\lower0.5em\hbox{$\smash{\scriptscriptstyle\frown}$}}{Z}  = \left[ {\begin{array}{*{20}{c}}
  { - {Z_1}^{ - 1}}&0&0&0 \\
   * &{ - {Z_2}^{ - 1}}&0&0 \\
   * & * &{ - {Z_3}^{ - 1}}&0 \\
   * & * & * &{ - {Z_4}^{ - 1}}
\end{array}} \right]$\\
Using Schur supplement, it can get
\be
\left[ {\begin{array}{*{20}{c}}
  {{\prod _{1,1}}}&{{\prod _{1,2}}} \\
  *&{{\prod _{2,2}}}
\end{array}} \right] < 0
\label{eqA20}
\ee
where
${\prod _{1,1}} = \overset{\lower0.5em\hbox{$\smash{\scriptscriptstyle\frown}$}}{\Lambda } ,{\prod _{1,2}} = {\Delta ^T},{\prod _{2,2}} = \overset{\lower0.5em\hbox{$\smash{\scriptscriptstyle\frown}$}}{Z} $.

Substituting $\bar A = A + DF(t)E$ into \eqref{eqA20}, and pre- and post-multiplying \eqref{eqA20} with
\[diag\left\{ {\underbrace {{P^{ - 1}}, \ldots ,{P^{ - 1}}}_8,\underbrace {Z_1^{ - 1}, \ldots ,Z_4^{ - 1}}_4} ,I \right\}\]
and its transpose respectively, and setting
\[X = {P^{ - 1}},{\tilde Q_i} = {P^{ - 1}}{Q_i}{P^{ - 1}}\left( {i = 1,2, \ldots ,7} \right)\]
\[{\tilde Z_j} = {P^{ - 1}}{Z_j}{P^{ - 1}}\left( {j = 1,2, \ldots ,4} \right),Y = KX,\]
we can convert most nonlinear variables into linear variables.

Since $\left( {\theta X - R} \right){R^{ - 1}}\left( {\theta X - R} \right) > 0$, where $R = {R^T}$ is a real symmetric matrix and $X$ is an appropriate dimensions matrix, it follows that
\[\left( {{\theta _j}{P^{ - 1}} - Z_j^{ - 1}} \right)Z_j\left( {{\theta _j}{P^{ - 1}} - Z_j^{ - 1}} \right) > 0\]
i.e.,
\[- Z_j^{ - 1} <  - 2{\theta _j}{P^{ - 1}} + \theta _j^2{P^{ - 1}}{Z_j}{P^{ - 1}}\left( {j = 1,2, \ldots ,4} \right).\]
It completes the proof.
\endproof

\begin{remark}
Adverse effects of F2SIE algorithm and image attacks have been analysed, which will produce extra computational delay and extra computational errors. According to these adverse effects, a new closed-loop uncertain time-delay model \eqref{eq3} of new NIPVSS is established. For \eqref{eq3},  an error-as-parameter-uncertainty robust controller is designed by Theorem 1, which can achieve system asymptotic stability. Specially, unlike \cite{8}, new robust controller can handle the following two cases: 1) When uncertainty ranges of cart position and pendulum angle under image attacks are within $ 0.4 + \Delta^M_{1,a}$, $0.82 + \Delta^M _{2,a}$, the controller is robust to parameter uncertainty from computational errors, i.e., system remains stability; 2) When $\tau(t)$, $\lambda(t)$ are within $[0,\bar \tau]$, $[\underline \lambda, \bar \lambda]$, the controller is robust to these two delays partly from image encryption and decryption, i.e., system is still stable. Therefore, the challenge 2 is solved.
\end{remark}

\section{Experimental Results and Discussion}
\begin{figure}[!t]
     \centering
       \includegraphics[width=0.4\textwidth]{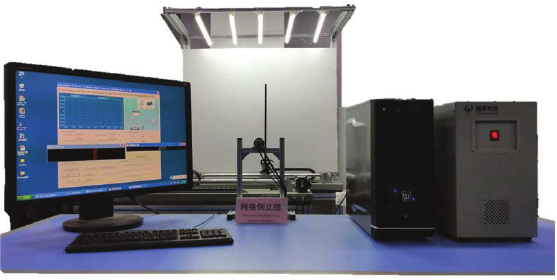}
     \caption{Experimental platform of  new NIPVSS fused with the proposed F2SIE algorithm and robust controller.}
     \label{fig10}
\end{figure}

To verify the proposed F2SIE algorithm and robust controller,  real-time control experiments are carried out, where new NIPVSS platform based on \cite{plat} is shown in Fig.~{\ref{fig10}}.

\subsection{Performance of F2SIE Algorithm}
The performance of F2SIE algorithm will be analysed by six aspects and the analysis results are summarized in Tab.~\ref{tabAII}. The real-time performance of encryption is firstly analysed, because only when real-time performance is satisfied, system can run stably.
\begin{table}[!t]
\caption{Performance of the F2SIE algorithm.}
\label{tabAII}
\centering
\begin{tabular}{ccc}
  \toprule
  ~ &  \cite{24}  & F2SIE \\
  \midrule
  Running Time(s) &  0.041  & 0.014 \\
  \midrule
  NPCR(\%) &  99.61   & 99.58 \\
  UACI(\%) &  33.42  & 33.53\\
  \midrule
  Horizontal(\%)  & 0.0038   & 0.0040 \\
  Vertical(\%) & 0.0048  & -0.0022\\
  Diagonal(\%) & 0.0044  & 0.0029\\
  \midrule
  $H$  & 7.9979  & 7.9898 \\
  \bottomrule
\end{tabular}
\end{table}

\subsubsection{Algorithm Running Time}
Real-time performance is one of key factors for stable operation of NIPVSS, where image scaling and selection in F2SIE algorithm can reduce amount of image data. Moreover, F2SIE algorithm reduces encryption time by generating row-column-level random numbers for encryption. The second row of Tab.~\ref{tabAII} shows running time of different algorithms, where it can be seen that F2SIE algorithm meets real-time requirement in comparison with \cite{24}.

\subsubsection{Histogram.} The histogram of an image is a function of intensity, which describes number or frequency of pixels in image. The more the effectiveness of encryption is, the smoother the gray level histogram of image is. The histograms of original image and the encrypted image are shown in  Fig.~\ref{figA10}. It can be seen that histogram of the encrypted area has a fairly uniform distribution obviously.

\subsubsection{Image sensitivity.} Usually, attackers will observe changes of the encrypted image according to a small point in image. For example, attackers discover relationship between normal image and the encrypted image by changing one pixel. This attack is called as differential attack. This attack can be avoided when the encrypted system is sensitive enough to modification of image. The sensitivity can be measured by two evaluation indicators \cite{S3}, i.e., rate of change of NPCR and rate of change of the average intensity UACI. The third row of Tab.~\ref{tabAII} shows comparison results and the proposed image encryption scheme can easily resist differential attacks.

\subsubsection{Correlation.} 
\begin{figure}[!t]
\centering
\subfigure[]{\includegraphics[width=3.8cm]{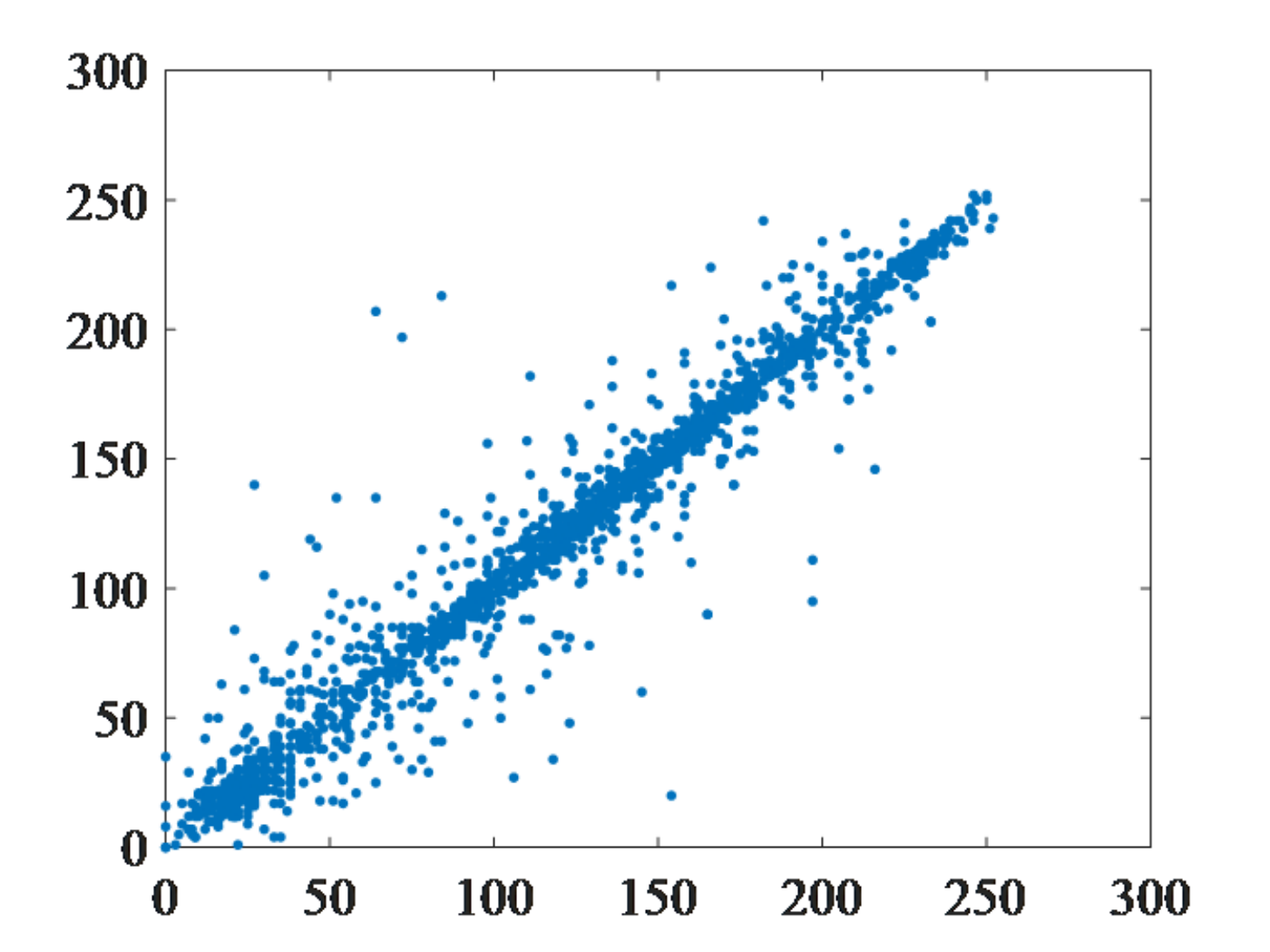}}\quad
\subfigure[]{\includegraphics[width=3.8cm]{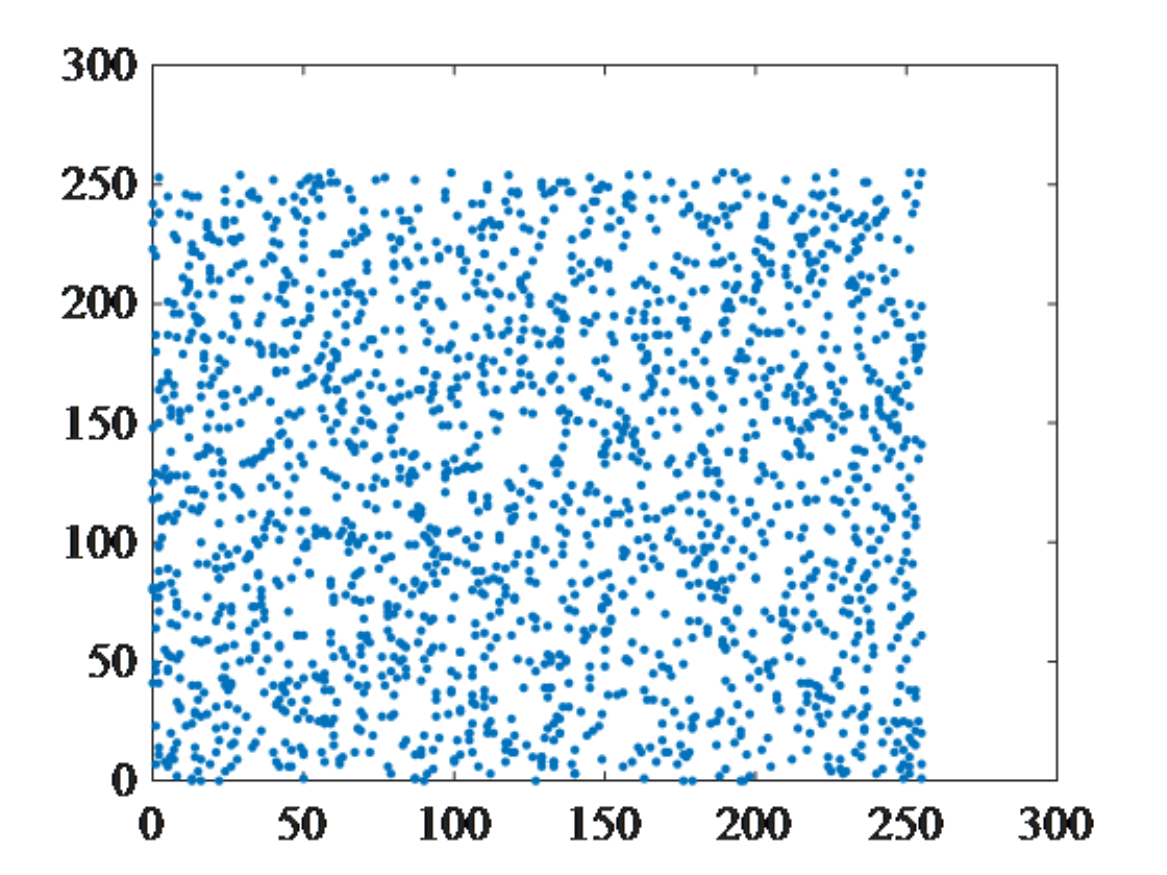}}\quad
\subfigure[]{\includegraphics[width=3.8cm]{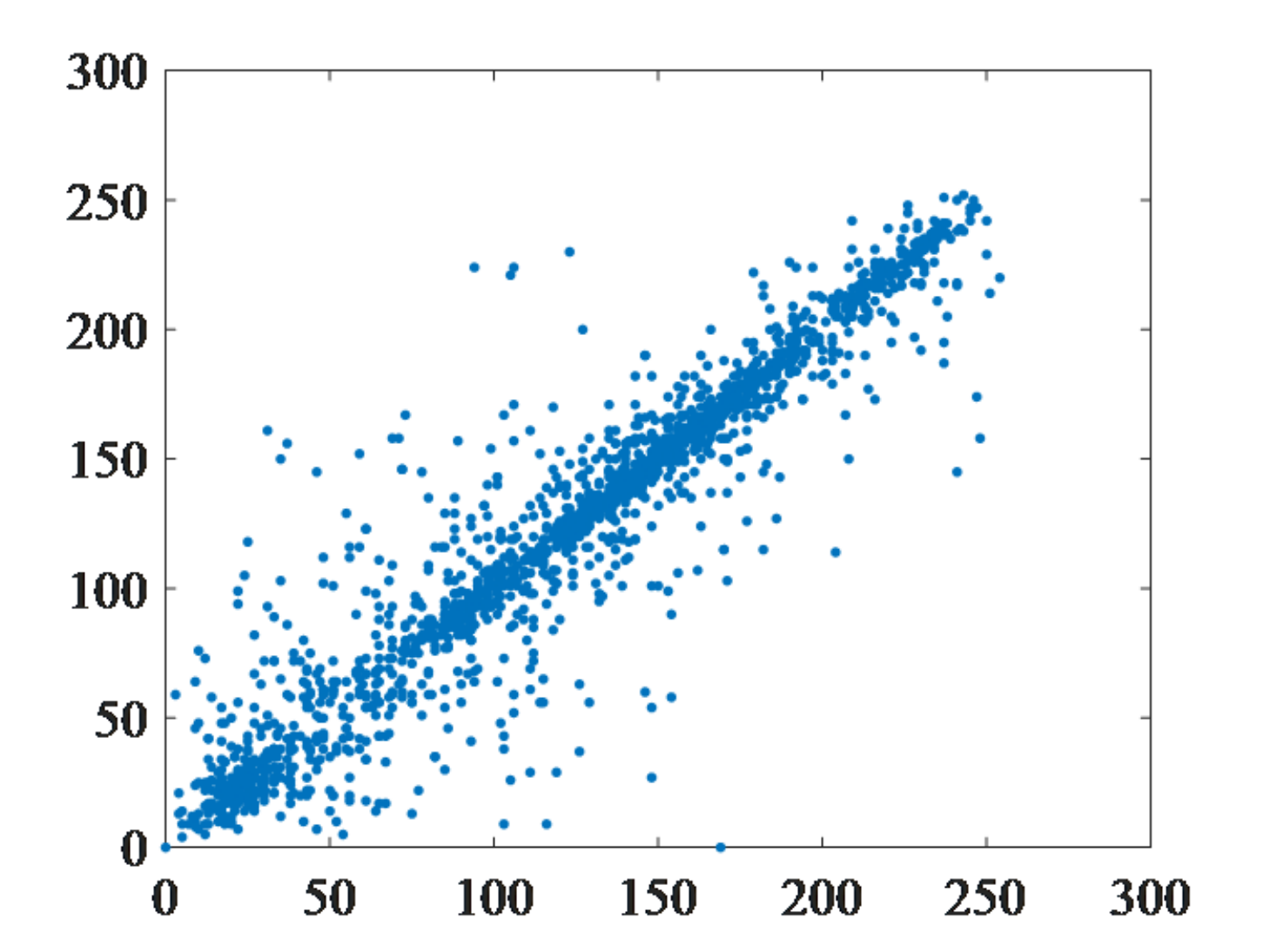}}\quad
\subfigure[]{\includegraphics[width=3.8cm]{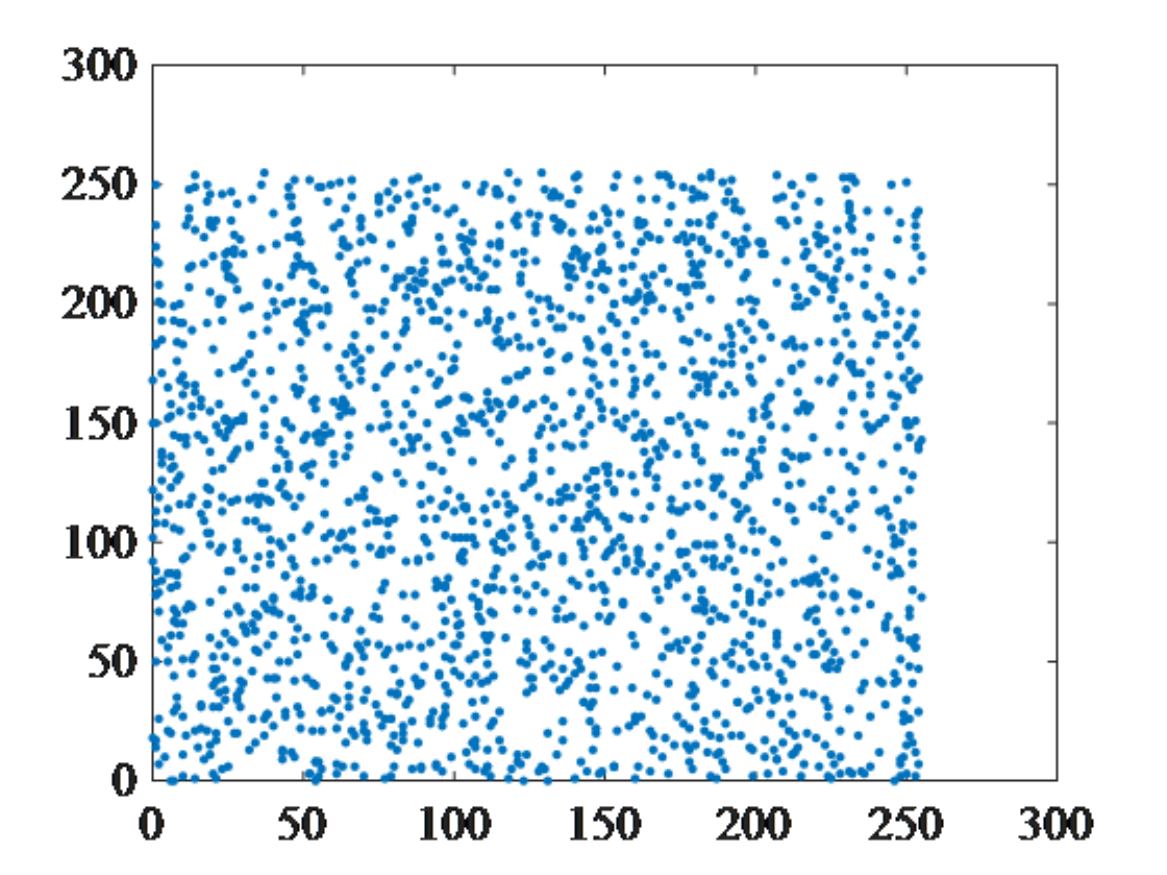}}\quad
\subfigure[]{\includegraphics[width=3.8cm]{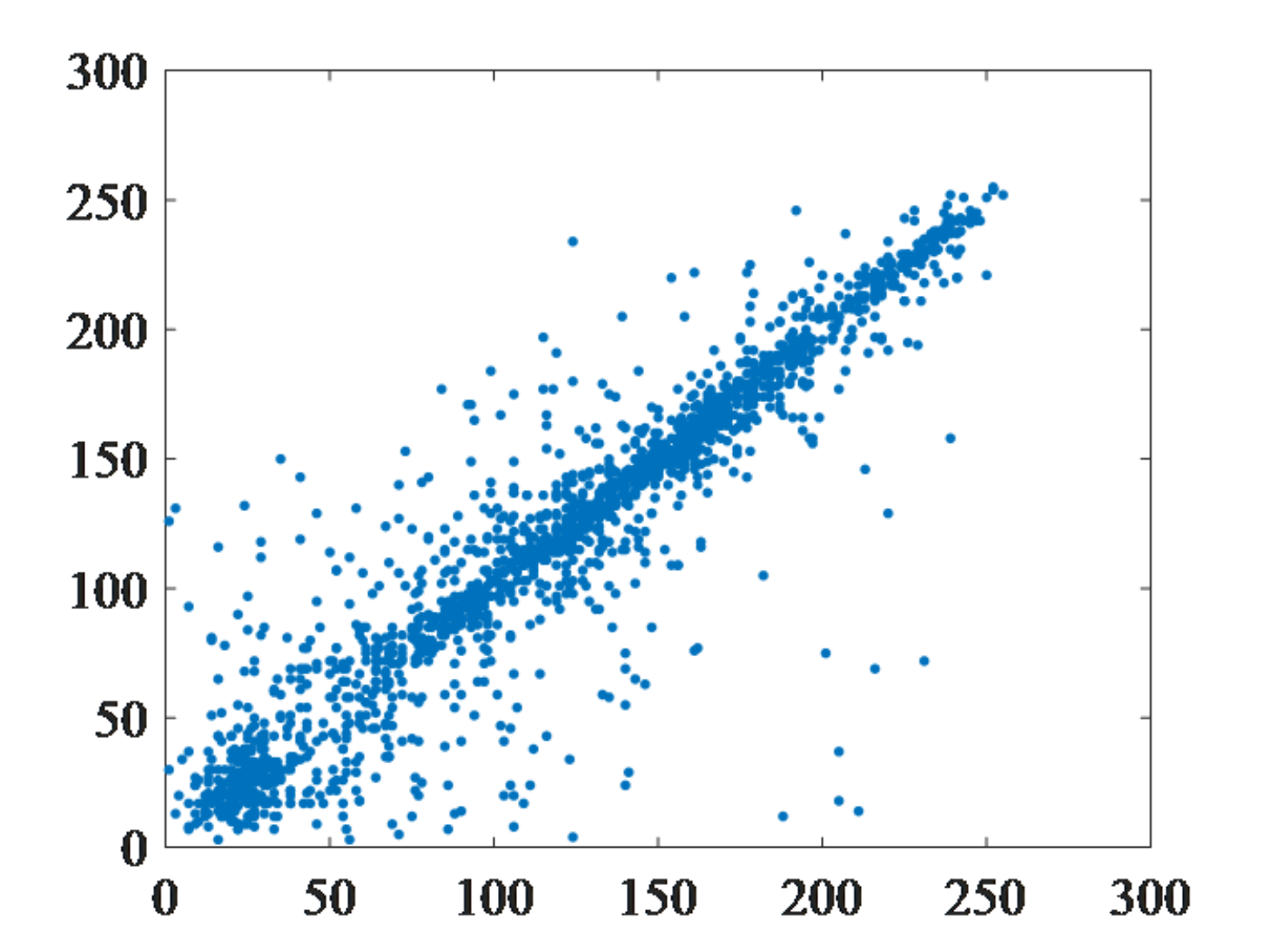}}\quad
\subfigure[]{\includegraphics[width=3.8cm]{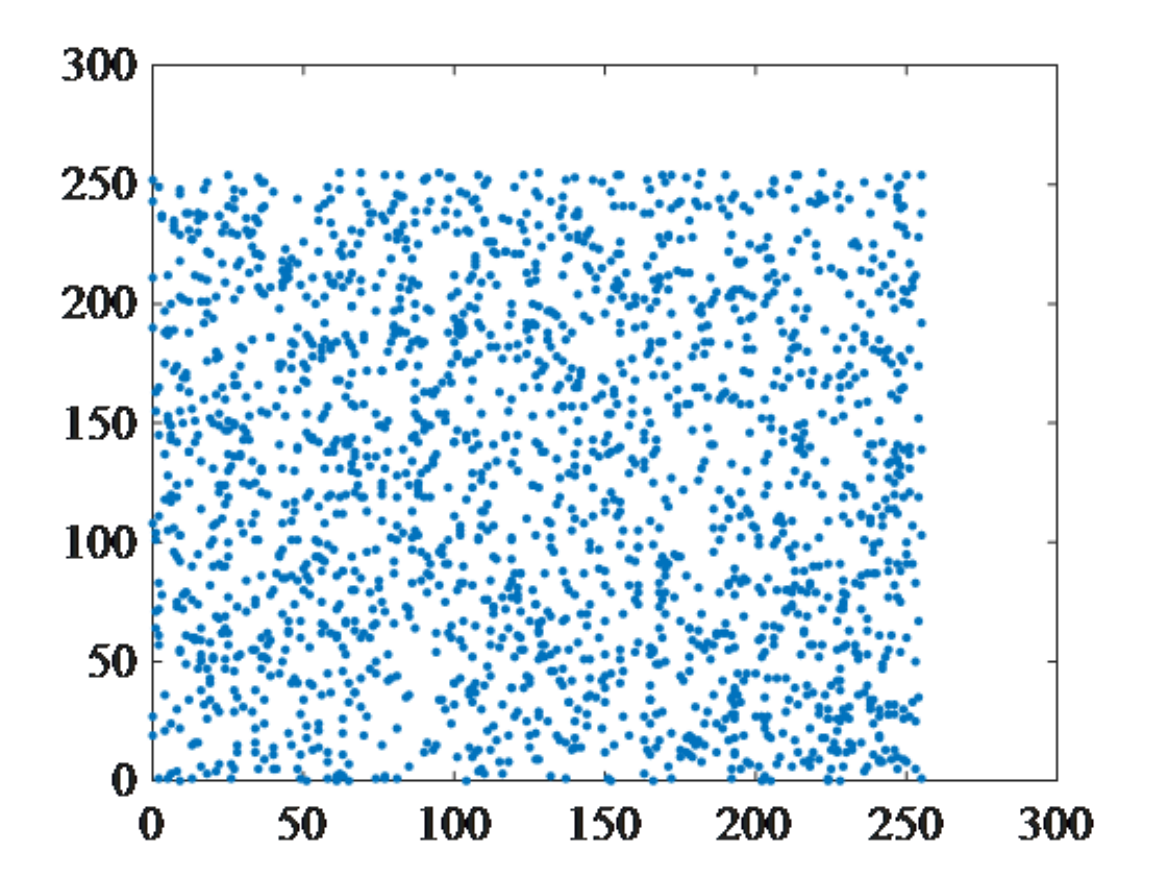}}
\caption{Correlation analysis of the original image and encrypted image: (a) (b) are horizontal direction; (c) (d) are vertical direction; (e) (f) are diagonal direction.}
\label{figA11}
\end{figure}

We compare relevant features of plain text image with its cipher text image. Generally, plain text image has a strong correlation between adjacent pixels in horizontal, vertical and diagonal directions, but there should be no correlation between adjacent pixels in an encrypted image. To measure correlation of image, 2000 pairs of adjacent pixels are randomly selected, and correlation coefficients of these pairs are calculated. It can be seen from Fig. \ref{figA11} that adjacent pixels of original image are highly correlated in horizontal, vertical and diagonal directions, while the encrypted image is distributed with little correlation. Furthermore, different correlation coefficients of the encrypted image are shown in the fourth row of Tab.~\ref{tabAII}, which show that correlation coefficients of horizontal, vertical and diagonal are all close to 0, indicating that security of image is improved.

\subsubsection{Information entropy.} Information entropy $H$ reflects uncertainty of image. It is generally believed that the greater the entropy, the greater the amount of information and the less visible information. The fifth row of Tab.~\ref{tabAII} shows results of information entropy and comparison with other encryption schemes. The result shows that the proposed encryption scheme has similar values.

\subsubsection{Image Robustness}
\begin{figure}[!t]
\centering
\subfigure[]{
\includegraphics[width=1.7cm]{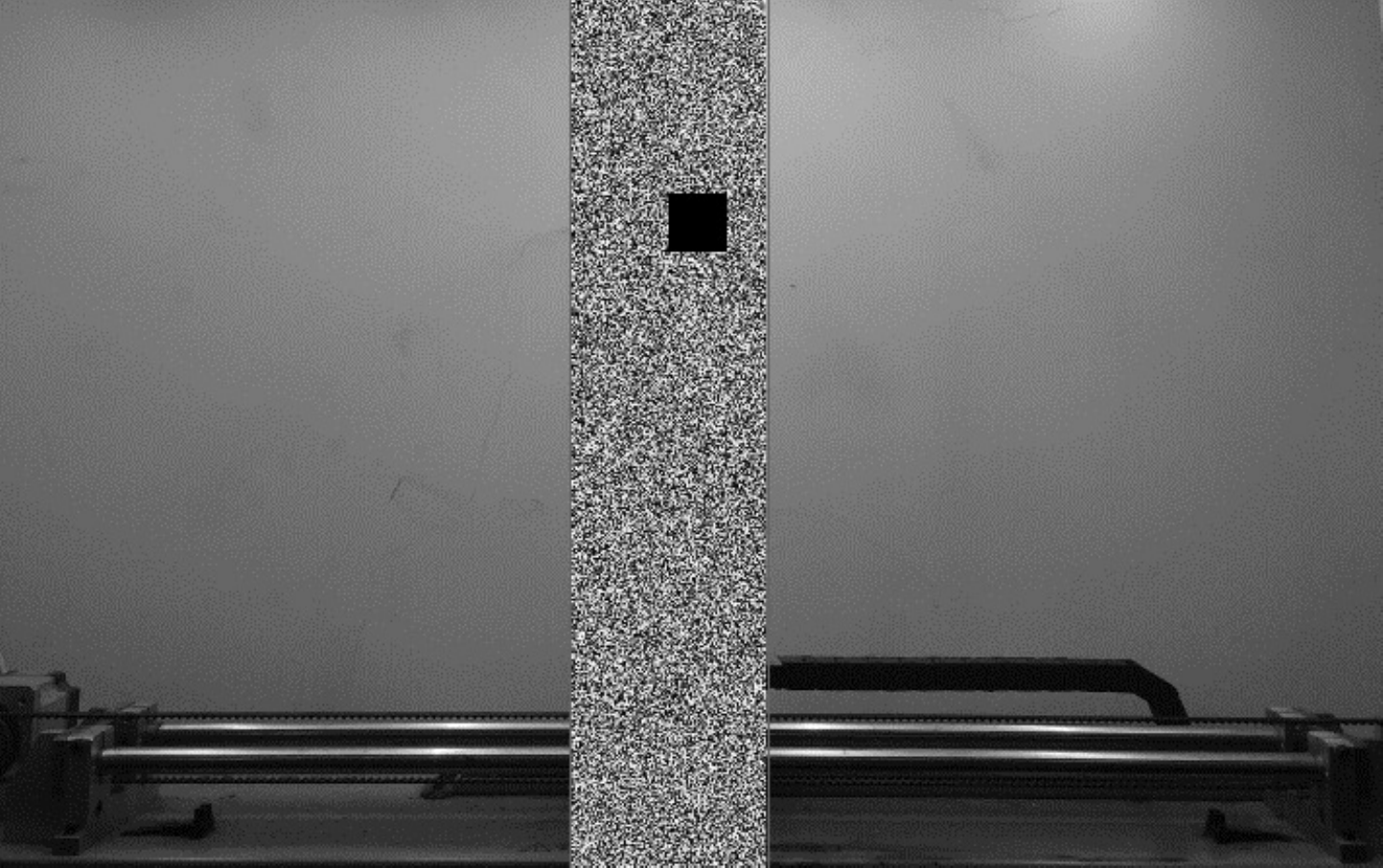}}
\hspace{-6.5mm}
\quad
\subfigure[]{
\includegraphics[width=1.7cm]{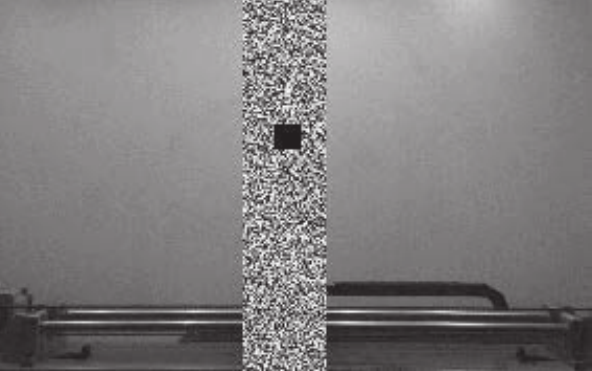}}
\hspace{-6.5mm}
\quad
\subfigure[]{
\includegraphics[width=1.7cm]{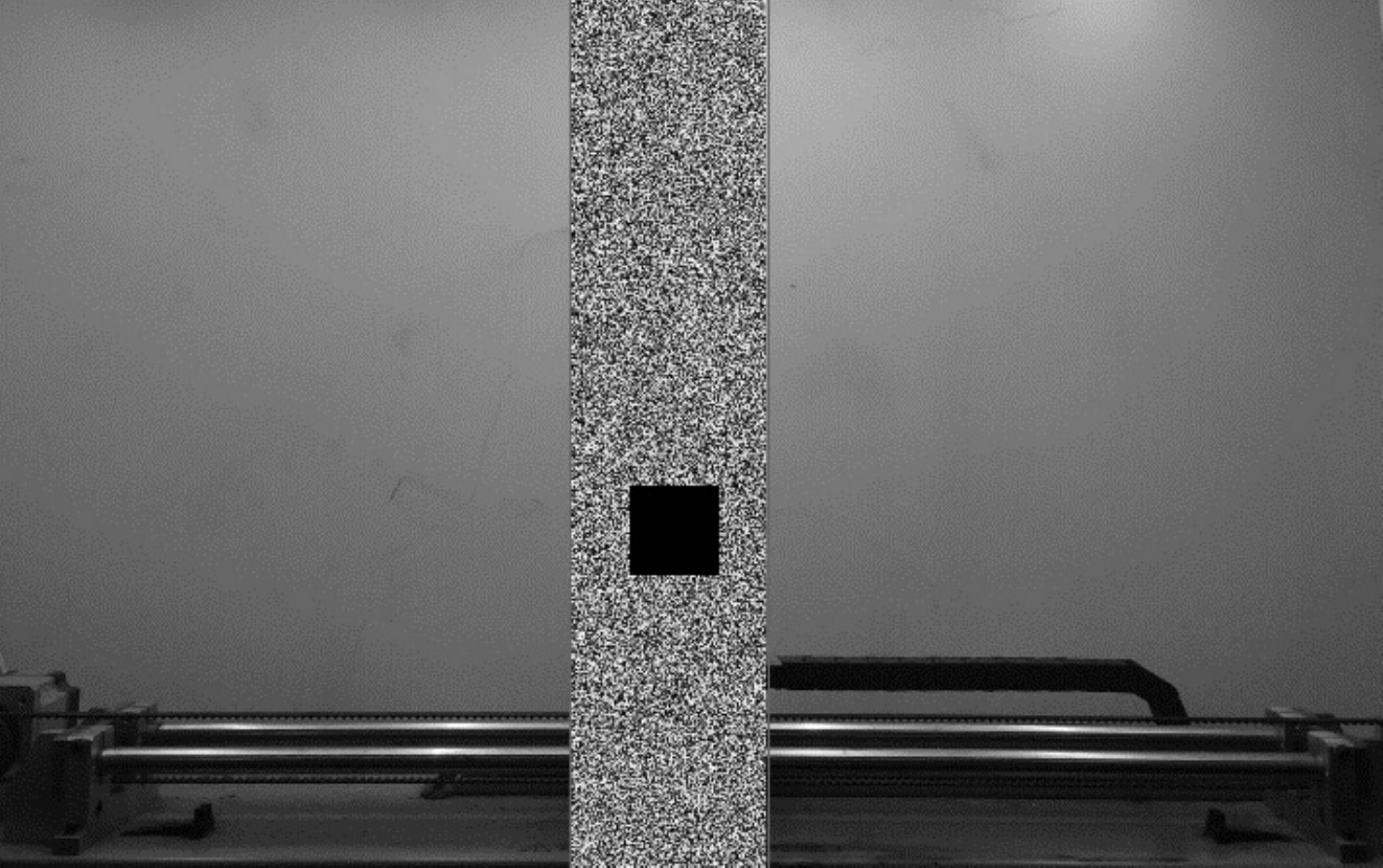}}
\hspace{-6.5mm}
\quad
\subfigure[]{
\includegraphics[width=1.7cm]{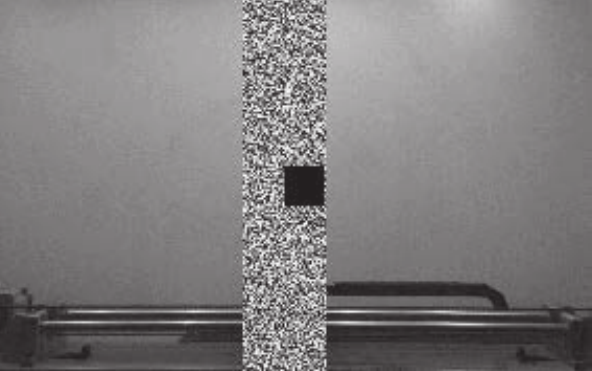}}
\hspace{-6.5mm}
\quad
\subfigure[]{
\includegraphics[width=1.7cm]{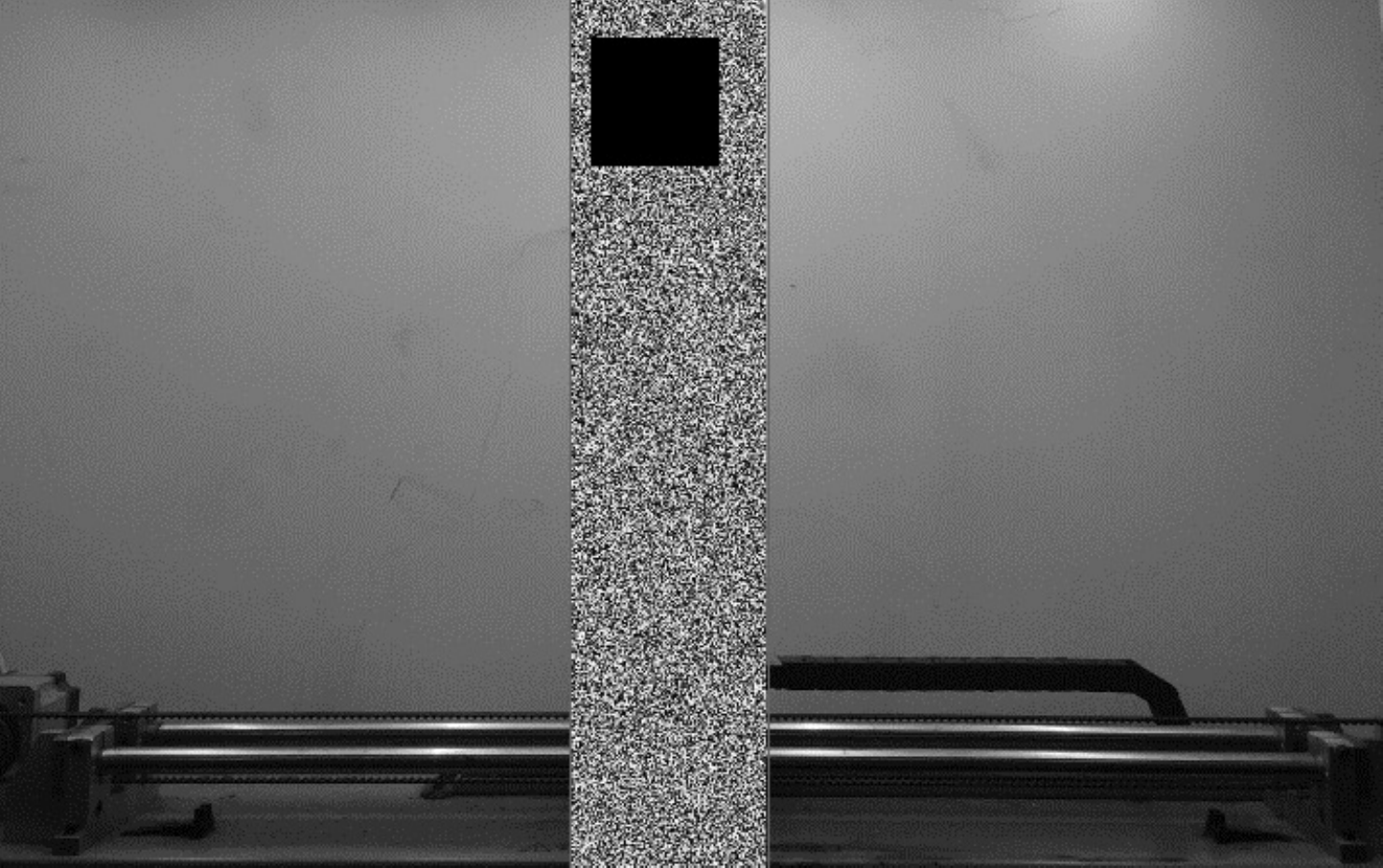}}
\hspace{-6.5mm}
\quad
\subfigure[]{
\includegraphics[width=1.7cm]{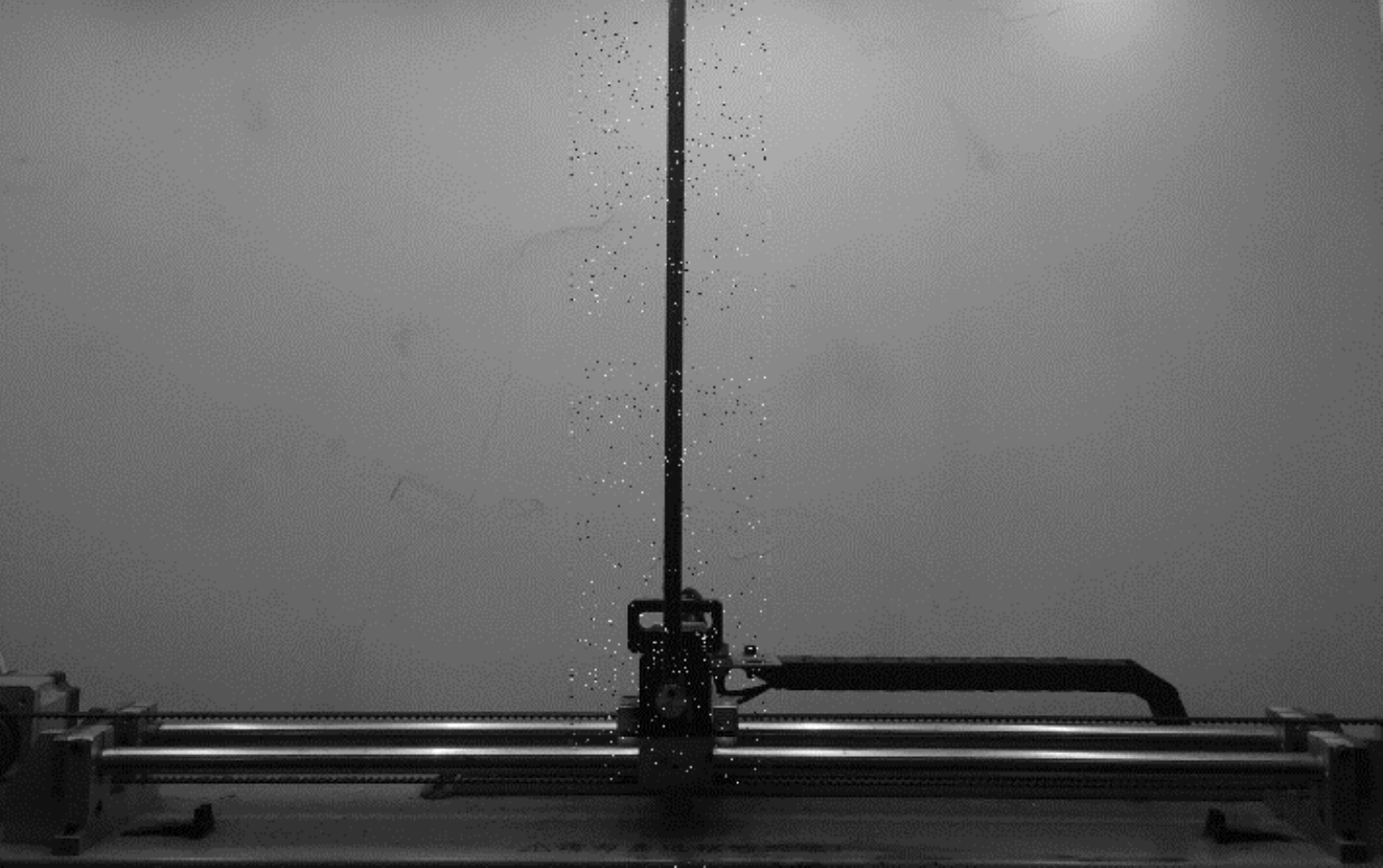}}
\hspace{-6.5mm}
\quad
\subfigure[]{
\includegraphics[width=1.7cm]{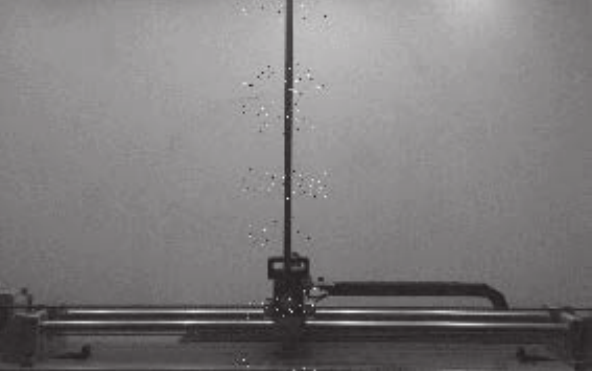}}
\hspace{-6.5mm}
\quad
\subfigure[]{
\includegraphics[width=1.7cm]{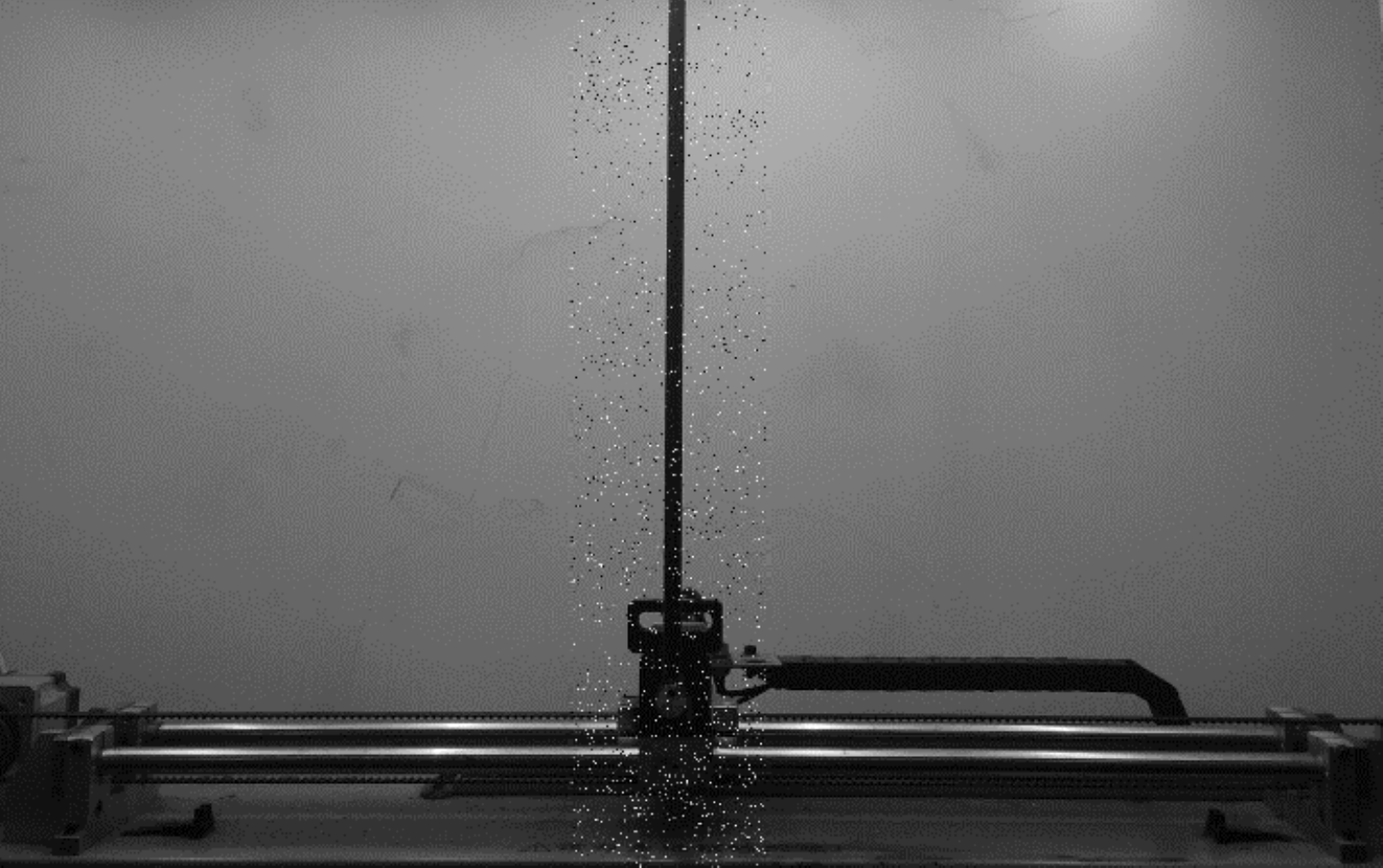}}
\hspace{-6.5mm}
\quad
\subfigure[]{
\includegraphics[width=1.7cm]{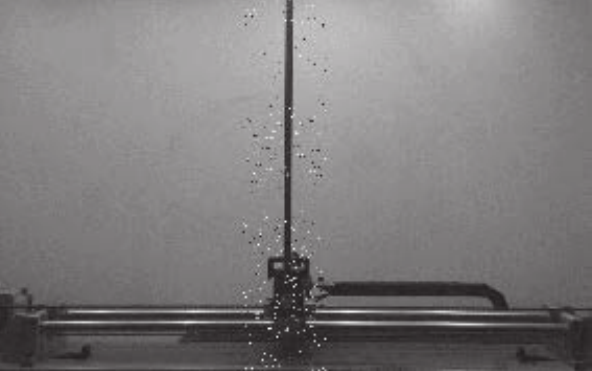}}
\hspace{-6.5mm}
\quad
\subfigure[]{
\includegraphics[width=1.7cm]{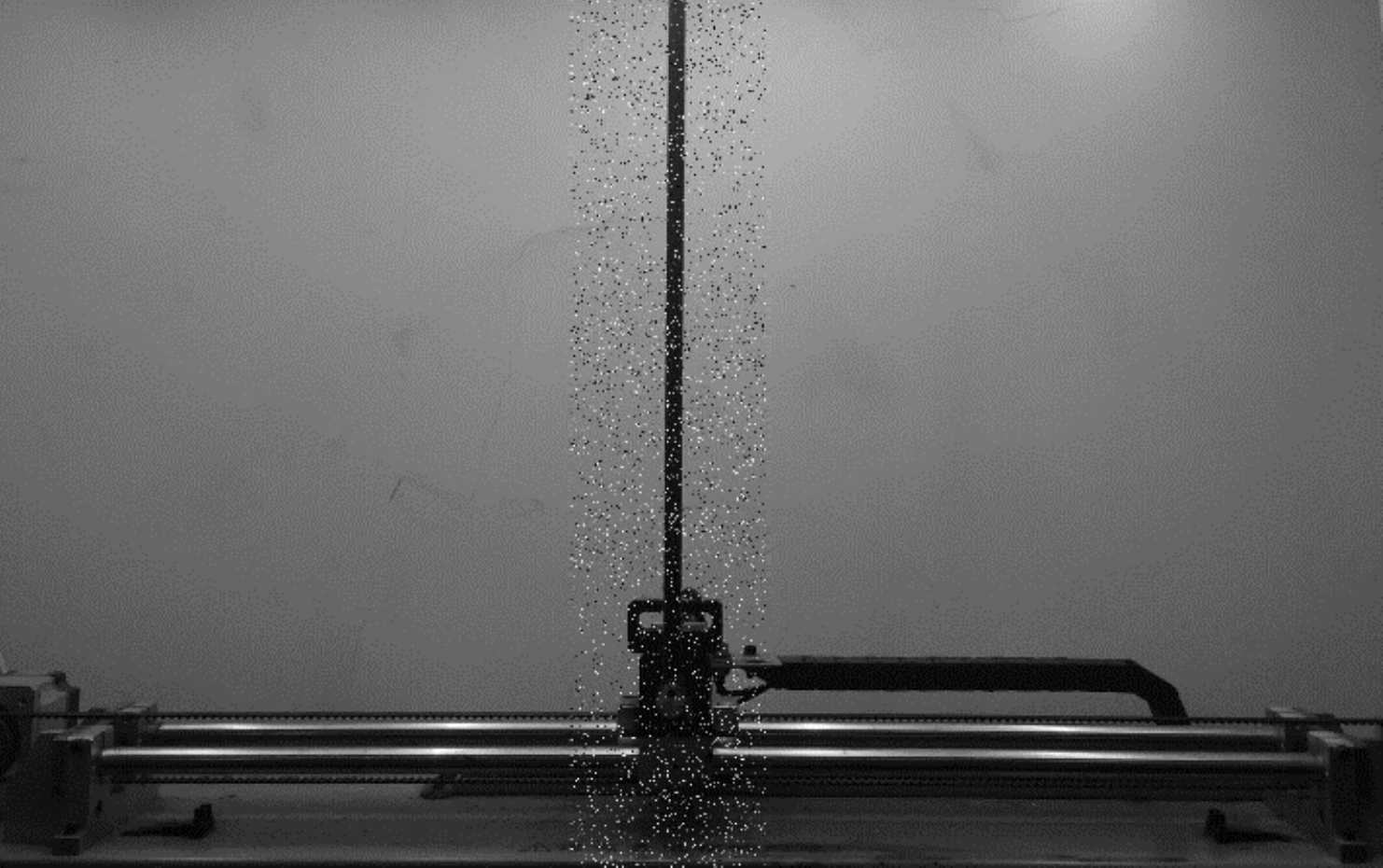}}
\caption{(a)-(e): The encrypted images under shearing attacks with different shear rates (1\%, 2\%, 4\%, 5\%, 6\%). (f)-(j): The decrypted images of (a)-(e).}
\label{fig4}
\end{figure}

\begin{figure}[!t]
\centering
\subfigure[]{
\includegraphics[width=1.5cm]{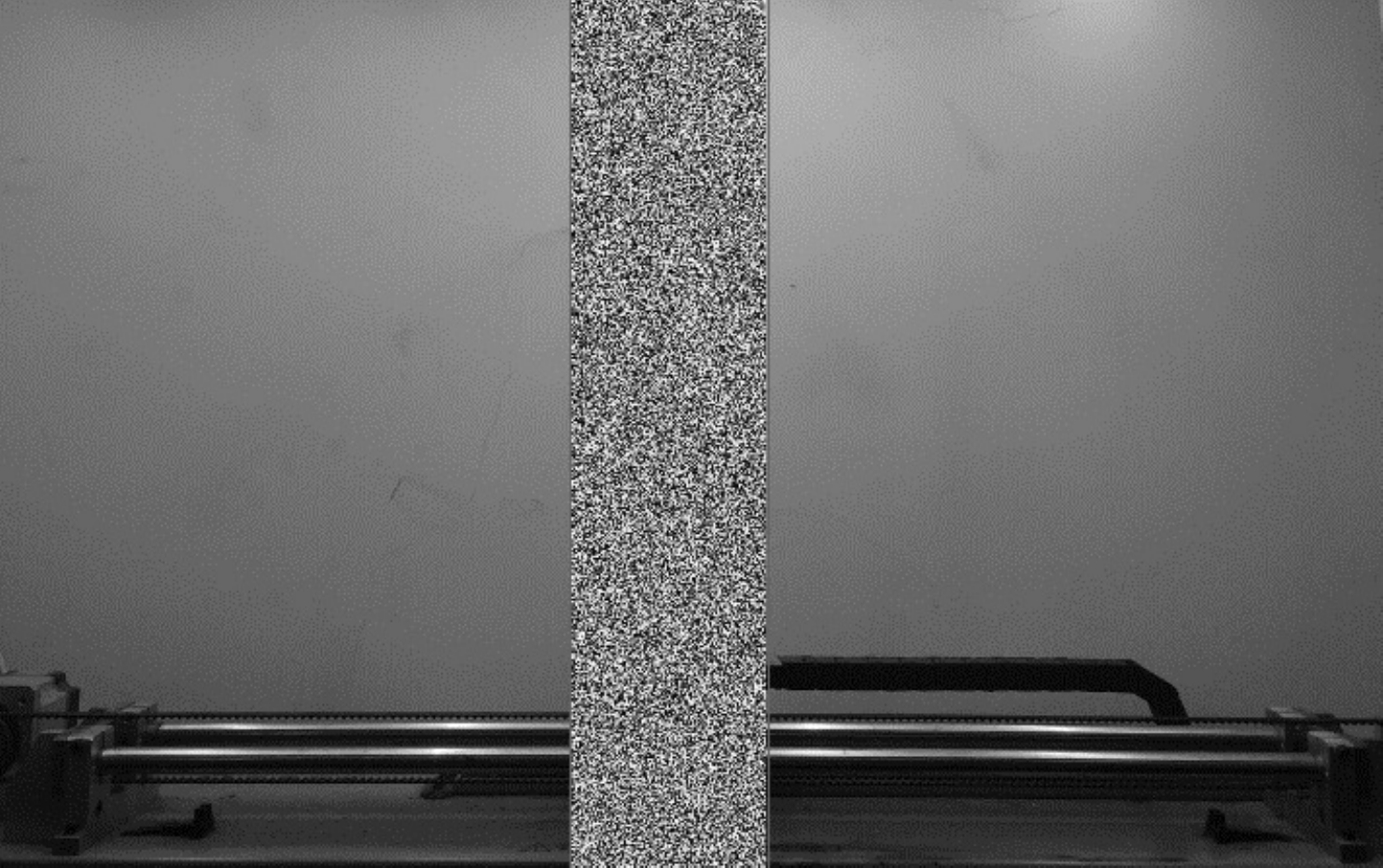}
}\hspace{-6.5mm}
\quad
\subfigure[]{
\includegraphics[width=1.5cm]{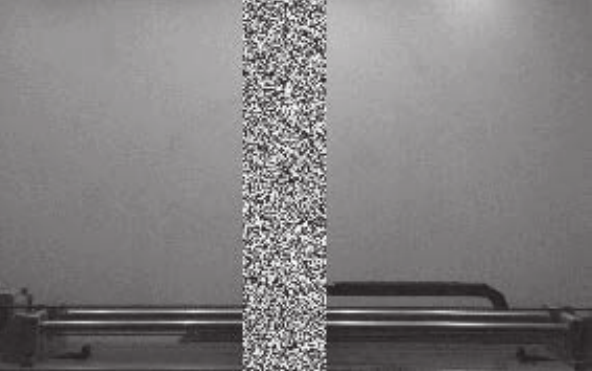}
}\hspace{-6.5mm}
\quad
\subfigure[]{
\includegraphics[width=1.5cm]{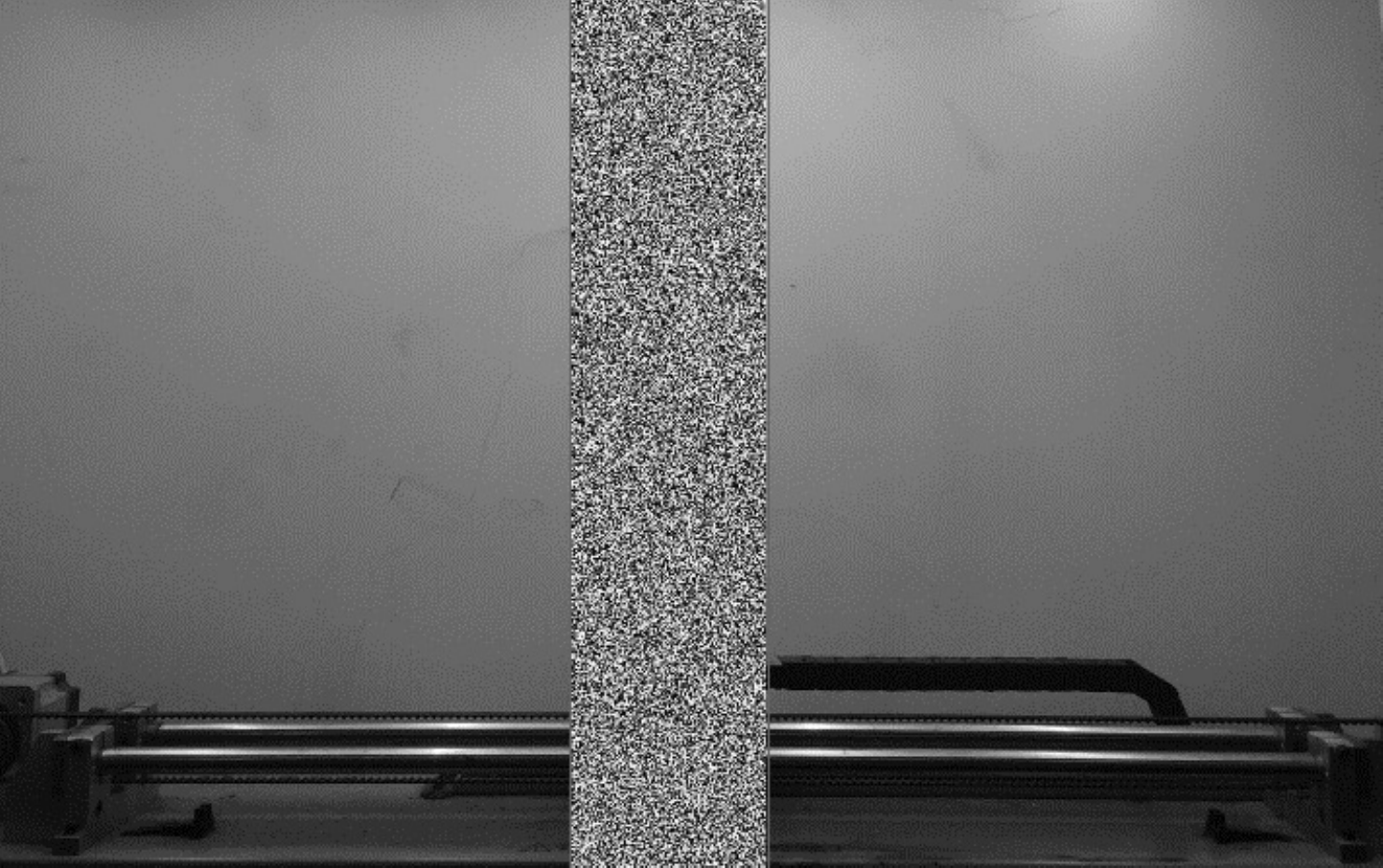}
}\hspace{-6.5mm}
\quad
\subfigure[]{
\includegraphics[width=1.5cm]{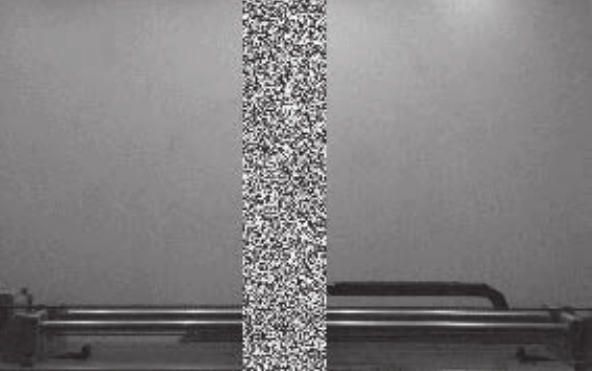}
}\hspace{-6.5mm}
\quad
\subfigure[]{
\includegraphics[width=1.5cm]{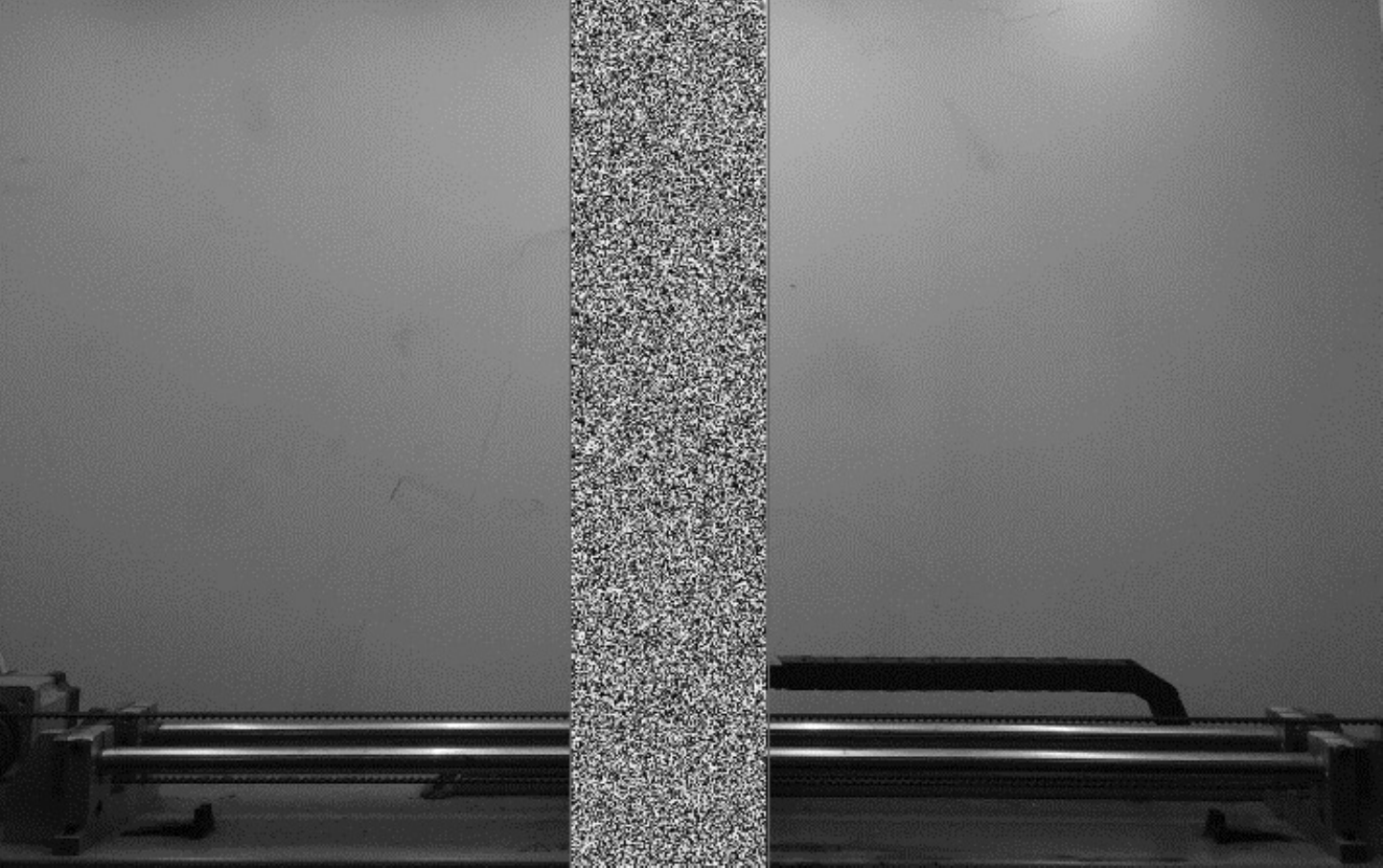}
}\hspace{-6.5mm}
\quad
\subfigure[]{
\includegraphics[width=1.5cm]{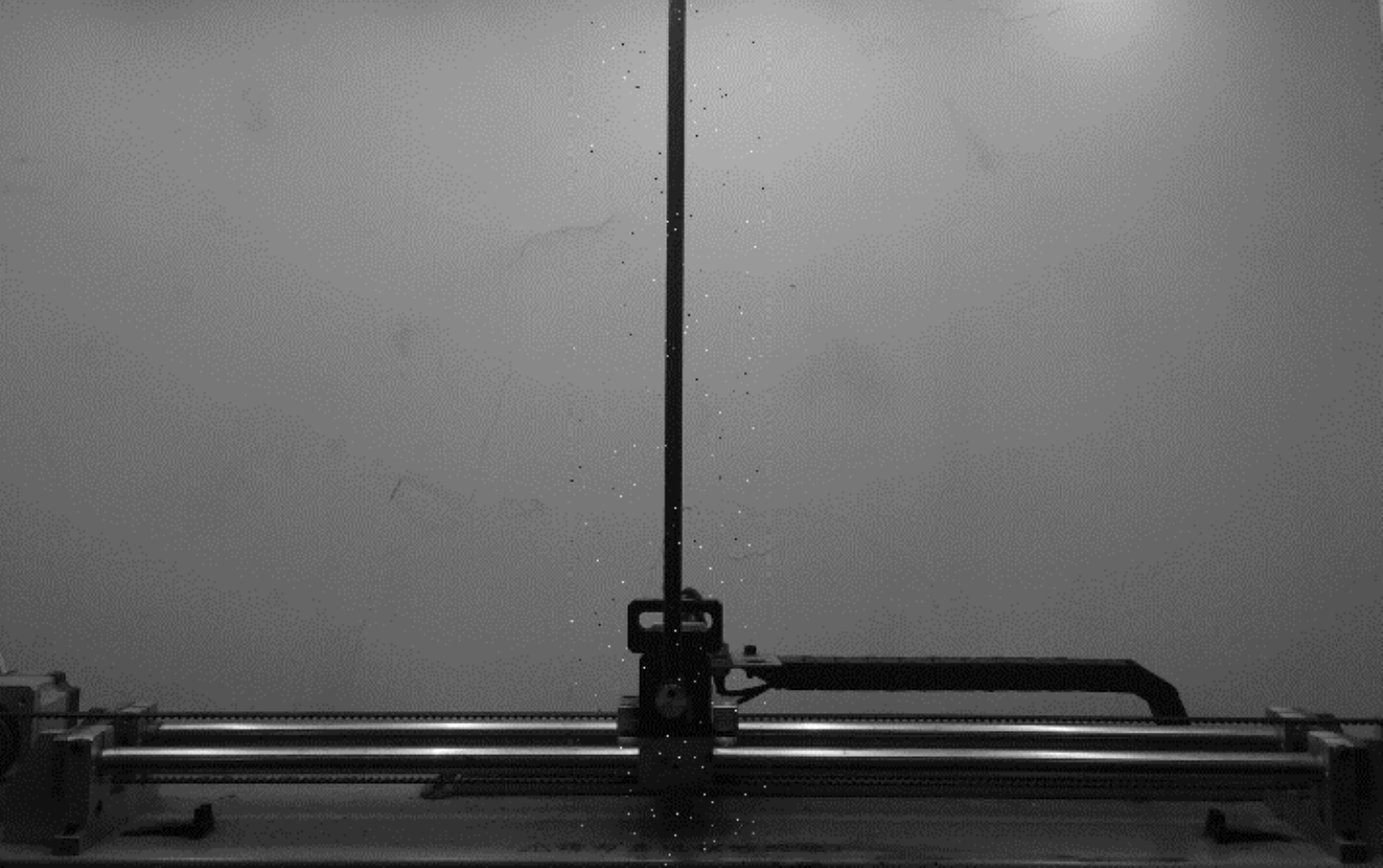}
}\hspace{-6.5mm}
\quad
\subfigure[]{
\includegraphics[width=1.5cm]{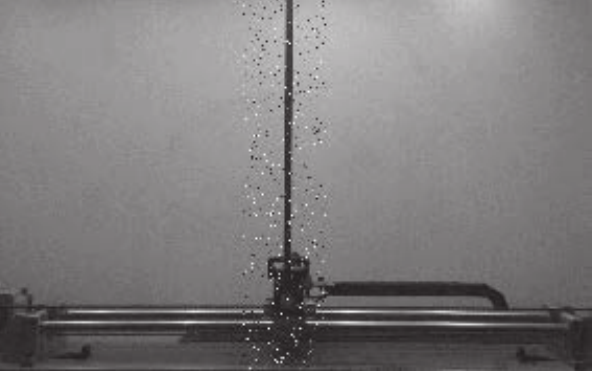}
}\hspace{-6.5mm}
\quad
\subfigure[]{
\includegraphics[width=1.5cm]{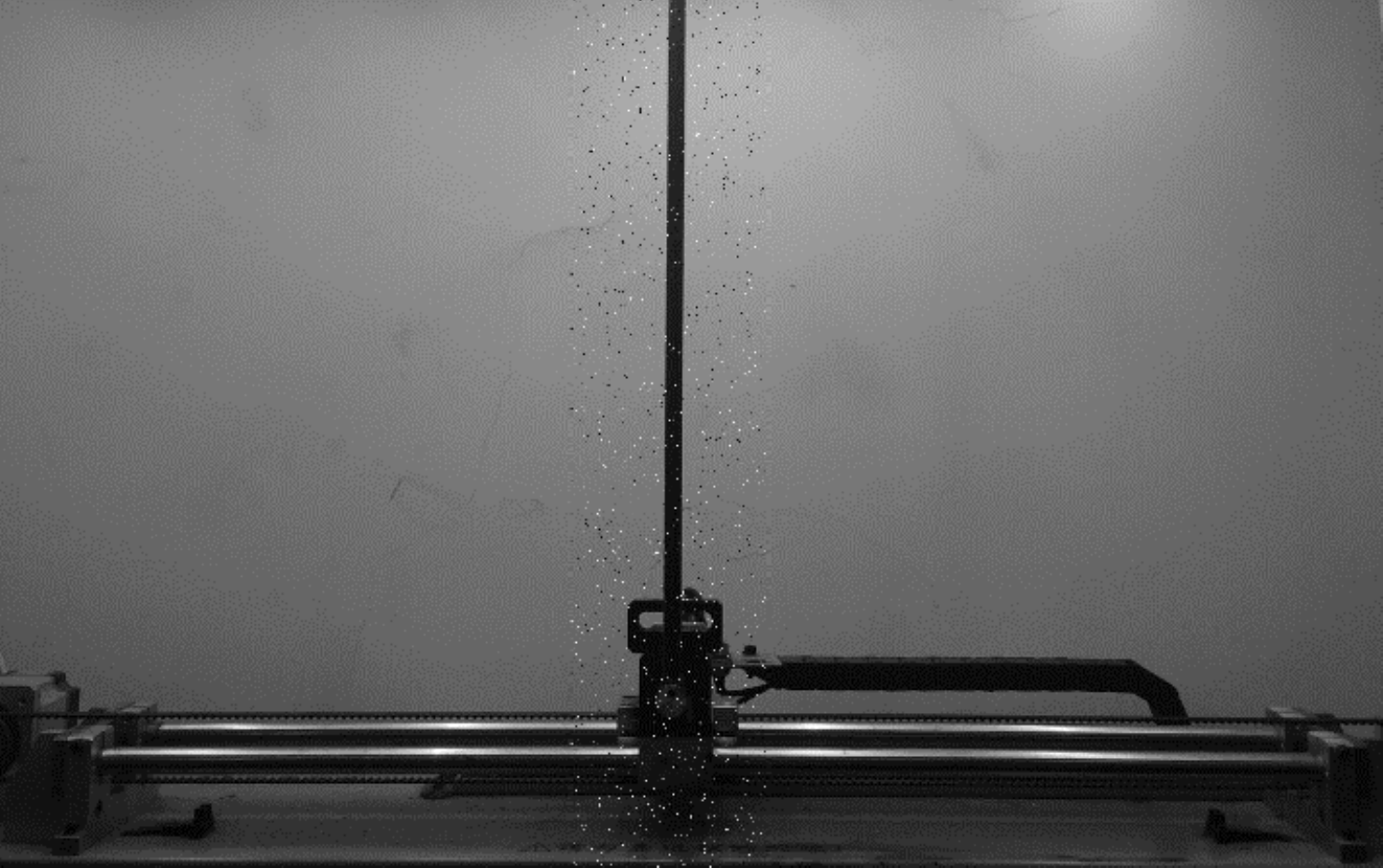}
}\hspace{-6.5mm}
\quad
\subfigure[]{
\includegraphics[width=1.5cm]{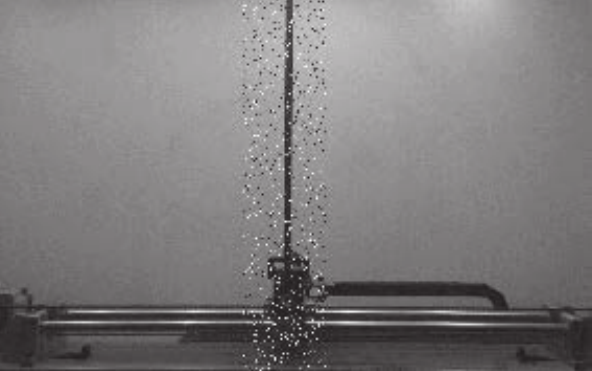}
}\hspace{-6.5mm}
\quad
\subfigure[]{
\includegraphics[width=1.5cm]{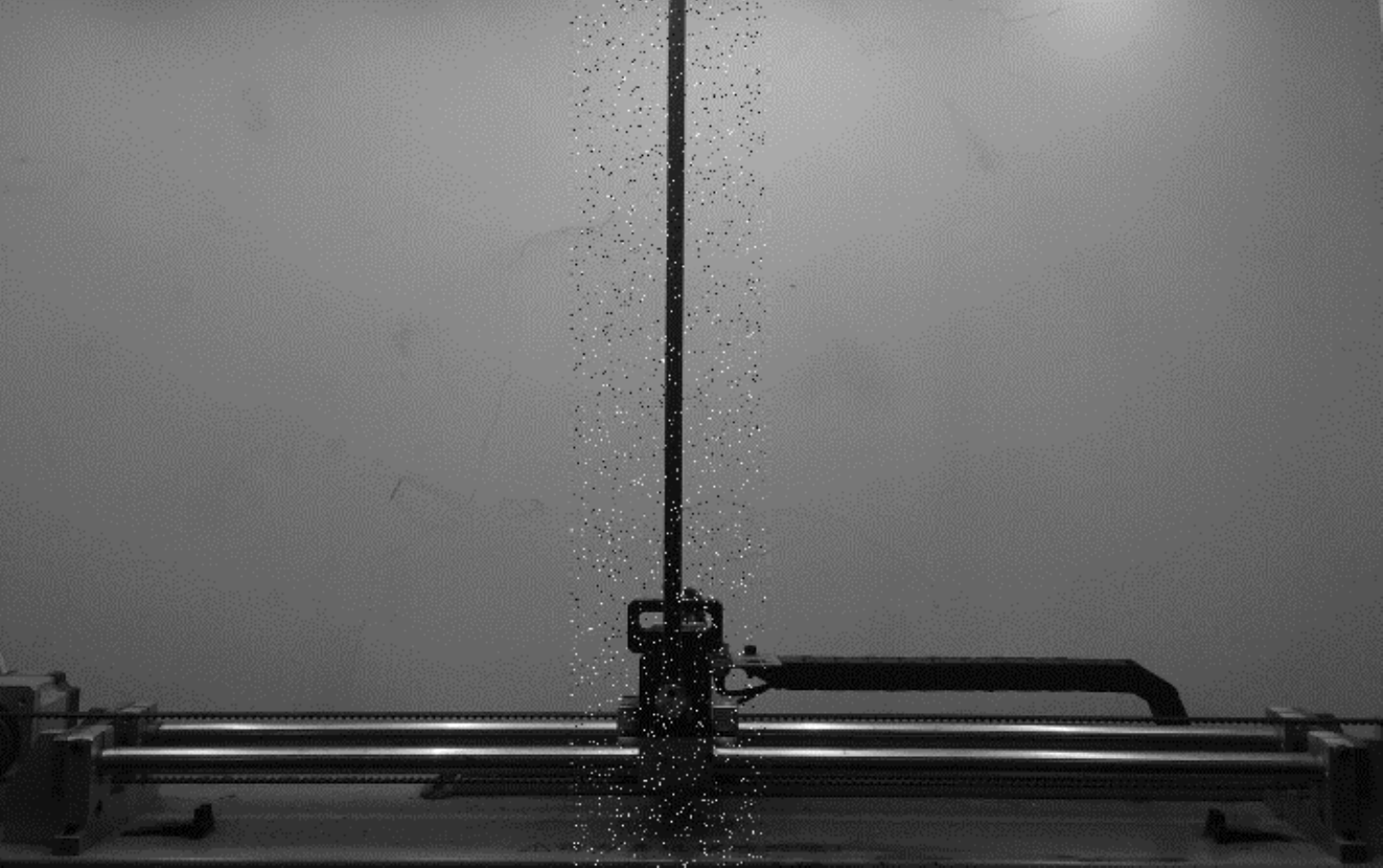}
}
\caption{ Attack tests (a) (b) (c) (d) (e) are salt and pepper attack with (0.1\%, 0.5\%, 0.9\%, 1.0\%, 1.1\%) intensity, respectively; (f) (h) (i) (j) (k) are the decrypted image of (a) (b) (c) (d) (e), respectively.}
\label{figA13}
\end{figure}

\begin{figure}[!t]
\centering
\subfigure[]{
\includegraphics[width=1.5cm]{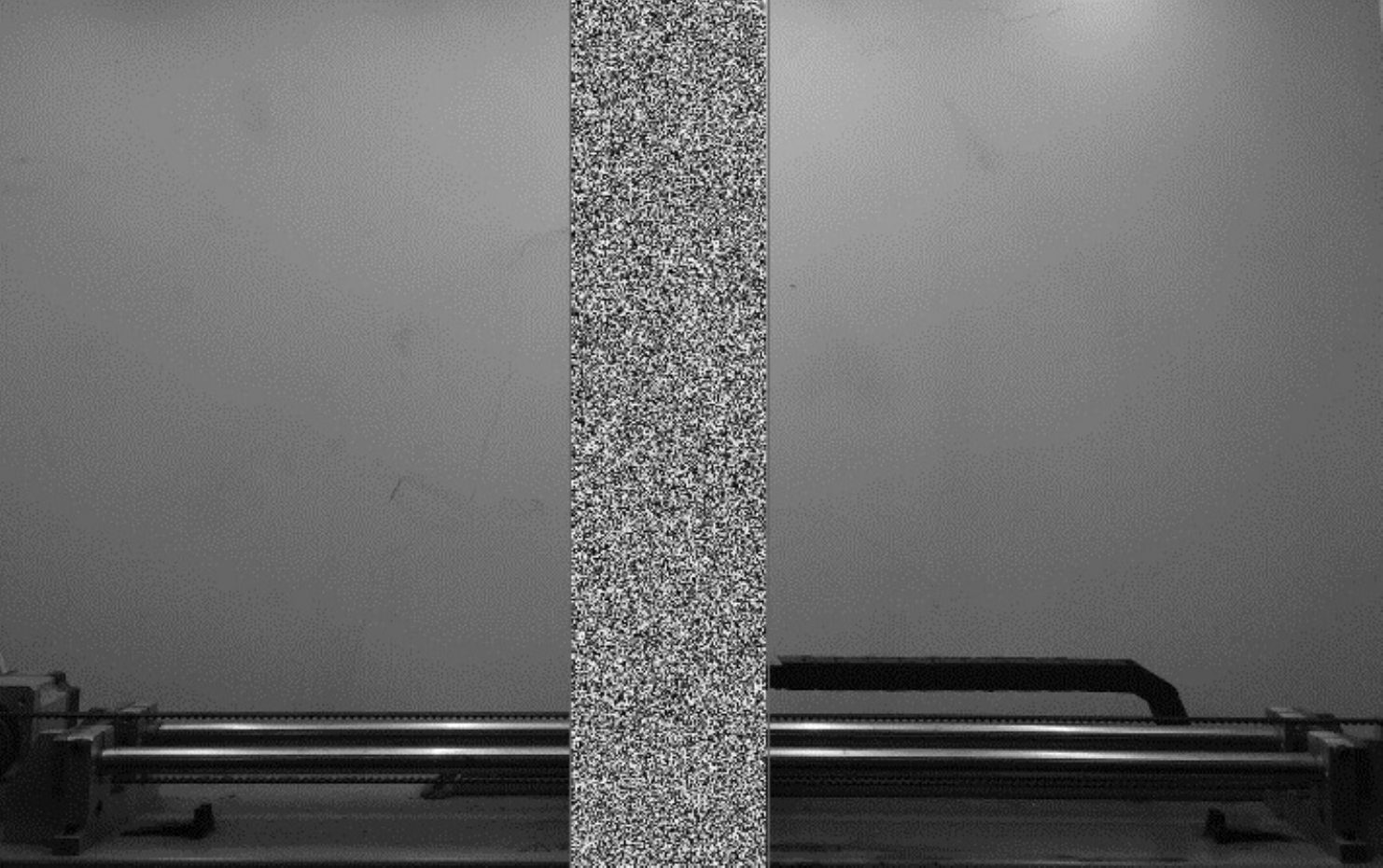}
}\hspace{-6.5mm}
\quad
\subfigure[]{
\includegraphics[width=1.5cm]{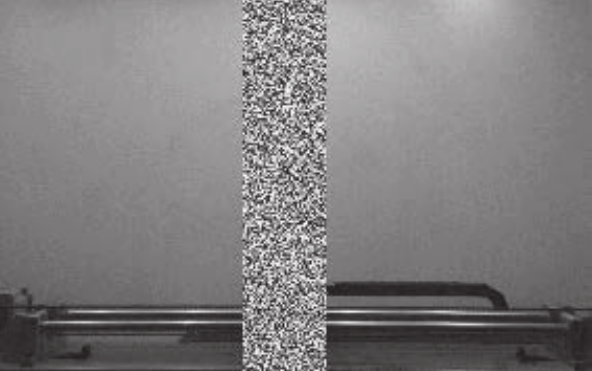}
}\hspace{-6.5mm}
\quad
\subfigure[]{
\includegraphics[width=1.5cm]{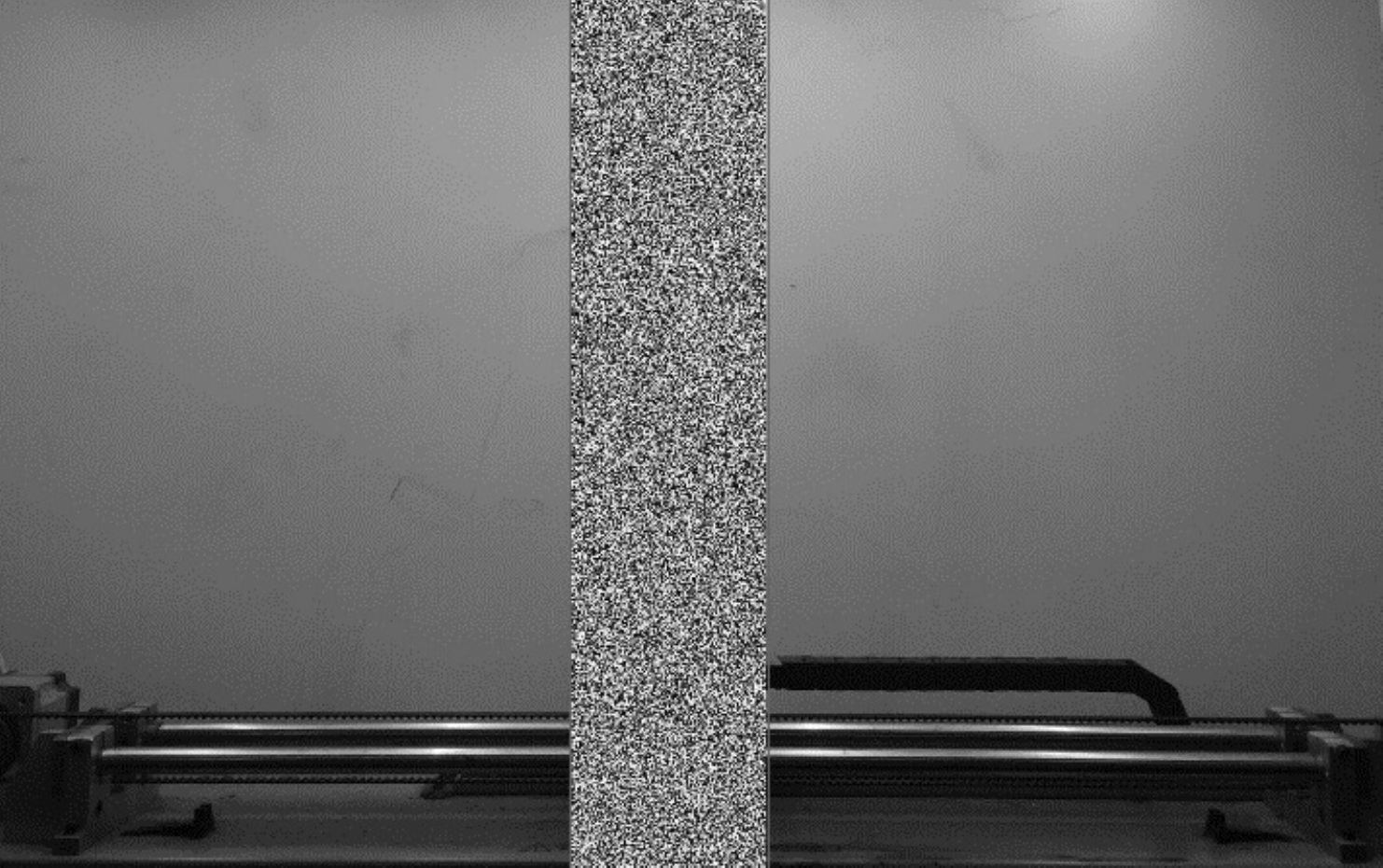}
}\hspace{-6.5mm}
\quad
\subfigure[]{
\includegraphics[width=1.5cm]{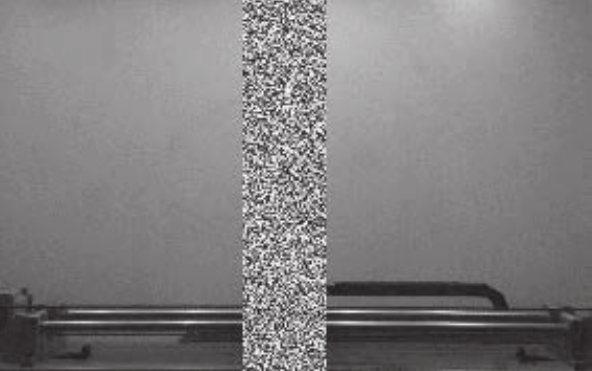}
}\hspace{-6.5mm}
\quad
\subfigure[]{
\includegraphics[width=1.5cm]{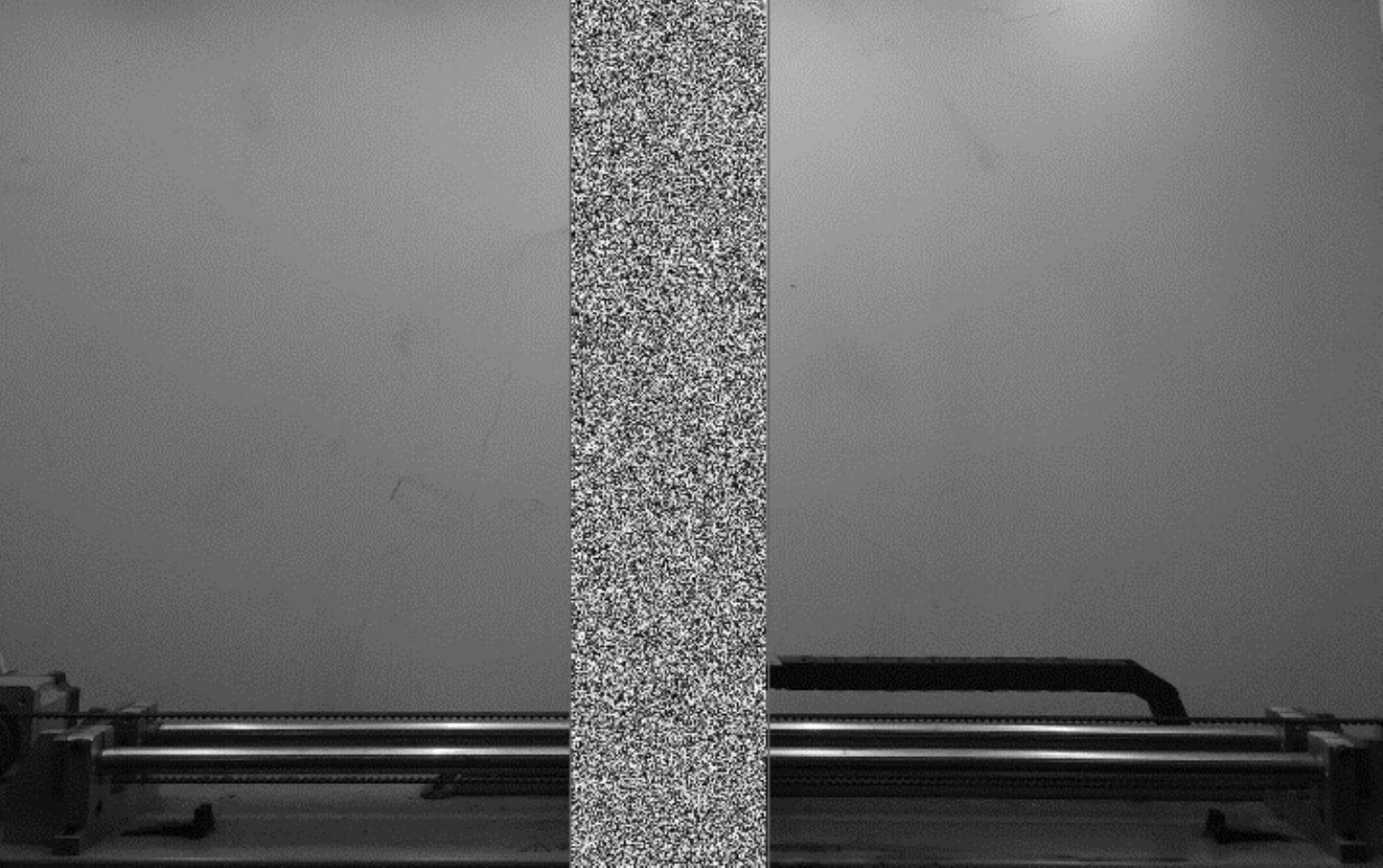}
}\hspace{-6.5mm}
\quad
\subfigure[]{
\includegraphics[width=1.5cm]{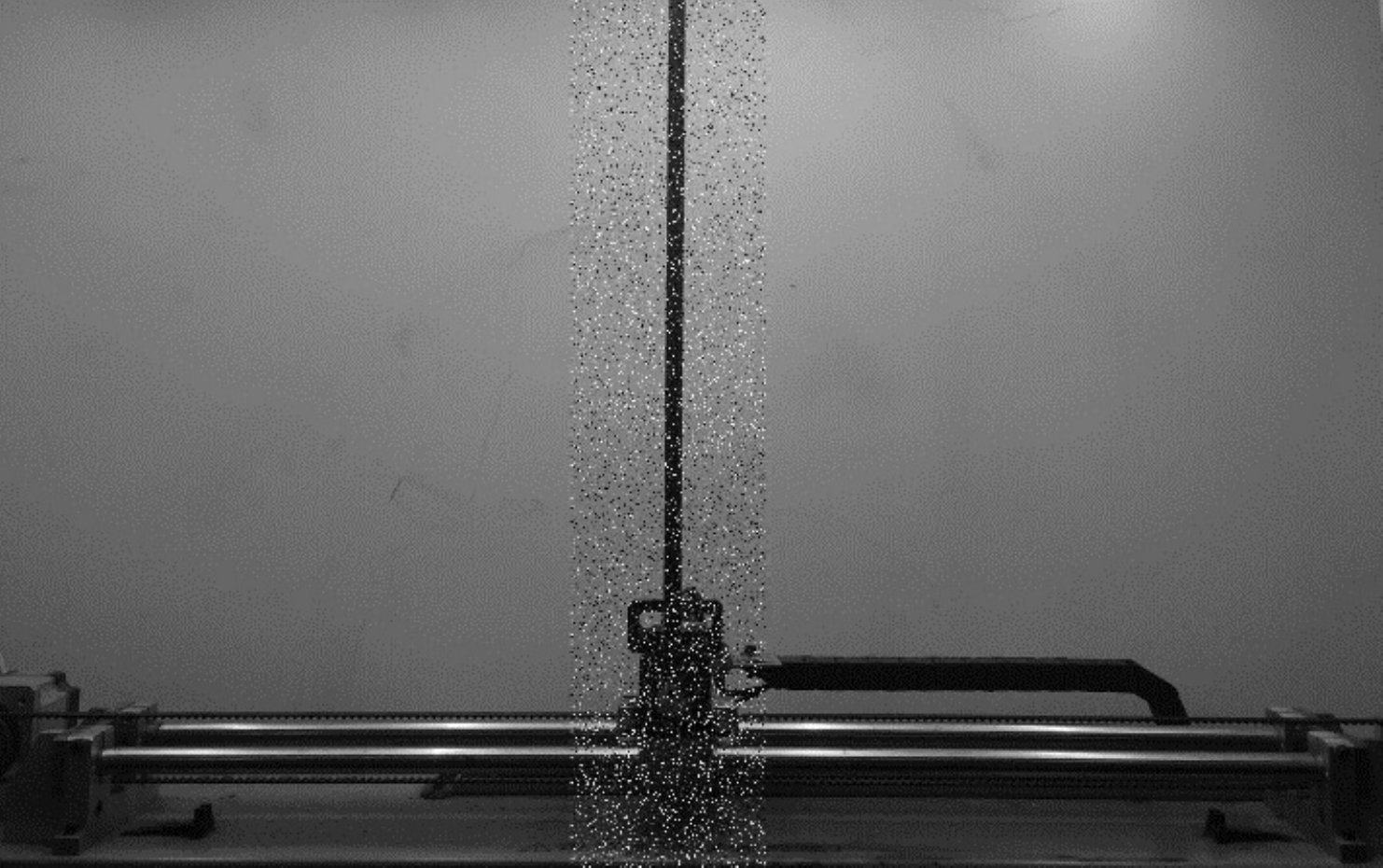}
}\hspace{-6.5mm}
\quad
\subfigure[]{
\includegraphics[width=1.5cm]{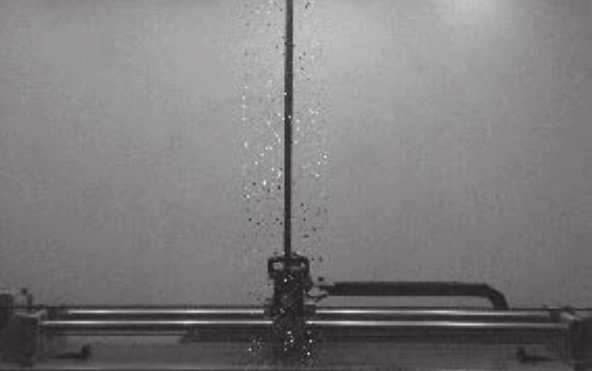}
}\hspace{-6.5mm}
\quad
\subfigure[]{
\includegraphics[width=1.5cm]{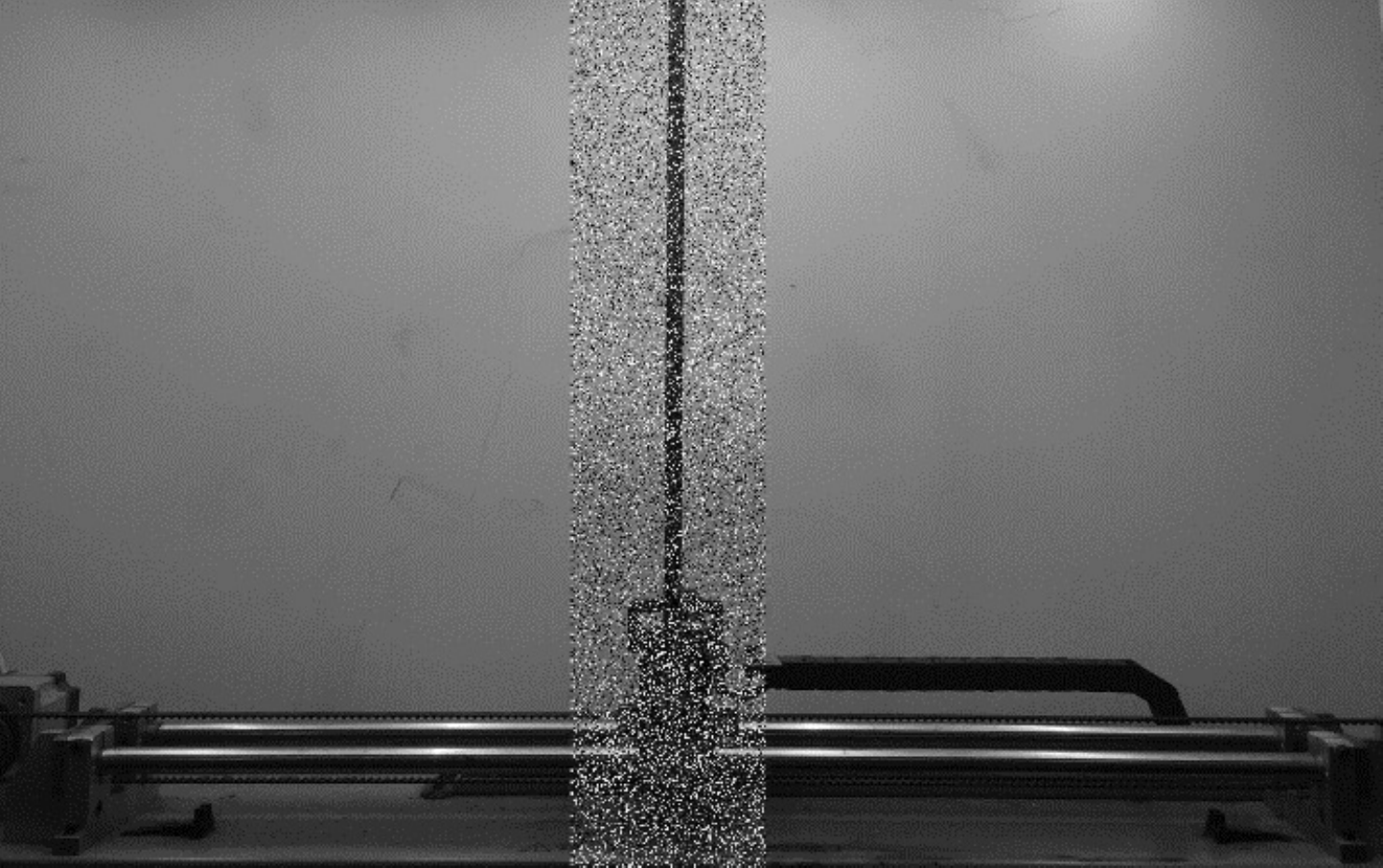}
}\hspace{-6.5mm}
\quad
\subfigure[]{
\includegraphics[width=1.5cm]{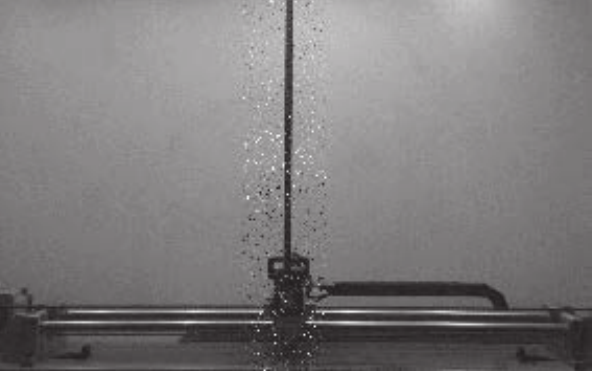}
}\hspace{-6.5mm}
\quad
\subfigure[]{
\includegraphics[width=1.5cm]{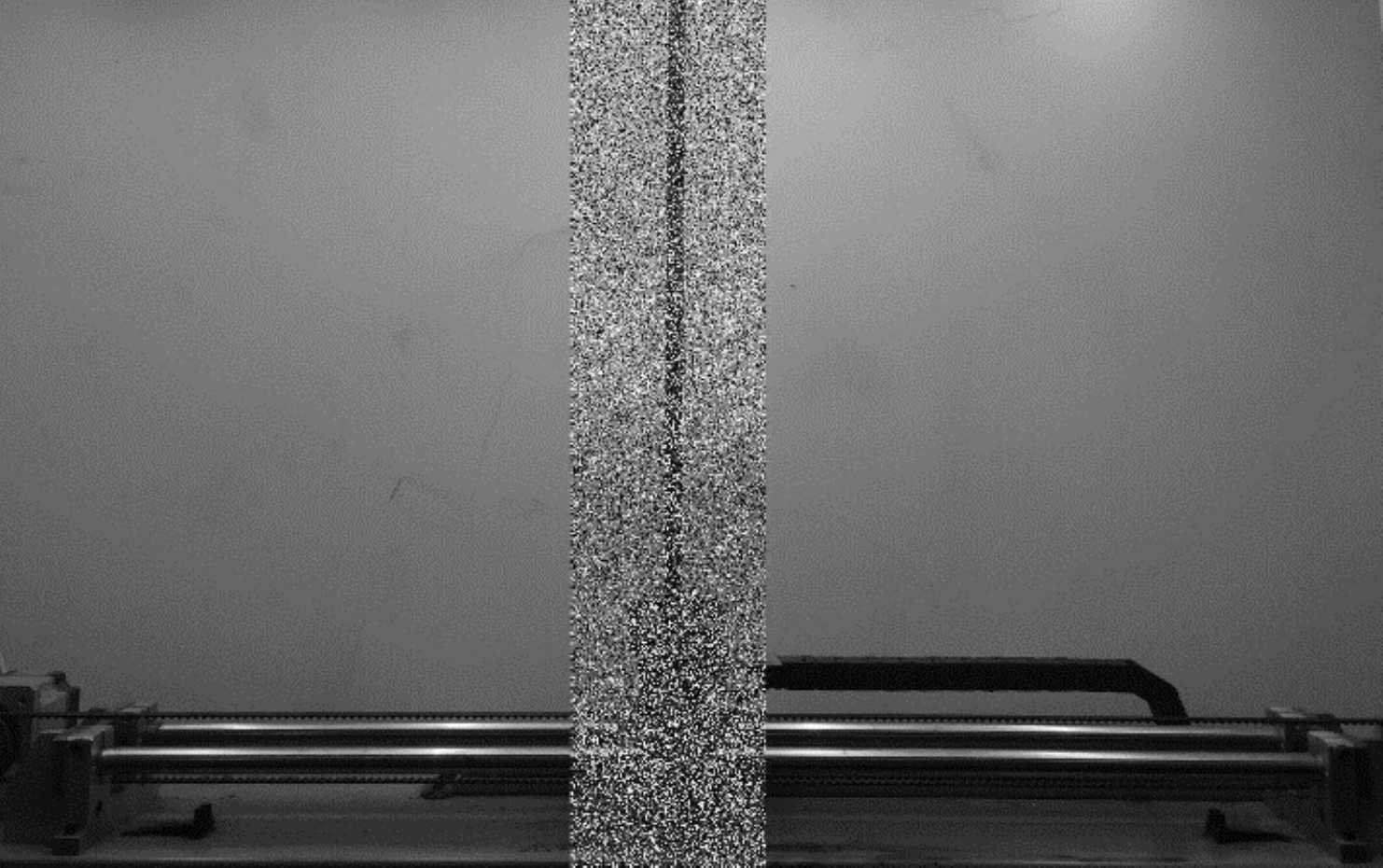}
}
\caption{Attack tests (a) (b) (c) (d) (e) are Gaussian attack obeying ((0,1), (0,5), (2,2), (2,5), (5,5)), respectively; (f) (h) (i) (j) (k) are the decrypted image of (a) (b) (c) (d) (e), respectively.}
\label{figA14}
\end{figure}

It is a common phenomenon that image is often subject to shearing, salt and pepper or Gaussian attacks during network transmission. For new NIPVSS, image decryption will be affected by these image attacks. Peak signal-to-noise ratio (PSNR) is an objective standard \cite{28}, and generally when PSNR is lower than 20dB, image quality is poor to unacceptable and feature extraction cannot be performed. Tab.~\ref{tabAIII} shows PSNR of the decrypted IPS image under these image attacks, where it can be seen that the stronger image attacks are, the lower PSRN are. But PSNR are all more than 20dB, which indicates that F2SIE algorithm has ability to keep image under these image attacks with good quality. Meanwhile, Figs.~\ref{fig4} (slight shearing attacks) and  \ref{figA13}(slight salt and pepper attack) and  \ref{figA14}(slight Gaussian attack) shows that F2SIE algorithm can recover  effective information of pendulum and cart after being affected by these slight image attacks. However, when attack is added too much, the recovered image contour will gradually become blurred. 
\begin{table}[!t]
\caption{The PSNR value of the recovered image under different image attacks. }
\label{tabAIII}
\centering
\begin{tabular}{cccccc}
  \toprule
  Shear rate &  1\% &  2\% &  4\% &  5\% &  6\%\\
  \midrule
  PSNR & 37.016 & 33.999 & 30.930 & 33.110 & 29.150 \\
  \midrule
  Intensity & 0.1\% & 0.5\% & 0.9\% & 1.0\% & 1.1\% \\
  \midrule
  PSNR & 38.948 & 32.54 & 28.609 & 27.616 & 25.323\\
  \midrule
  $(\mu,\sigma)$ & (0, 1) & (0, 5) & (2, 2) & (2, 5) & (5, 5)\\
  \midrule
  PSNR & 39.979 & 33.007 & 32.458 & 30.590 & 28.138 \\
  \bottomrule
  \multicolumn{6}{l}{Shear rate represents shear rate of shear attacks}\\
  \multicolumn{6}{l}{Intensity represents the intensity of salt and pepper attack}\\
  \multicolumn{6}{l}{$\mu$ and $\sigma$ represent mean and variance of Gaussian attack}\\
\end{tabular}
\end{table}

\subsection{Controller Robustness}
The proposed F2SIE algorithm has been confirmed, then controller robustness will be analysed. Using Theorem 1, some parameters are given as ${\underline \eta }^{en} = 0.004s$, ${{\bar \eta }^{en}} = 0.007s$, ${\underline \eta } ^{de} = 0.004s$, ${\bar \eta }^{de} = 0.007s$, $\bar \tau = 0.005s$, ${\underline \lambda} = 0.026s$, ${\bar\lambda} = 0.054s$, ${\theta _1} = 0.01$, ${\theta _2} = 0.75$, ${\theta _3} = 1.2$, ${\theta _4} = 0.022$, ${\varepsilon _1} = 0.9$, ${\varepsilon _2} = 0.1$, ${\Delta _{1,a}} = {\Delta _{2,a}} = 0$. Before image is encrypted, each image has an exposure time of 0.01s. Thus, an additional 0.01s is needed for all parameters involved in image encryption. Using the above parameters and solving \eqref{eq4}, control gain $K$ of closed-loop system \eqref{eq3} can be calculated, and its value is $ {K} =[3.7633,- 29.9925,4.0355,- 5.4562]$

The controller robustness will be analysed from two aspects.

\subsubsection{Controller Robustness against Multiple Delay}
The relationship between $\tau (t)$ and ${\lambda(t)}$ is firstly analysed. By using Theorem 1, theoretical $\bar \tau $ and $\bar \lambda $ can be obtained while closed-loop system \eqref{eq3} is stable. Tab.~\ref{tabAIV} lists this relationship between $\bar \tau $ and $\bar \lambda $, where theoretical $\bar \tau$ is from $ 0s$ to $0.013s$, and the corresponding theoretical ${\bar \lambda}$ is from $0.067s$ to $0.054s$.  To analyse impact of image encryption-decryption time $\eta^{en}_k + \eta^{de}_k$ on the system, value of $\bar \eta^{en}+ \bar \eta^{de}$ is calculated. Recalling ${\bar\lambda} = 2{\bar \eta ^{en}} + \bar \eta^{de}  + {\bar \tau ^{sc}} + \bar d$ with fixed ${\bar \tau ^{sc}}$ and $\bar d{\text{ }}$, the corresponding values of ${\bar \eta ^{en}} + \bar \eta^{de}$ can be calculated (${\bar \eta ^{en}}= \bar \eta^{de}$), which is shown in the third row of Tab.~\ref{tabAIV}. For instance, when $\bar \tau=0.005s$, the theoretical ${\bar \eta ^{en}} + \bar \eta^{de}$ is 0.018s. 
\begin{table}[!t]
\caption{Given the Upper Bound $\bar\tau $ of Network-Induced Delay, the Upper Bound ${\bar\lambda}$ of Image-Processing-Induced Delay and the Upper Bound ${\bar \eta^{en}}+\bar \eta^{de}$ of Image Encryption and Decryption Time.}
\label{tabAIV}
\centering
\begin{tabular}{cccccc}
  \toprule
  Given $\bar\tau (s) $ & 0 & 0.003 & 0.005  & 0.011 & 0.013  \\
  \midrule
  ${\bar\lambda}$ (s) & 0.067 & 0.064 & 0.062  & 0.056 & 0.054  \\
  \midrule
  ${\bar \eta^{en}}+\bar \eta^{de}$ (s)  & 0.022 & 0.019 & 0.018  & 0.015 & 0.013   \\
  \bottomrule
\end{tabular}
\end{table}

\begin{figure}[!t]
\centering
\subfigure[Cart position]{\includegraphics[width=0.46\textwidth]{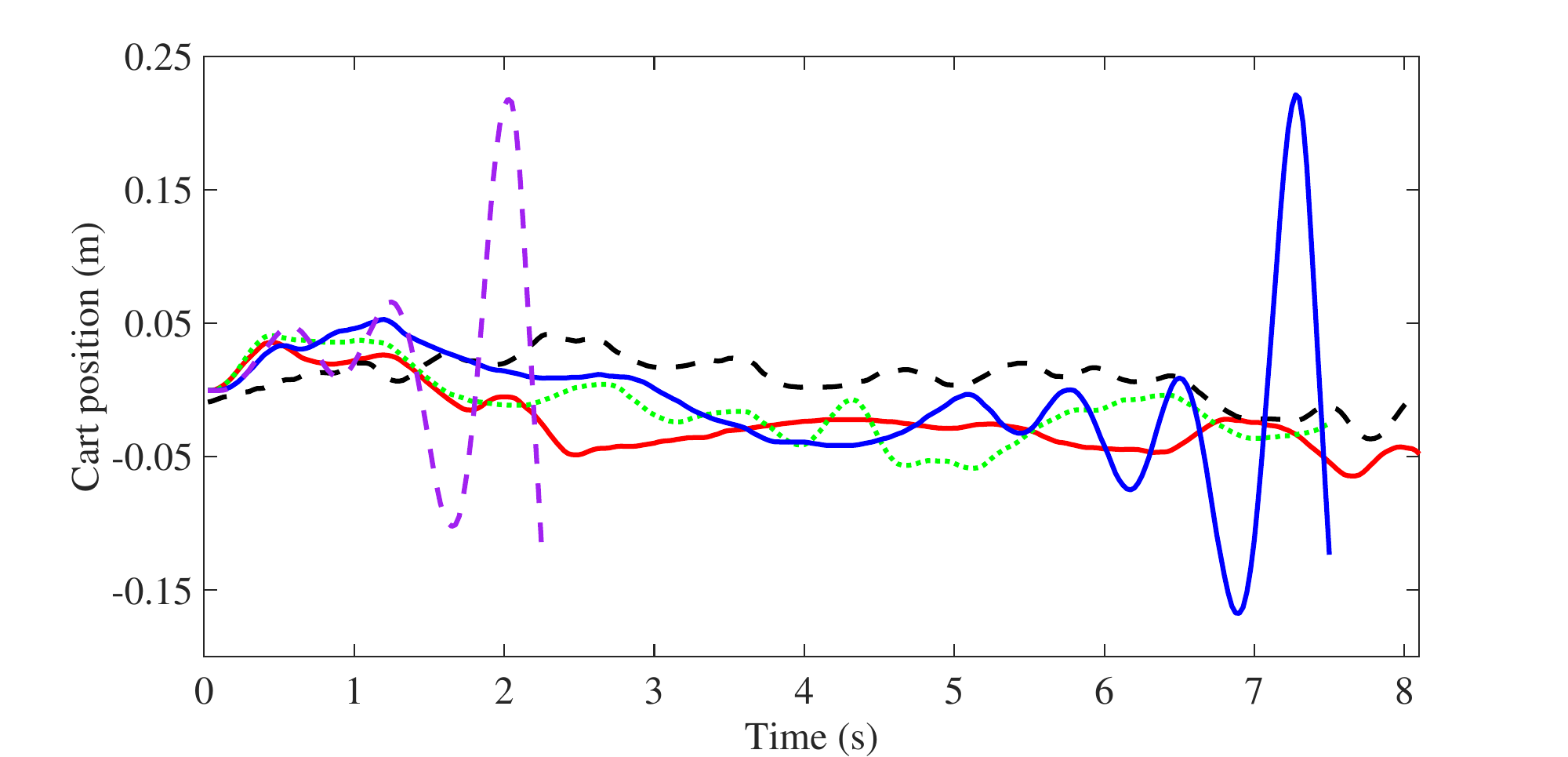}} \\
\subfigure[Pendulum angle]{\includegraphics[width=0.46\textwidth]{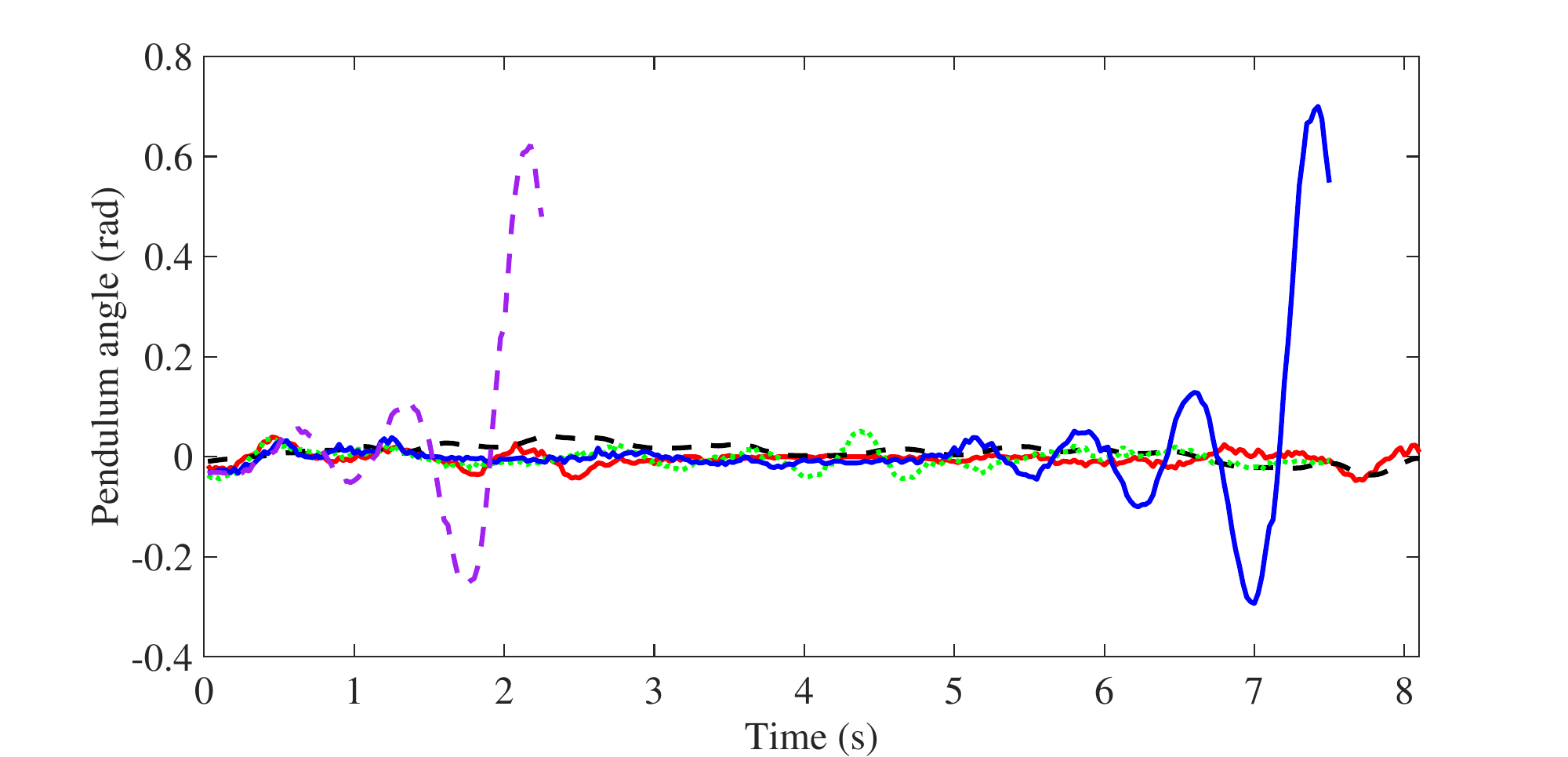}}
\caption{Cart position and pendulum angle of the new NIPVSS with different upper bounds ${\bar \eta ^{en}} + \bar \eta^{de}$ of image encryption and decryption time: \textcolor[rgb]{1,0,0}{---}, ${\bar \eta ^{en}} + \bar \eta^{de} = 0.014s$; \textcolor[rgb]{0,0,0}{- -}, ${\bar \eta ^{en}} + \bar \eta^{de} = 0.015s$; \textcolor[rgb]{0,1,0}{$\cdot  \cdot  \cdot  \cdot  \cdot $}, ${\bar \eta ^{en}} + \bar \eta^{de} = 0.016s$, \textcolor[rgb]{0,0,1}{---}, ${\bar \eta ^{en}} + \bar \eta^{de} = 0.017s$;  \textcolor[rgb]{0.63,0.13,0.94}{- -}, ${\bar \eta ^{en}} + \bar \eta^{de} = 0.018s$.}
\label{fig16}
\end{figure}
To further explore the maximum experimental value of ${\bar \eta ^{en}} + \bar \eta^{de}$ that system can tolerate when $\bar \tau=0.005s$, values of ${\bar \eta ^{en}} + \bar \eta^{de}$ are set as 0.014s, 0.015s, 0.016s, 0.017s, 0.018s. Real-time curves of cart position and pendulum angle are shown in Fig. \ref{fig16}, where is can be seen that as ${\bar \eta ^{en}} + \bar \eta^{de}$ increases, the fluctuation of cart position and pendulum angle increases. When ${\bar \eta ^{en}} + \bar \eta^{de}$ goes to $0.017s$ or $0.018s$, cart position and pendulum angle diverge. Hence, when $\bar \tau=0.005s$, the maximum experimental value of ${\bar \eta ^{en}} + \bar \eta^{de}$ that system can tolerate is $0.016s$.

\subsubsection{Controller Robustness against Parameter Uncertainty}
The relationship between  ${\Delta _1}$ and ${\Delta _2}$ is analysed. According to Theorem 1, for the given $|{\Delta _1}|$,  $|{\Delta _2}|$ can be obtained while ensures closed-loop system \eqref{eq3} stable, where $|{\Delta _1}|$ is from 0 to 0.67 and the corresponding $|{\Delta _2}|$ is from 3.24 to 0. Furthermore, to analyse ${\Delta _{1,a}}$ and ${\Delta _{2,a}}$ caused by image attacks, we add shearing attack, salt and pepper attack and Gaussian attack in images.

\begin{figure}[!t]
\centering
\subfigure[Cart position]{\includegraphics[width=0.46\textwidth]{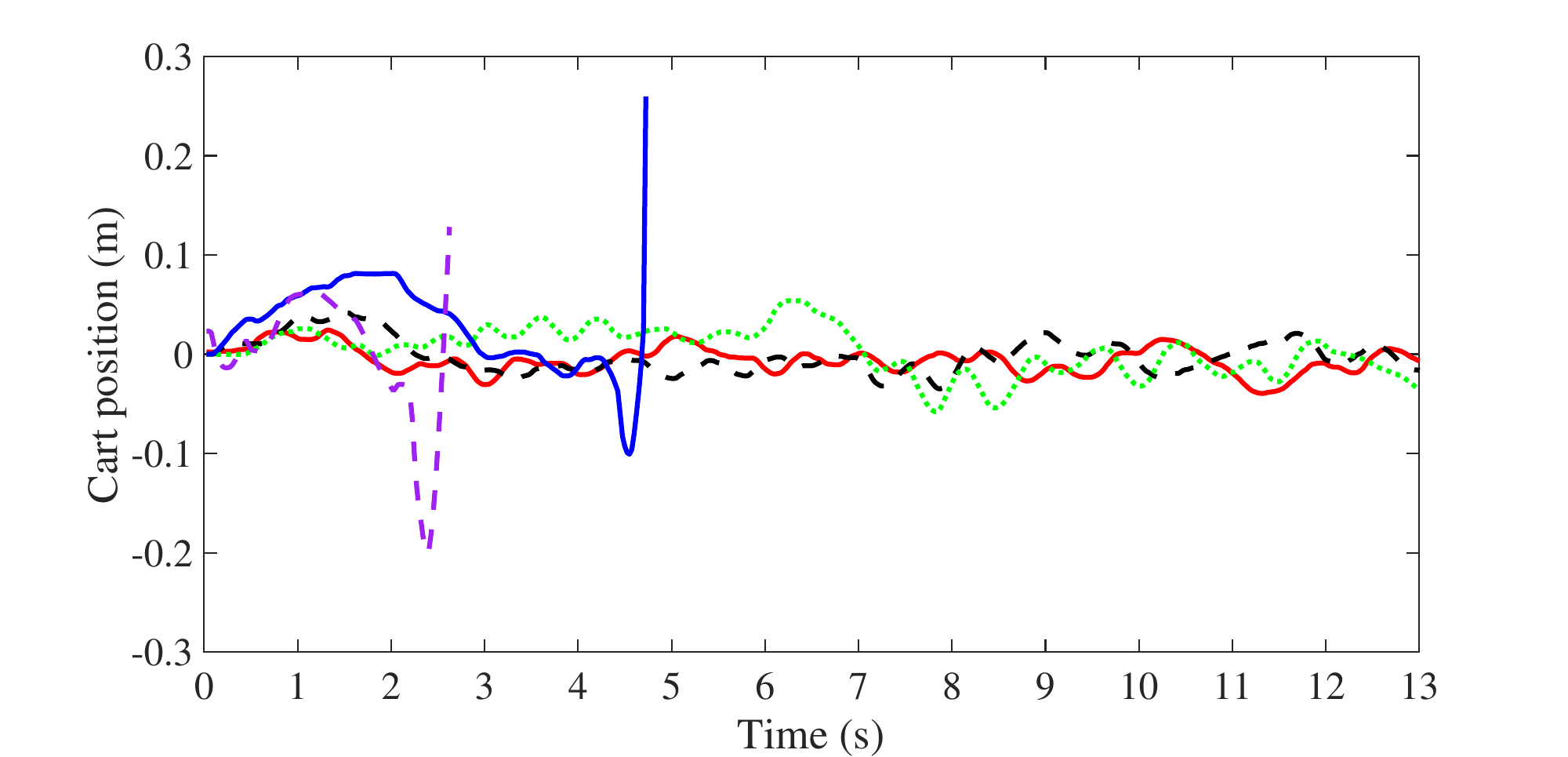}}\\
\subfigure[Pendulum angle]{\includegraphics[width=0.46\textwidth]{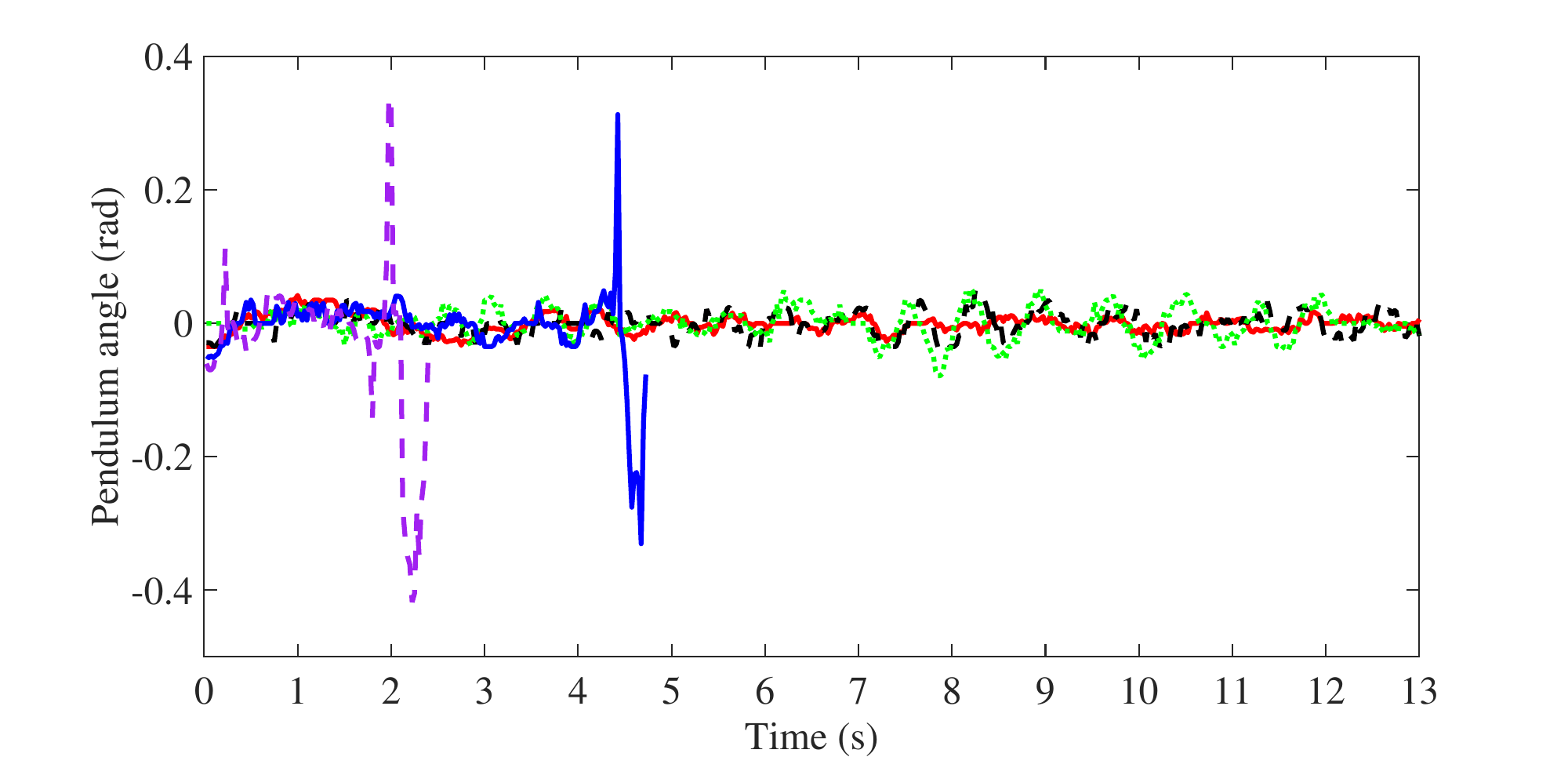}}
\caption{Cart position and pendulum angle of the new NIPVSS under shearing attacks with different shearing rates:  \textcolor[rgb]{1,0,0}{---} represents 1\% shearing rate, \textcolor[rgb]{0,0,0}{- -} represents 2\% shearing rate, \textcolor[rgb]{0,1,0}{$\cdot  \cdot  \cdot  \cdot  \cdot $} represents 4\% shearing rate, \textcolor[rgb]{0,0,1}{---} represents 5\% shearing rate and  \textcolor[rgb]{0.63,0.13,0.94}{- -} represents 6\% shearing rate.}
\label{fig20}
\end{figure}
\textbf{Influence of shearing attack on the IPS performance:} Different shearing rates (1.0$\%$, 2.0$\%$, 4.0$\%$, 5.0$\%$) of shearing attacks are added, and the corresponding real-time curves of cart position and pendulum angle are shown in Fig.~\ref{fig20}. It can be seen from Fig.~\ref{fig20} that when shearing rate is within 4.0$\%$, cart position and pendulum angle tend to be stable and diverge when shearing rate exceeds 5.0$\%$. However, when shearing rate goes to 5.0$\%$, cart position and pendulum angle diverge rapidly. This is because information of the cut picture is seriously damaged, which leads to increase of computational error and makes system unstable. Hence, it means that the maximum shearing rate that system can withstand is 4.0$\%$.

\begin{table}[!t]
\caption{Values of ${\underline \Delta  _{1,a}}$, ${\underline \Delta  _{2,a}}$, ${\overline \Delta  _{1,a}}$ and ${\overline \Delta  _{2,a}}$ under Different Shearing Rates of Shearing Attacks.}
\label{tabAVII}
\centering
\begin{tabular}{cccccc}
  \toprule
  Shear rate & 1.0$\%$ & 2.0$\%$ & 4.0$\%$ & 5.0$\%$ & 6.0$\%$\\
  \midrule
   ${\underline \Delta  _{1,a}}$ & -0.25 & -0.28 & -0.32 & -0.45  & -0.52 \\
   \specialrule{0em}{3pt}{3pt}
   ${\overline \Delta  _{1,a}}$&  0.21 & 0.32 & 0.40 & 0.46 & 0.58\\
  \midrule
   ${\underline \Delta  _{2,a}}$ & -0.52 & -0.63 & -0.74 & -1.00 & -2.02 \\
   \specialrule{0em}{3pt}{3pt}
   ${\overline \Delta  _{2,a}}$ & 0.50 & 0.63 &0.75 & 1.03 & 1.32\\
  \midrule
  Stability  & $\surd$ & $\surd$ & $\surd$ & $\times$ & $\times$ \\
  \bottomrule
  \multicolumn{6}{l}{ $\surd$ represents the IPS is stable, $\times$ represents the IPS is unstable}\\
\end{tabular}
\end{table}
Values of ${\underline \Delta  _{1,a}}$, ${\underline \Delta  _{2,a}}$, ${\overline \Delta  _{1,a}}$, ${\overline \Delta  _{2,a}}$ under shearing attacks with different shearing rates are shown in Table~\ref{tabAVII}, where it can be seen that when IPS runs stably, ${\Delta _{1,a}} \in \left[ {-0.32,0.40} \right]$ and ${\Delta _{2,a}}\in \left[ {-0.74,0.75} \right]$. When IPS starts to diverge, $|{\Delta _{1,a}}|$ is beyond 0.45 and $|{\Delta _{2,a}}|$ is beyond 1.00. Hence, the designed controller can be able to achieve stable control of NIPVSS under slight shearing attacks, but cannot prevent from severe shearing attacks.

To intuitively show performance change of system under different shearing rates of shearing attack, mean and standard deviation (SD) of cart position and pendulum angle are calculated, which are illustrated in Tab.~\ref{tabAVIII}. It can be seen from Tab.~\ref{tabAVIII} that as shearing rate increase, mean and SD of cart position and pendulum angle increase; the greater the shearing rate is, the greater the influence of cart position and pendulum angle is; the performance slightly deteriorates with gradually increasing shearing rates. Tab.~\ref{tabAVIII} also shows that shearing attack has a significant influence on performance of IPS and an efficient image encryption algorithm is required to further improve image security.
\begin{table}[!t]
\caption{Mean and SD of the Cart Positions and Pendulum Angles under Different Shearing Rates of Shearing Attack.}
\label{tabAVIII}
\centering
\begin{tabular}{cccccc}
  \toprule
  Shear rate & 1$\%$ & 2$\%$ & 4$\%$ & 5$\%$ & 6$\%$\\
  \midrule
   MCP$^1$(m) & -0.0017 & -0.0035 & \textbf{-0.0053} & -0.0234  & 0.0266 \\
   SCP$^2$(m) &  0.0153 & 0.0214 & \textbf{0.0239} & 0.0444 & 0.0642\\
  \midrule
   MPA$^3$(rad)  & -0.0006 & 0.0.0008 & \textbf{-0.0010} & -0.0063 & -0.0274 \\
   SPA$^4$(rad) & 0.0197 & 0.0216 & \textbf{0.0268} & 0.0568 & 0.1166\\
  \bottomrule
  \multicolumn{6}{l}{$^1$ Mean of cart position. $^2$ SD of cart position.}\\
  \multicolumn{6}{l}{$^3$ Mean of pendulum angle. $^4$ SD of pendulum angle.}\\
\end{tabular}
\end{table}

\begin{figure}[!t]
\centering
\subfigure[]{\includegraphics[width=0.46\textwidth]{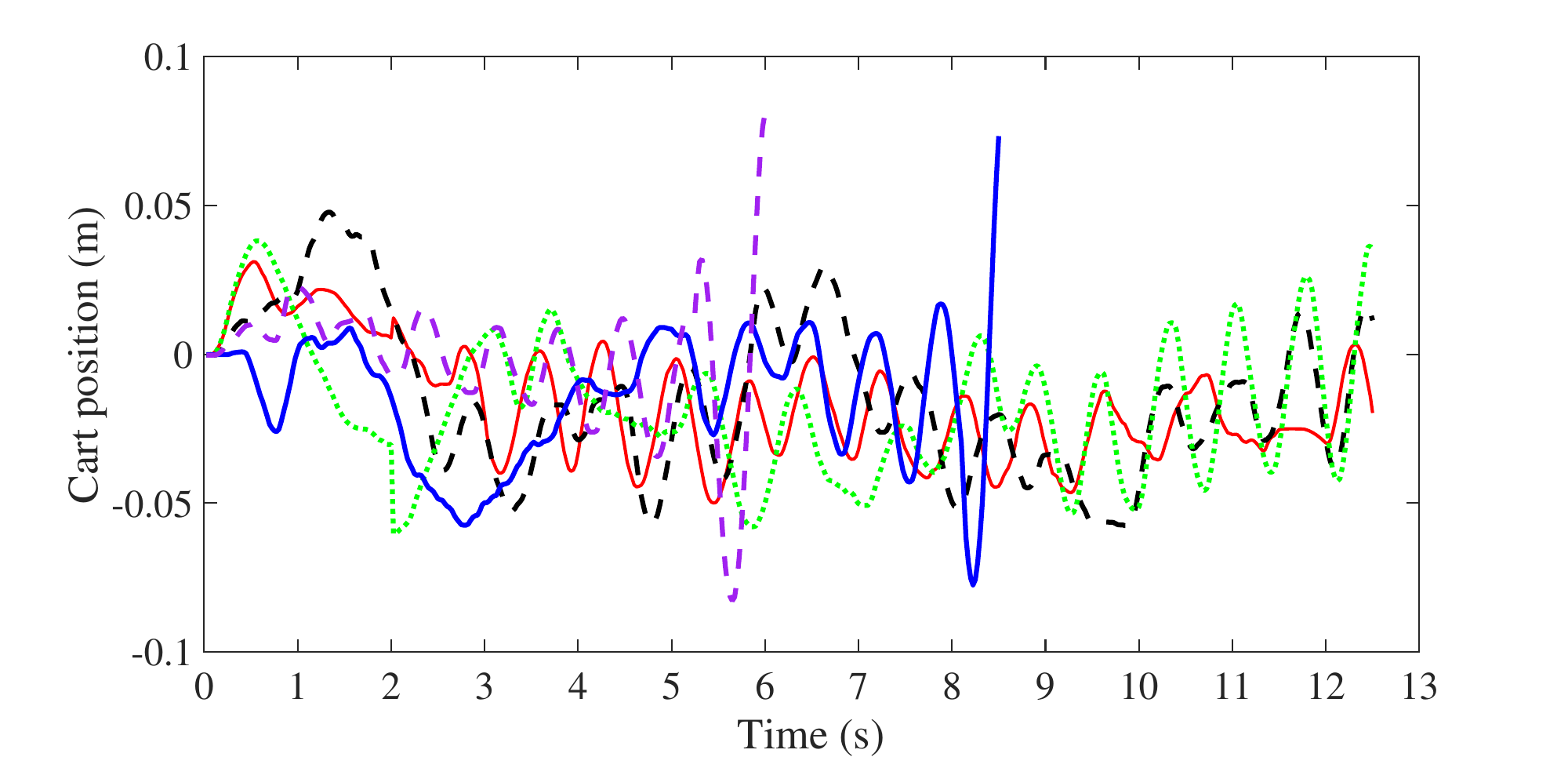}}\quad
\subfigure[]{\includegraphics[width=0.46\textwidth]{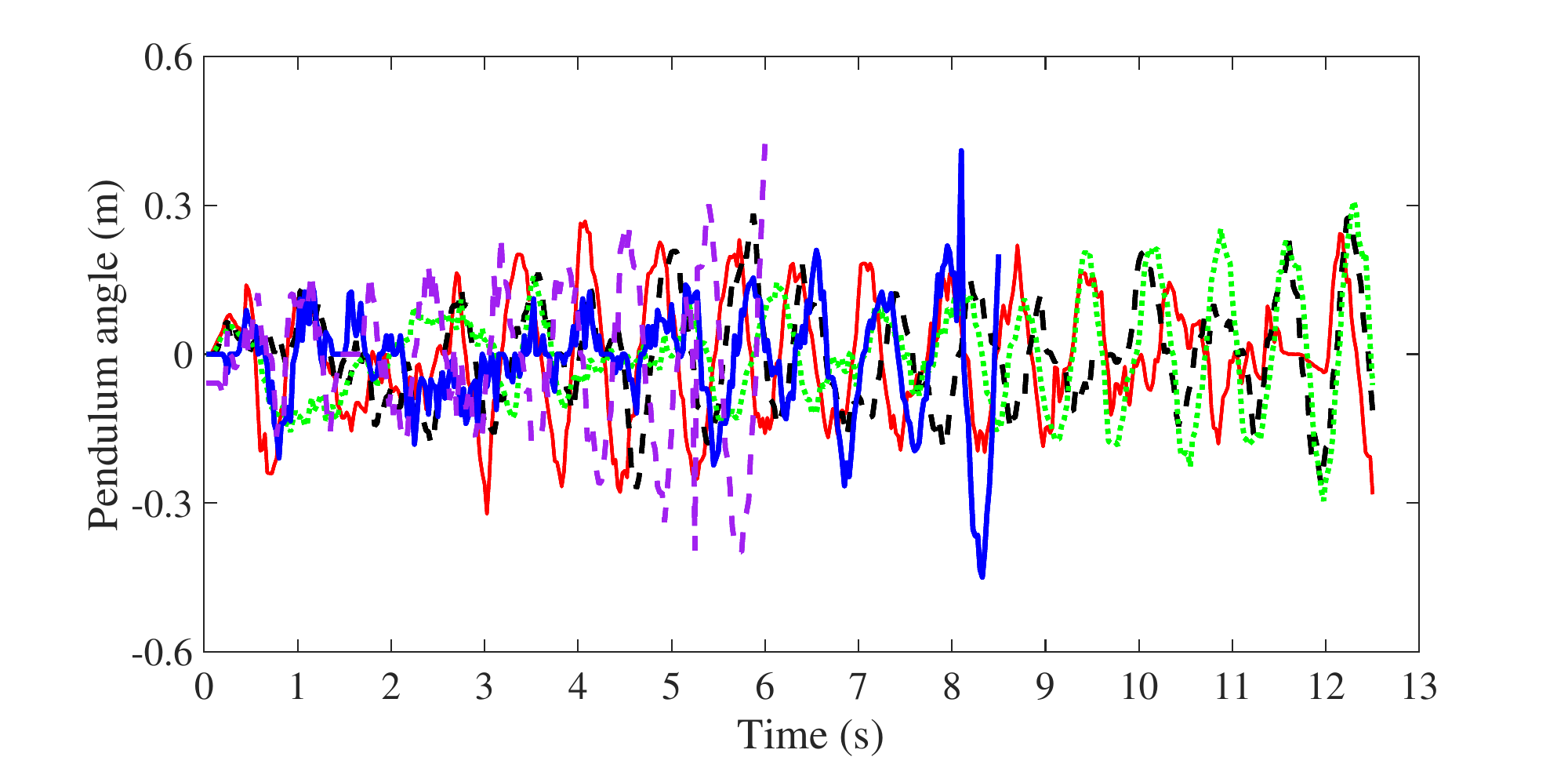}}
\caption{Cart position and pendulum angles with different intensity of salt and pepper attack: \textcolor[rgb]{1,0,0}{---} represents 0.1\% intensity, \textcolor[rgb]{0,0,0}{- -} represents 0.5\% intensity, \textcolor[rgb]{0,1,0}{$\cdot  \cdot  \cdot  \cdot  \cdot $} represents 0.9\% intensity, \textcolor[rgb]{0,0,1}{---} represents 1.0\% intensity and  \textcolor[rgb]{0.63,0.13,0.94}{- -} represents 1.1\% intensity.}
\label{figA15}
\end{figure}

\textbf{Influence of salt and pepper attack on the IPS performance}
Different intensity of salt and pepper attack (0.1$\%$, 0.5$\%$, 0.9$\%$, 1.0$\%$ and 1.1$\%$) are added. Real-time curves of cart position and pendulum angle are shown in Fig.~\ref{figA15}, where it can be seen that as intensity of salt and pepper attack increases, fluctuation of curve of cart positions increases, i.e., system stability decreases. Obviously, when intensity of salt and pepper attack goes to 1.0$\%$, cart position diverges and becomes uncontrollable. When number of salt and pepper attack goes to 1.1$\%$, cart curve diverges very fast.

Furthermore, the lower bounds ${\underline \Delta  _{1,a}}$, ${\underline \Delta  _{2,a}}$ and the upper bounds ${\overline \Delta  _{1,a}}$, ${\overline \Delta  _{2,a}}$ under different intensity of salt and pepper attack are shown in Table~\ref{tabAV}, where it can be seen that when IPS runs stably, range of ${\Delta _{1,a}}$ is $\left[ {-0.41,0.42} \right]$ and ${\Delta _{2,a}}$ is $\left[ {-0.78,0.76} \right]$. When IPS starts to diverge, ${\Delta _{1,a}}$ is larger than $\left| {0.48} \right|$ and ${\Delta _{2,a}}$ is larger than $\left| {0.75} \right|$. Hence, the designed controller can be able to achieve stable control on IPS to some extent and stability performance slightly deteriorates with gradually increasing intensity of salt and pepper attack.
\begin{table}[!t]
\caption{Values of ${\underline \Delta  _{1,a}}$, ${\underline \Delta  _{2,a}}$, ${\overline \Delta  _{1,a}}$ and ${\overline \Delta  _{2,a}}$ under Different Intensities of Salt and Pepper Attack.}
\label{tabAV}
\centering
\begin{tabular}{cccccc}
  \toprule
  Intensity & 0.1$\%$ & 0.5$\%$ & 0.9$\%$ & 1.0$\%$ & 1.1$\%$ \\
  \midrule
  ${\underline \Delta  _{1,a}}$ & -0.3 & -0.35 & -0.41 & -0.48 & -0.54 \\
  \specialrule{0em}{3pt}{3pt}
  ${\overline \Delta  _{1,a}}$ &  0.34 & 0.40 & 0.42 & 0.55 & 0.61 \\
  \midrule
  ${\underline \Delta  _{2,a}}$  & -0.63 & -0.67 & -0.78 & -1.02 & -1.24 \\
  \specialrule{0em}{3pt}{3pt}
 ${\overline \Delta  _{2,a}}$  & 0.50 & 0.59 & 0.67 & 0.75 & 1.40 \\
  \midrule
  Stability  & $\surd$ & $\surd$ & $\surd$ & $\times$ & $\times$ \\
  \bottomrule
  \multicolumn{6}{l}{ $\surd$ represents the IPS is stable, $\times$ represents the IPS is unstable}\\
\end{tabular}
\end{table}

Furthermore, to intuitively show performance of system under different salt and pepper attack, mean and SD of cart position and pendulum angle are calculated to show change of performance. The mean of cart positions displays extent of cart deviating from initial position and that of pendulum angles displays degree of pendulum deviating from vertical direction. SD of cart positions and pendulum angles displays extent of cart and pendulum deviating from mean, i.e., fluctuation degree of cart and pendulum. Mean and SD of cart positions and pendulum angles are illustrated in Table~\ref{tabAVI}. Table~\ref{tabAVI} shows that as intensity of salt and pepper attack increase, mean and SD of cart position and pendulum angle are all increase, which indicates the greater the intensity of salt and pepper attack is, the greater the influence of cart position and pendulum angle is. This experimental results show that salt and pepper attack has a significant influence on performance of IPS and an efficient image encryption algorithm is required to further improve image security.
\begin{table}[!t]
\caption{Mean and SD of the Cart Positions and Pendulum Angles under Different Intensity of Salt and Pepper Attack.}
\label{tabAVI}
\centering
\begin{tabular}{cccccc}
  \toprule
  Intensity & 0.1$\%$ & 0.5$\%$ & 0.9$\%$ & 1.0$\%$ & 1.1$\%$\\
  \midrule
   MCP$^1$(m) & -0.0119 & -0.0141 & \textbf{-0.0166} & -0.0176  & -0.0194 \\
   SCP$^2$(m) &  0.0259 & 0.0270 & \textbf{0.0280} & 0.0315 & 0.0322\\
  \midrule
   MPA$^3$(rad)  & 0.0062 & 0.0067 & \textbf{-0.0075} & -0.0126 & -0.0130 \\
   SPA$^4$(rad) & 0.1041 & 0.1044 & \textbf{0.1075} & 0.1200 & 0.1476\\
  \bottomrule
  \multicolumn{6}{l}{$^1$ Mean of cart position. $^2$ SD of cart position.}\\
  \multicolumn{6}{l}{$^3$ Mean of pendulum angle. $^4$ SD of pendulum angle.}\\
\end{tabular}
\end{table}

\begin{figure}[!t]
\centering
\subfigure[]{\includegraphics[width=0.46\textwidth]{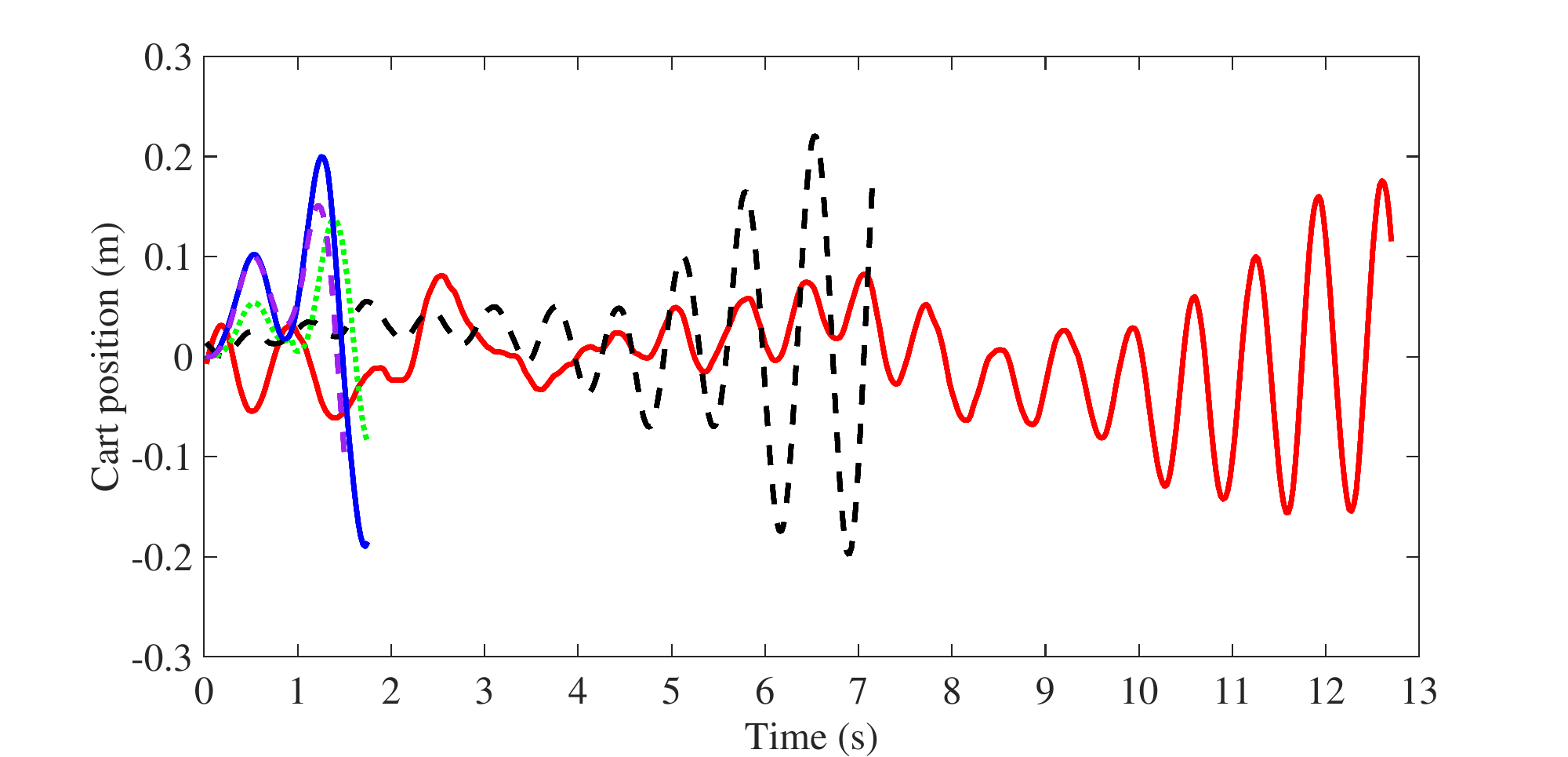}}\quad
\subfigure[]{\includegraphics[width=0.46\textwidth]{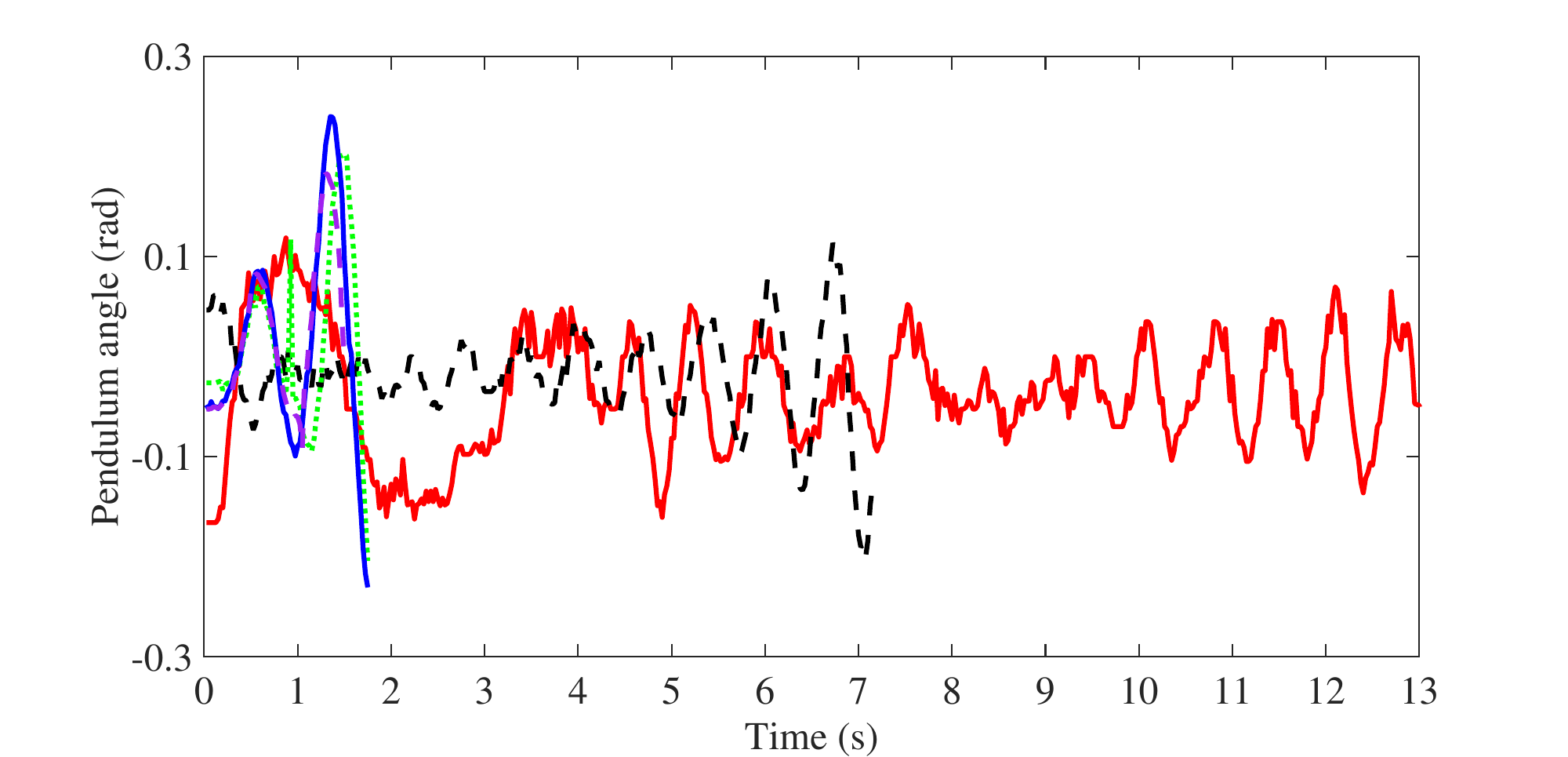}}
\caption{Cart position and pendulum angle under different Gaussian attack distribution:  \textcolor[rgb]{1,0,0}{---} represents ($\mu$=0, $\sigma$=1), \textcolor[rgb]{0,0,0}{- -} represents ($\mu$=0, $\sigma$=5), \textcolor[rgb]{0,1,0}{$\cdot  \cdot  \cdot  \cdot  \cdot $} represents ($\mu$=2, $\sigma$=2), \textcolor[rgb]{0,0,1}{---} represents ($\mu$=2, $\sigma$=5) and  \textcolor[rgb]{0.63,0.13,0.94}{- -} represents ($\mu$=5, $\sigma$=5).}
\label{figA16}
\end{figure}

\textbf{Influence of Gaussian attack on IPS performance}
In addition to adding salt and pepper attack and shearing attacks, different mean $\mu$ and variance $\sigma$  of Gaussian attack is also added. Real-time curves of cart positions and pendulum angles are shown in Fig.~\ref{figA16}. However, system is unstable no matter how small the Gaussian attack is. According to analysis in Fig.~\ref{figA16}, it can be seen that the maximum delays of image encryption and decryption in experiment is 0.016s, which means that the average maximum upper bound of image encryption or decryption is 0.008s. However, injection time of Gaussian attack is 0.001s and the actual bound of image encryption or decryption is 0.007s, so injection time of Gaussian attack is not primary reason. The primary reason is that Gaussian attack greatly affects pixel value of judging cart position, which leads to the bigger computational error. The uncertainties of cart position and pendulum angle are in Table~\ref{tabAIX}. Table~\ref{tabAIX} depicts that the smallest $\left| {{\Delta _{1,a}}} \right|$ and $\left| {{\Delta _{1,a}}} \right|$ are 0.52 and 0.88. Compared with salt and pepper attack and shear attack, the smallest $\left| {{\Delta _{1,a}}} \right|$ and $\left| {{\Delta _{1,a}}} \right|$ all exceed  maximum uncertainty that system can tolerate under image attacks. Hence, these experimental results show that Gaussian attack has a large influence on performance of IPS and an efficient image encryption algorithm is urgently required to further improve image security.
\begin{table}[!t]
\caption{Values of ${\underline \Delta  _{1,a}}$, ${\underline \Delta  _{2,a}}$, ${\overline \Delta  _{1,a}}$ and ${\overline \Delta  _{2,a}}$ under  Gaussian attacks with different distributions.}
\label{tabAIX}
\centering
\begin{tabular}{cccccc}
  \toprule
  $(\mu,\sigma)$  & (0,1) & (0,5) & (2,2) & (2,5) & (5,5)\\
  \midrule
   ${\underline \Delta  _{1,a}}$ & -0.50 & -0.81 & -1.04 & -1.38  & -1.96 \\
   \specialrule{0em}{3pt}{3pt}
   ${\overline \Delta  _{1,a}}$ &  0.52 & 0.79 & 0.92 & 1.46 & 2.05\\
  \midrule
   ${\underline \Delta  _{2,a}}$  & -0.73 & -0.92 & -1.34 & -1.96 & -2.35 \\
   \specialrule{0em}{3pt}{3pt}
   ${\overline \Delta  _{2,a}}$ & 0.88 & 1.13 &1.48 & 2.03 & 2.46\\
  \midrule
  Stability  & $\times$ & $\times$ & $\times$ & $\times$ & $\times$ \\
  \bottomrule
  \multicolumn{6}{l}{ $\times$ represents the IPS is unstable}\\
\end{tabular}
\end{table}

\section{Conclusion}
This paper has introduced secure control of NIPVSS with adverse effect of image computation. To meet real-time and security requirements, an F2SIE algorithm was proposed. Adverse effects of image information security were analysed, and the closed-loop model of NIPVSS has been established with computational errors and multiple time-varying delays. Furthermore, a robust controller was designed to tolerate computational errors and multiple time-varying delays, and system stability was examined. Experimental results confirms the effectiveness of the proposed image encryption algorithm and the controller on the NIPVSS platform. In the future, it is a challenge to further reduce image encryption algorithm time, and how to improve performance of encryption algorithm is very interesting and meaningful research venues.

\end{document}